\documentclass[11pt,a4paper]{article}
\pdfoutput=1
\usepackage{jheppub}
\usepackage{hyperref} 
\usepackage{xspace}
\usepackage[tight]{subfigure}
\usepackage{amsmath}
\usepackage{amssymb}
\usepackage{amsfonts}
\usepackage{mathrsfs}
\usepackage{comment}
\usepackage{afterpage}
\usepackage{booktabs}
\usepackage{array}

\usepackage[title,titletoc]{appendix}

\newcommand{\bea}{\begin{eqnarray}}
\newcommand{\eea}{\end{eqnarray}}
\newcommand{\beanon}{\begin{eqnarray*}}
\newcommand{\eeanon}{\end{eqnarray*}}
\newcommand{\ba}{\begin{array}}
\newcommand{\ea}{\end{array}}
\newcommand{\bd}{\begin{description}}
\newcommand{\ed}{\end{description}}
\newcommand{\bi}{\begin{itemize}}
\newcommand{\ei}{\end{itemize}}
\newcommand{\ben}{\begin{enumerate}}
\newcommand{\een}{\end{enumerate}}
\newcommand{\bc}{\begin{center}}
\newcommand{\ec}{\end{center}}

\newcommand{\ordEW}{\mathcal{O}(\alpha_{\scriptscriptstyle EM}^6)\xspace}
\newcommand{\ordQCD}{\mathcal{O}(\alpha_{\scriptscriptstyle EM}^4
  \alpha_{\scriptscriptstyle S}^2)\xspace}

\newcommand{\eqn}[1]{eq.(\ref{#1})}
\newcommand{\eqns}[2]{eqs.(\ref{#1}--\ref{#2})}

\newcommand{\tbn}[1]{tab.~\ref{#1}}
\newcommand{\Tbn}[1]{Tab.~\ref{#1}}

\newcommand{\fig}[1]{fig.~\ref{#1}}
\newcommand{\Fig}[1]{Fig.~\ref{#1}}

\newcommand{\figsc}[2]{figs.~\ref{#1},~\ref{#2}}
\newcommand{\sect}[1]{sect.~\ref{#1}}

\newcommand{\append}[1]{Appendix~\ref{#1}}

\newcommand{\rf}[1]{ref.~\cite{#1}}

\newcommand{\rfs}[1]{refs.~\cite{#1}}  

\newcommand{\Phantom}{{\tt PHANTOM}\xspace}

\newcommand{\al}{\alpha}

\newcommand{\hzero}{\ensuremath{h}} 
\newcommand{\Hzero}{\ensuremath{H}} 

\title{
$W$ boson polarization in vector boson scattering at the LHC
}

\author[a]{Alessandro Ballestrero,}
\author[a,b]{Ezio Maina}
\author[a,b]{and Giovanni Pelliccioli}

\affiliation[a]{INFN, Sezione di Torino,\\
Via Giuria 1, 10125 Torino, Italy}
\affiliation[b]{Dipartimento di Fisica, Universit\`a di Torino,\\
Via Giuria 1, 10125 Torino, Italy}

\emailAdd{ballestrero@to.infn.it}
\emailAdd{maina@to.infn.it}
\emailAdd{gpellicc@to.infn.it}

\abstract{
Measuring the scattering of longitudinally-polarized vector bosons represents a fundamental test of
ElectroWeak Symmetry Breaking.

In addition to the challenges provided by low rates and large backgrounds,
there are conceptual issues which need to be clarified for the definition of a suitable signal.
Since vector bosons are unstable and can only be observed through their decay products,
the polarization states interfere among themselves.
Moreover, already at tree level, there are diagrams which cannot be interpreted as production
times decay of EW bosons but are necessary for gauge invariance.

We discuss a possible way to define a cross section for polarized $W$'s, dropping all non resonant
diagrams, and projecting on shell the resonant ones, thus preserving gauge invariance.
In most cases, the sum of polarized distributions reproduces the full results.
In the absence of cuts, the ratios of the polarized cross sections to the
full one agree with the results of a
standard projection on Legendre polynomials. While the latter cannot be employed in a realistic environment,
a comparison of the data with the shapes of the angular distributions for polarized
vector bosons allows the extraction of the polarization fractions in the presence of selection
cuts on the charged leptons.
}

\begin{document}

\maketitle

\section{Introduction}
\label{sec:intro}

The discovery of a scalar particle \cite{Aad:2012tfa,Chatrchyan:2012ufa},
whose properties are compatible with those of the Higgs Boson \cite{Aad:2015zhl,Khachatryan:2016vau},
represents an
historic confirmation of the Standard Model description of ElectroWeak Symmetry Breaking (EWSB).
In the SM, vector bosons acquire mass through their coupling to the Higgs field.
At the same time, the Higgs is essential in the scattering of vector bosons (VBS),
avoiding the divergence at high energy, in the longitudinally-polarized sector,
which would be triggered by their very masses.
In the SM, the cross section for VBS processes is very small because of cancellations among different contributions.
Processes related to new physics can disturb this delicate balance and lead to
potentially large enhancements of the VBS rate, making it the ideal process for searches of deviations from the SM
and hints of New Physics
\cite{Duncan:1985vj,Dicus:1986jg,Kleiss:1987cj,Barger:1990py,Baur:1990xe,Dicus:1990fz,Dicus:1991im,
Bagger:1995mk,Iordanidis:1997vs,Butterworth:2002tt,Accomando:2005hz,Alboteanu:2008my,
Englert:2008tn,Ballestrero:2008gf,Ballestrero:2009vw,Ballestrero:2010vp,Dittmaier:2011ti,Dittmaier:2012vm,
Heinemeyer:2013tqa,deFlorian:2016spz}.

The experimental results on VBS obtained in Run 1 at the LHC have very low statistics.
In the near future, Run 2 will collect data with much better precision. Still, the effects
searched for are expected to be very small, therefore improvements of theoretical predictions and experimental
strategies will be required.

Understanding from a theoretical point of view how to separate the polarizations of the $W$'s in VBS is a necessary
prerequisite to any effort to perform the separation in the data. In particular, it would be useful to isolate the 
longitudinal component where deviations from the SM  are larger and easier to reveal.

\begin{figure}[!tbh]
\centering
\subfigure[{Resonant diagrams: signal.}\label{fig:VBSsgn_diag}]
{\qquad\includegraphics[scale=0.3]{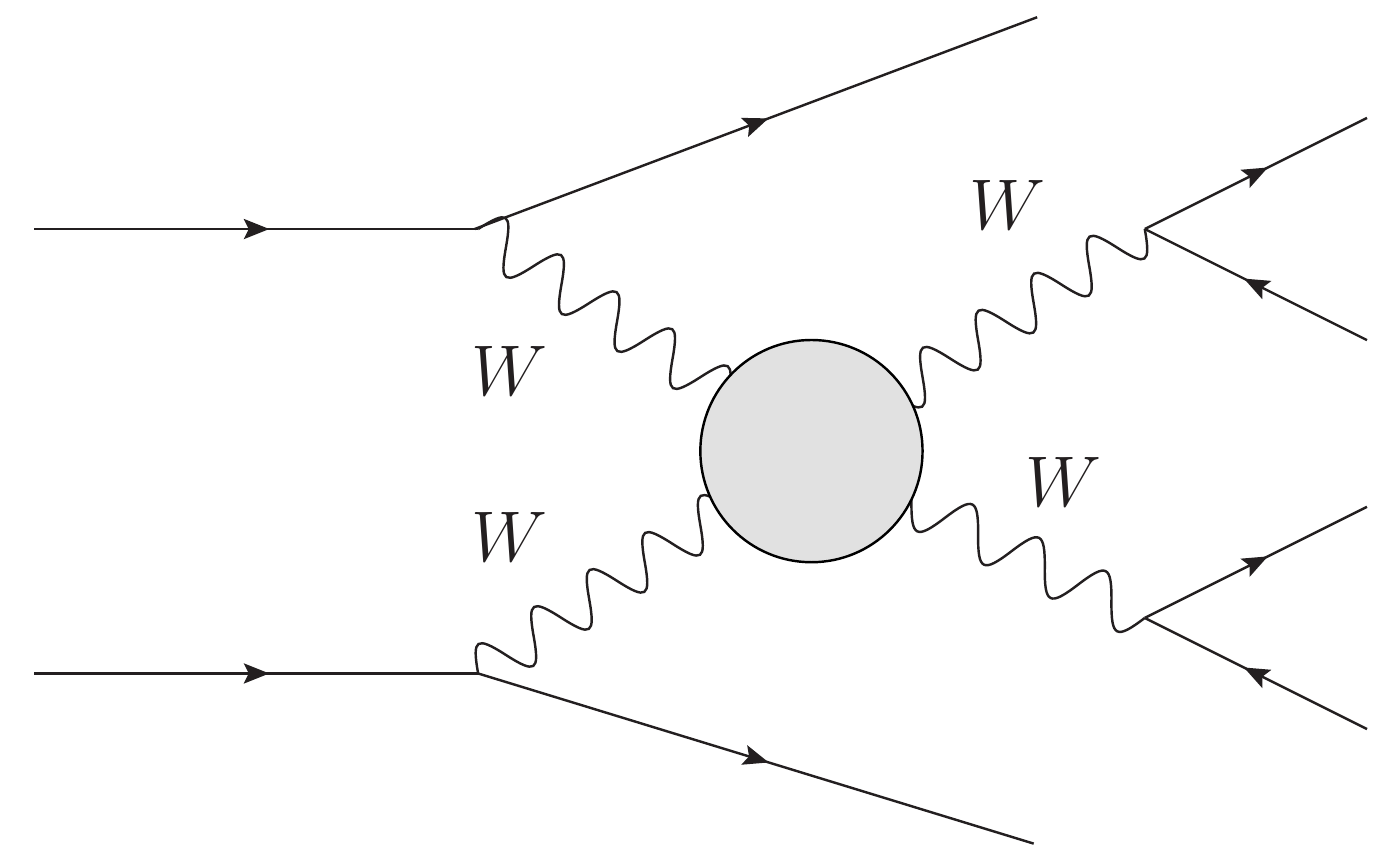}}
\qquad\qquad
\subfigure[{Resonant diagrams: irreducible background.}\label{fig:VBSbkgRes_diag}]
{\includegraphics[scale=0.27]{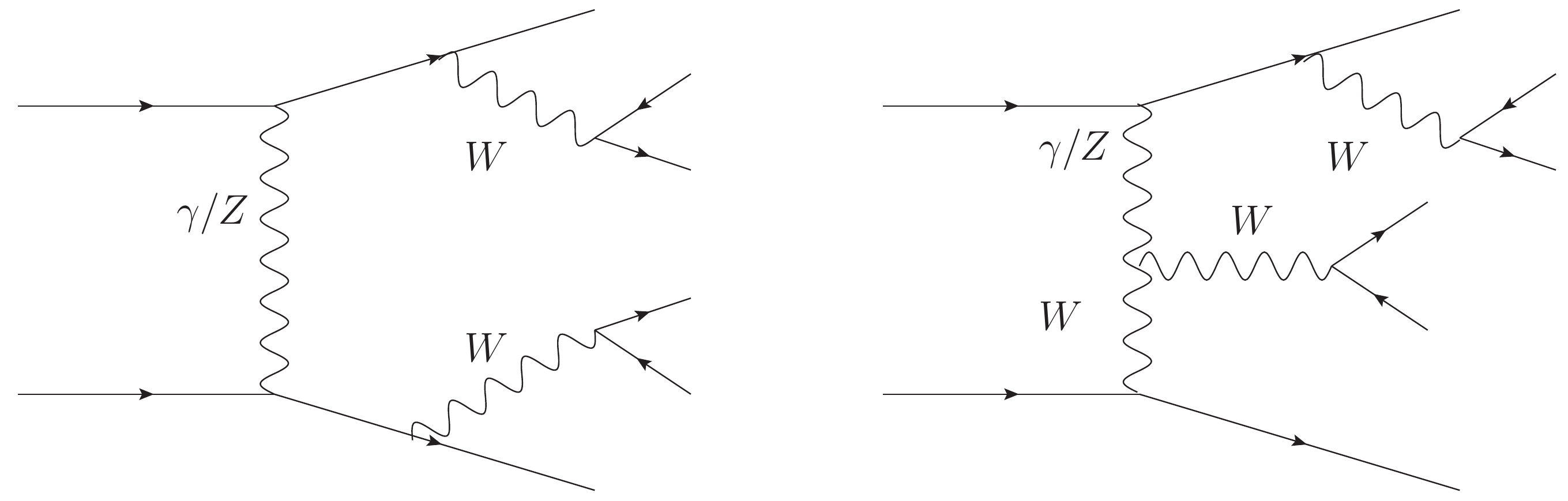}}\\
\subfigure[{Non resonant diagrams.}\label{fig:VBSbkgNonRes_diag}]
{\includegraphics[scale=0.28]{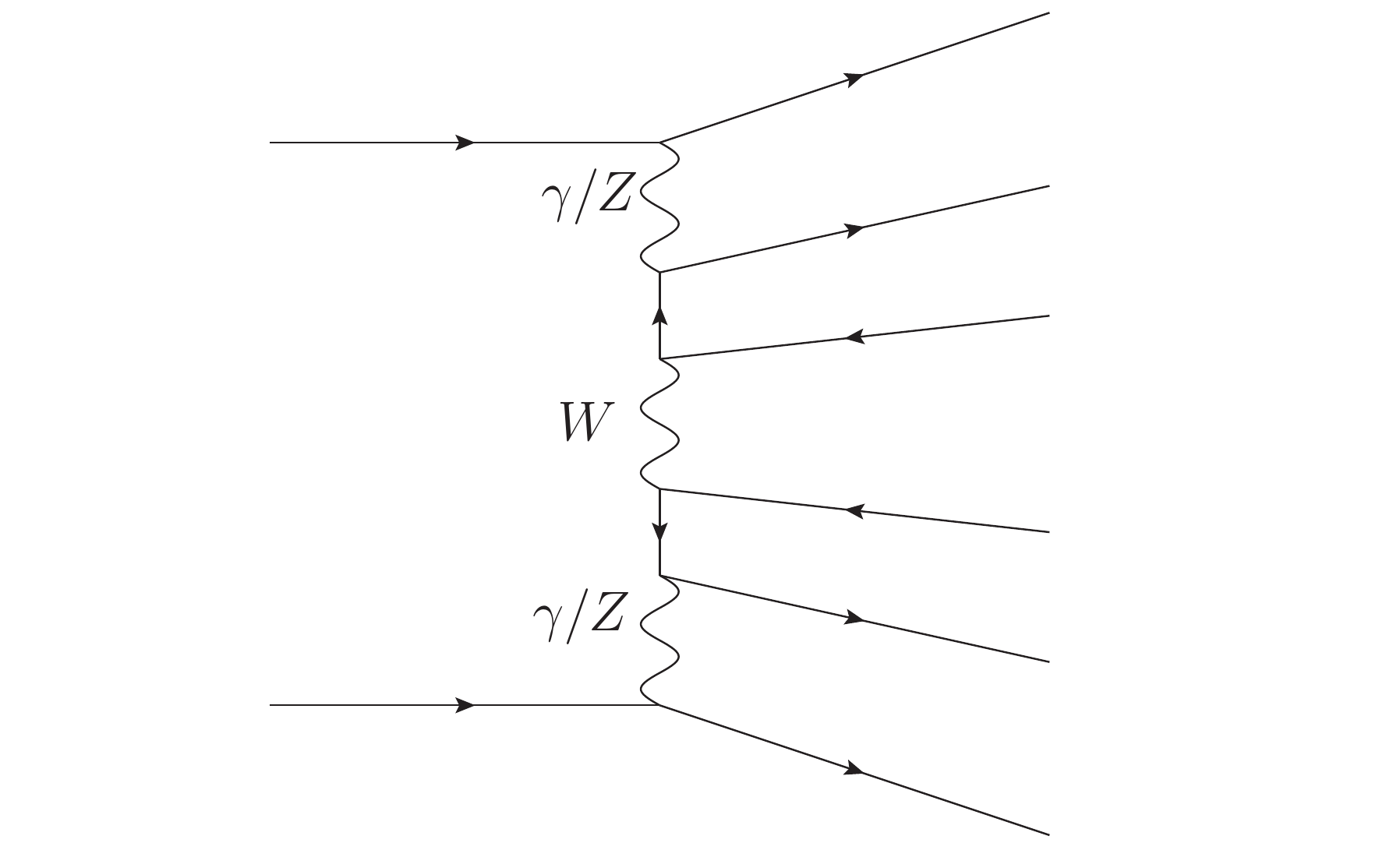}}
\caption{Representative diagrams for $WW$ scattering.}\label{fig:diag}
\end{figure}

The basic tool has been known for a long time:
each of the three on shell polarizations results in a specific decay distribution of the charged leptons.
Two obstructions, however, hamper progress along this path:

\begin{itemize}

\item Since the $W$'s are unstable particles, the decays of the indidual
polarizations interfere among themselves, even in the limit of narrow width. These interference contributions cancel
exactly only when an integration over the full azimuth of the lepton is performed. Acceptance cuts,
however, inhibit collecting data over the full angular range.
In addition, cuts affect differently the three polarizations and modify the angular distributions. 
Notice that the effects discussed here are present in every $W$ production channel, not only in VBS.

\item With few exceptions, ElectroWeak boson production processses are described by amplitudes
including non resonant diagrams, which cannot be interpreted as
production times decay of any vector boson, as shown in \fig{fig:VBSbkgNonRes_diag} for the VBS case.
These diagrams are essential for gauge invariance and cannot be ignored.
For them, separating polarizations is simply unfeasible. Furthermore, as it will be shown later, their contribution
is, in general, not negligible.

\end{itemize}

The polarization fractions in the SM for $W + jets$ processes, without cuts on the charged leptons, have been
discussed in \rf{Bern:2011ie}. The effects of selection cuts have been studied in \rf{Stirling:2012zt} for
$W + jets$ and a number of other $W$ production mechanisms. The analysis in \rf{Stirling:2012zt} showed how
the decay distributions of the charged leptons get distorted by the presence of cuts and how the simple methods
that allow to measure the polarization components when no cut on the charged leptons is imposed, fail
when cuts are introduced.
The interplay between interference among $W$ polarizations and selection cuts has been also examined in
\rf{Belyaev:2013nla}.

A number of measurements of the polarization fractions of the $W$ have been performed at the LHC, both by
CMS \cite{Chatrchyan:2011ig} and ATLAS \cite{ATLAS:2012au}, in the $W + jets$ channel. Both collaborations
have also
studied the polarization of the $W$'s \cite{Aaboud:2016hsq,Khachatryan:2016fky} in top--antitop events.

It should be mentioned that the charged lepton decay angles cannot be measured exactly
because of the difficulties in reconstructing the center of mass frame of the $W$.
Therefore, in practice, other, directly
observable, quantities are studies as proxies. Examples are $L_P$ \cite{Chatrchyan:2011ig},
$\cos\theta_{2D}$ \cite{ATLAS:2012au} and  $R_{p_T}$ \cite{Doroba:2012pd}, which is mostly useful for the
$W^+W^+$ channel.

In this paper, we discuss under which conditions it is possible to define VBS cross sections for polarized $W$'s, and
study their basic properties at the LHC. We also show how to overcome the difficulties related to the presence of 
acceptance cuts for the charged leptons.
We believe these preliminary steps to be essential for this kind of measurement.

We are far from being fully realistic. Nonetheless, our results pave the way to future phenomenological analyses.
We limit ourselves to the simple case of $W^+W^-$ production
in VBS when both vector bosons decay leptonically. We also consider only the case in which the two leptons
are an $e^-\mu^+$ pair, which avoids the small complication of the $Z$ contribution which appears when
the two leptons belong to the same family. A cut on the mass of the two leptons would suffice to completely
eliminate this channel.

While the leptonic decay of the $W$'s leads to a cleaner environment, the presence of two neutrinos makes the
estimate of the invariant mass of the boson boson system more involved.
Some of these difficulties might be alleviated studying the semileptonic channel instead of the fully leptonic one.
However, in the semileptonic case the QCD background is larger and  it is difficult to separate $WW$ from $WZ$
production. In addition, one of the two
vector bosons needs to be identified from its hadronic decay.

The present study could be easily extended to include NLO QCD corrections
\cite{Jager:2006zc,Jager:2013mu} since they do not modify the $W$ decay.
EW corrections, which have been recently calculated \cite{Biedermann:2016yds} for same sign $W$'s,
potentially mix the production and decay part of the amplitudes and will require an additional effort.

We are confident that if
separation between the different polarizations can be achieved in Monte Carlo simulations it will also be
within reach of the experiments, given sufficient luminosity.

The structure of the paper is the following: in the next section we recall how the polarizations of the $W$'s
enter
the amplitudes when the decay is taken into account exactly and how interferences between polarizations arise.
Next, we present the approximations which we propose in order to separate the different polarizations and
show how they
reproduce the full result in the absence of cuts on the leptons. In particular we discuss how the full
differential distributions are reproduced, in most
cases, by the sum of singly polarized distributions with the exception of those variables, like the charged lepton
transverse momentum, which constrain the available angular range for the decay.
In \sect {sec:cuts} we introduce acceptance leptonic cuts and discuss how they spoil the cancellation of the
interference contributions and modify the simple form of the decay angular distribution.
Finally, in \sect{sec:measuring} we show that the shapes of the angular 
decay distributions are sufficiently universal to
allow an almost model independent measurement of the polarization fractions. We extract the polarization
components in
the Higgsless model and in one instance of a Singlet extension of the SM, fitting the full distribution
with a sum of singly polarized SM shapes.

\section{$W$ boson polarization and angular distribution of its decay products}
\label{sec:Wpol_decay}
As already mentioned, a vector boson production tree level reaction, in general, receives contribution from different 
classes of diagrams, both resonant and non resonant. If we concentrate on hadronic processes 
at ${\mathcal{O}(\alpha_{\scriptscriptstyle EM}^2
\alpha_{\scriptscriptstyle S}^n)\xspace}$, involving a single intermediate $W^{+}$ which decays leptonically,
each diagram includes one ElectroWeak propagator of timelike momentum.
The amplitude can be written, in the Unitary Gauge, as

\begin{equation}\label{eq:Mlep}
\mathcal{M} = \mathcal{M}_{\mu} \frac{i}{k^2 - M^2 + i \Gamma M}\left(-g^{\mu\nu}+\frac{k^{\mu}
k^{\nu}}{M^2}\right)\left(\frac{-i\,g}{2\sqrt{2}}\bar{\psi}_{l} \gamma_{\nu}({1 - \gamma^5})\psi_{\nu_{l}}
\right)\,,
\end{equation}
where $M$ and $\Gamma$ are the $W$ mass and width, respectively.

The polarization tensor can be expressed in terms of four polarization vectors \cite{Kadeer:2005aq}
\begin{equation}
-g^{\mu\nu} + \frac{k^{\mu}k^{\nu}}{M^2} = \sum_{\lambda = 1}^4 \varepsilon^{\mu}_\lambda(k)
\varepsilon^{\nu^*}_{\lambda}(k)\,\,.
\end{equation}

In a frame in which the off shell $W$ boson propagates along the $z$-axis, with momentum $\kappa$,
energy $E$ and invariant mass $\sqrt{Q^2}=\sqrt{E^2-\kappa^2}$, the polarizations read:

\begin{align}
\varepsilon^{\mu}_{L} &= \frac{1}{\sqrt 2}(0, +1, -i, 0)\,\, \textrm{(left)} \,,\nonumber \\
\varepsilon^{\mu}_{R} &= \frac{1}{\sqrt 2}(0, -1, -i, 0)\,\, \textrm{ (right)} \,,\\
\varepsilon^{\mu}_{0} &= (\kappa,0,0,E)/\sqrt{Q^2}\,\, \textrm{(longitudinal)}\,, \nonumber \\
\varepsilon^{\mu}_{A} &= \sqrt{\frac{Q^2 - M^2}{Q^2\,M^2}}(E, 0, 0 ,\kappa)\,\, \textrm{(auxiliary)}\,.
\nonumber
\end{align}
The longitudinal and transverse polarizations obey the standard constraints
$\varepsilon_{i}\cdot k = 0$, $\varepsilon_{i}\cdot \varepsilon_{j}^* = -\delta_{i,j}$, $ i,j=0,L,R$.
The auxiliary polarization $\varepsilon^{\mu}_{A} $
satisfies $\varepsilon_{A}\cdot \varepsilon_{i}^* = 0$, $ i=0,L,R$,
$\varepsilon_{A}\cdot \varepsilon_{A}^* = (Q^2 - M^2)/M^2$,
$\varepsilon_{A}\cdot k = \sqrt{(Q^2 - M^2)\,Q^2/M^2}$.

On shell, the auxiliary polarization is zero and the longitudinal polarization reduces to the usual expression:
$\varepsilon^{\mu}_{0} = (\kappa,0,0,E)/M_W$. The most general case, in which the $W$ propagates along a
generic direction, can easily be obtained by a rotation.

The decay amplitudes of the $W$,
\begin{equation}\label{eq:Mdec}
\mathcal{M^D}_\lambda =\frac{-i\,g}{2\sqrt{2}}\bar{\psi}_{l} \varepsilon^{\mu *}_{\lambda} \gamma_{\mu}
({1 - \gamma^5}) \psi_{\nu_{l}},
\end{equation}
 depend on its polarization.
In the rest frame of the $\ell\nu$ pair, they are:
\begin{equation}\label{eq:longamp}
\mathcal{M^D}_0 = ig\,\sqrt{2}E \,\sin\theta \,\,,
\end{equation}
\begin{equation}\label{eq:transvamp}
\mathcal{M^D}_{R/L} = ig\,E \,(1 \pm \cos\theta)e^{\pm i\phi} \,\,,
\end{equation}
 where $(\theta, \phi)$ are the charged lepton polar and azimuthal angles, respectively,  relative to the 
 boson direction in the laboratory frame.
 The decay amplitude for
 the auxiliary polarization is zero, for massless leptons, because $\varepsilon^{\mu}_{A}$ is proportional to
 the four--momentum of the virtual boson.
 Hence, each physical polarization is uniquely associated with a specific angular distribution of the charged
 lepton, even when the $W$ boson is off mass shell.

Defining a polarized production amplitude,
\begin{equation}\label{eq:Mprod}
\mathcal{M^P}_{\lambda} = \mathcal{M}_{\mu} \varepsilon^{\mu}_{\lambda}\,,
\end{equation}
the full amplitude can be written as:
\begin{equation}\label{eq:Msum}
\mathcal{M} = \sum_{\lambda = 1}^3\mathcal{M^P}_{\lambda} \,\, \frac{i}
{k^2 - M^2 + i \Gamma_w M} \,\mathcal{M^D}_\lambda = \sum_{\lambda = 1}^3
\mathcal{M^F}_\lambda\,,
\end{equation}
where $\mathcal{M^F}_\lambda$ is the full amplitude with a single polarization for the intermediate $W$.
Notice that in each $\mathcal{M^F}_\lambda$ all correlations between production and decay are exact.

The squared amplitude becomes:
\begin{equation}\label{eq:interfpol}
\underbrace{\left|\mathcal{M}\right|^2}_{\textrm{coherent sum}} = \underbrace{\sum_{\lambda}\left|
\mathcal{M^F}_{\lambda}\right|^2}_{\textrm{incoherent sum}} + \underbrace{\sum_{\lambda \neq \lambda'}
\mathcal{M^F}_{\lambda}^{ *}\mathcal{M^F}_{\lambda'}}_{\textrm{interference terms}}\,.
\end{equation}

The interference terms in \eqn{eq:interfpol} are not, in general, zero. They cancel only when the squared
amplitude
is integrated over the full range of the angle $\phi$, or,
equivalently, when the charged lepton can be observed for any value of $\phi$. 
This remains true in the Narrow Width
Approximation in which $1/((k^2 - M^2)^2 + \Gamma_w^2 M^2)$ is replaced by
$\pi\,\delta(k^2 - M^2)/(\Gamma M)$. With this substitution, the integration over the invariant mass of the
 intermediate state becomes trivial, but the angular integration is unaffected.

If we denote by $d\sigma(\theta,\phi,X)/d\,Lips$ the fully differential cross section,
where $\theta,\,\phi$ are the
$W$ decay  variables in the boson rest frame and $X$ stands for all additional phase space variables,
by $d\sigma(\theta,X)/ d\cos\theta/dX$ its integral over $\phi$,
\begin{equation}\label{eq:dsigmahat}
\frac{d\sigma(\theta,X)}{d\cos\theta\, d X} = \int d\phi \,\, \frac{ d \sigma(\theta,\phi,X)}{dLips}\,,
\end{equation}
and by $d\sigma(X)/d X$ the integral of $d\sigma(\theta,X)/d\cos\theta/d X$ over $\cos\theta$,
\begin{equation}\label{eq:sigmahat}
\frac{d \sigma(X)}{d X} = \int d \cos\theta \,\, \frac{d\sigma(\theta,X)}{d\cos\theta\, d X}\,,
\end{equation}

one can write, using \eqn{eq:longamp} and \eqn{eq:transvamp},
\begin{equation}
\frac{1}{\frac{d\sigma(X)}{d X}} \,\,\frac{d\sigma(\theta,X)}{d\cos\theta\, d X}
\ =\ \frac{3}{8} (1 \mp \cos\theta)^2 \,f_L(X)
+ \frac{3}{8} (1 \pm \cos\theta)^2 \,f_R(X)
+ \frac{3}{4} \sin^2\theta \, f_0(X)\,,
\label{eq:dcdist}
\end{equation}

where the upper sign is for $W^+$ and the lower sign for $W^-$. In general the $f_i$ depend
on the variables $X$ which are not integrated over.
The normalizations are chosen so that
\begin{eqnarray}
\frac{1}{\frac{d\sigma(X)}{d X}}  \,\,\int_{-1}^{1} d\cos\theta\,\, \frac{d\sigma(\theta,X)}{d\cos\theta\, d X}
= f_L + f_0 + f_R = 1 \,
\end{eqnarray}
and $f_L,\, f_0$ and $f_R$ represent the left, longitudinal and right polarization fractions, respectively.
The steps which lead to \eqn{eq:dcdist} can be repeated even when a partial or complete integration over the $X$
variables is performed.

If \eqn{eq:dcdist} holds, the polarized components can be extracted from the differential
angular distribution by a projection on the first three Legendre polinomials:
\begin{equation}
\frac{1}{\sigma} \frac{d\sigma}{d\cos\theta}
= \sum_{l=0}^2 \alpha_l P_l(\cos\theta) \, ,\qquad
\alpha_l = \frac{2l+1}{2} \, \int_{-1}^{1} d\cos\theta \,\frac{1}{\sigma}
\frac{d\sigma}{d\cos\theta} P_l (\cos\theta) \,.
\label{eq:LegendreExpansion}
\end{equation}
This procedure is completely equivalent to the method used in \rfs{Bern:2011ie,Stirling:2012zt} based on the
calculation of the first few moments of the angular distribution.

The polarization fractions $f_0,f_L,f_R$ can be obtained as:
\begin{align}\label{eq:inversion_eq}
f_0 &= \frac{2}{3}(\alpha_0 - 2\alpha_2)\,,\nonumber\\
f_L &= \frac{2}{3}(\alpha_0 \mp \alpha_1 + \alpha_2)\,,\nonumber\\
f_R &= \frac{2}{3}(\alpha_0 \pm \alpha_1 + \alpha_2)\,,
\end{align}
where the upper/lower sign refers to the $W^+$/$W^-$.
We note that the sum $f_0 + f_R +f_L = 2\alpha_0$ is bound to be one.

A word of caution is necessary when acceptance cuts are imposed on the charged leptons, as is unavoidable
in practice.
While the cancellation of the interference terms in \eqn{eq:interfpol} is a necessary condition for the validity
of \eqn{eq:dcdist}, this is by no means sufficient.
A generic cut, think of the lepton transverse momentum, will depend on both the angular variables and the variables 
$X$. Integrating over $X$, in the presence of cuts, results in a different  theta dependence.
As a consequence, the measured lepton decay distribution is not described any more by the simple formula
\eqn{eq:dcdist} and the polarization fractions cannot be computed as in \eqn{eq:inversion_eq}.

In order to separate the polarized components in the data, it is necessary to
compute the individual amplitudes
$\mathcal{M^F}_{\lambda}$ in \eqn{eq:Msum}, which requires making the substitution
\begin{equation}
\sum_{\lambda'} \varepsilon_{\lambda'}^{\mu}\varepsilon_{\lambda'}^{\nu *}\,\,\,\rightarrow\,\,\,
\varepsilon_{\lambda}^{\mu}\varepsilon_{\lambda}^{\nu *}\,\,.
\label{eq:substitution}
\end{equation}
in the $W$ propagator. For the present analysis, this possibility has been introduced in
\Phantom\cite{Ballestrero:2007xq}.

\section{Separating the resonant contribution: On Shell Projection}
\label{sec:resonant}
After our discussion of $W$ boson polarization in processes with a single $W$, we now turn to reactions which 
contain non resonant diagrams, in particular to those in which
two lepton pairs, $e^-\bar{\nu_e}$ and $\mu^+\nu_\mu$, are produced. For the set of diagrams in which each
leptonic line is connected to a single intermediate $W$, as in \figsc{fig:VBSsgn_diag}{fig:VBSbkgRes_diag},
which we will call doubly resonant or just resonant for short, one can proceed as in the previous section.
However, there are many diagrams, like the one shown in \fig{fig:VBSbkgNonRes_diag},  which cannot be expanded
in a similar fashion. 
The only way to proceed, in order to define amplitudes with definite $W$ polarization,
is to devise an approximation to the full result that only involves doubly
resonant diagrams.

There are, obviously, several conceivable approximations. 
The simplest one is simply to drop all non resonant diagrams,
possibly restricting the mass of the lepton pair in the decay to lie close to the $W$ mass.  Since this procedure 
clearly violates gauge invariance, 
one expects it would produce distinctly wrong cross sections, at least in particular regions of
phase space.
For this reason we have chosen an On Shell Projection (OSP) method, which is more commonly known as the pole scheme or pole approximation in the literature. 
For instance, it has been employed for
the calculation of EW radiative corrections to $W^+W^-$ production in
refs.\cite{Aeppli:1993cb,Aeppli:1993rs,Denner:2000bj,Billoni:2013aba,Biedermann:2016guo}.
This can be realized as a completely gauge invariant approximation.

The procedure can be summarized as follows: 
\begin{align}
\mathcal{M} &= \mathcal{M}_{\textrm{res}} + \mathcal{M}_{\textrm{nonres}} = \nonumber \\
&\frac{\sum_{\lambda_1,\lambda_2} \mathcal{M}^{\mathcal{P}}_{\mu\nu}(k_1,k_2,X)\,
\varepsilon_{\lambda_1}^{\mu}(k_1)\,\varepsilon_{\lambda_2}^{\nu}(k_2)\,
\varepsilon_{\lambda_1}^{*\alpha}(k_1)\,\varepsilon_{\lambda_2}^{*\beta}(k_2)\,
\mathcal{M}^{\mathcal{D}}_{\alpha}(k_1,X_1) \, \mathcal{M}^{\mathcal{D}}_{\beta}(k_2,X_2)}
{(k_1^2-M_W^2 + i\Gamma_WM_W)(k_2^2-M_W^2 + i\Gamma_WM_W)} + \mathcal{M}_{\textrm{nonres}}
 \nonumber \\
&\rightarrow \frac{\sum_{\lambda_1,\lambda_2} 
\mathcal{M}^{\mathcal{P}}_{\mu\nu}(\overline{k}_1,\overline{k}_2,X)\,
\varepsilon_{\lambda_1}^{\mu}(\overline{k}_1)\,\varepsilon_{\lambda_2}^{\nu}(\overline{k}_2)\,
\varepsilon_{\lambda_1}^{*\alpha}(\overline{k}_1)\,\varepsilon_{\lambda_2}^{*\beta}(\overline{k}_2)\,
\mathcal{M}^{\mathcal{D}}_{\alpha}(\overline{k}_1,\overline{X}_1) \, 
\mathcal{M}^{\mathcal{D}}_{\beta}(\overline{k}_2,\overline{X}_2)}
{(k_1^2-M_W^2 + i\Gamma_WM_W)(k_2^2-M_W^2 + i\Gamma_WM_W)} \nonumber \\
&=\frac{\sum_{\lambda_1,\lambda_2} 
\mathcal{M}^{\mathcal{P}}_{\lambda_1,\lambda_2}(\overline{k}_1,\overline{k}_2,X)
\mathcal{M}^{\mathcal{D}}_{\lambda_1}(\overline{k}_1,\overline{X}_1) 
\mathcal{M}^{\mathcal{D}}_{\lambda2}(\overline{k}_2,\overline{X}_2)} 
{(k_1^2-M_W^2 + i\Gamma_WM_W)(k_2^2-M_W^2 + i\Gamma_WM_W)} =  \mathcal{M}_{\textrm{OSP}}\,,
\end{align}

where $\overline{k}_i$ is the on mass shell projection of $k_i$. Here $X_i,\,i=1,2$ stand for the lepton 
momenta, while $X$ refer to the initial and final quark momenta.  $\overline{X}_i$ refer to the 
lepton momenta after the projection,
when the momentum of each $\ell\nu$ pair in the calculation of $\mathcal{M}_{\textrm{res}}$ is on the $W$
mass shell  momentum.  $\mathcal{M}_{\textrm{nonres}}$, which includes all singly
resonant and non resonant diagrams, is dropped.
The denominator in each $W$ propagator is left untouched.

The projected production and decay amplitudes, 
\begin{align}
\mathcal{M}^{\mathcal{P}}_{\lambda_1,\lambda_2}(\overline{k}_1,\overline{k}_2,X) 
&=\mathcal{M}^{\mathcal{P}}_{\mu\nu}(\overline{k}_1,\overline{k}_2,X)\,
\varepsilon_{\lambda_1}^{\mu}(\overline{k}_1)\,\varepsilon_{\lambda_2}^{\nu}(\overline{k}_2)\, , \\
\mathcal{M}^{\mathcal{D}}_{\lambda_1}(\overline{k}_1,\overline{X}_1) &=
\varepsilon_{\lambda_1}^{*\alpha}(\overline{k}_1)\,
\mathcal{M}^{\mathcal{D}}_{\alpha}(\overline{k}_1,\overline{X}_1)
\end{align}

are ordinary, complete amplitudes with polarized, on shell,  external $W$ bosons. Therefore, the final
expression is gauge invariant provided $\mathcal{M}^{\mathcal{P}}$ and $\mathcal{M}^{\mathcal{D}}$
are both invariant.

However,
this projection is not uniquely defined. 
As the momentum of a vector boson is sent to mass shell at least  the momenta of its two decay products
need to be adjusted.
One can keep fixed the direction of particle one or of particle two in the overall center of mass. 
Alternatively one can keep fixed the
angles in the center of mass of the vector boson. These three procedures lead to different momenta of the 
decay products and therefore to
different matrix elements.

In order to have an unambiguous
prescription we have chosen to conserve:
\begin{enumerate}
\item the total four--momentum of the $WW$ system (thus, also $M_{WW}$ is conserved);
\item the direction of the two $W$ bosons in the $WW$ center of mass frame;
\item the angles of each charged lepton, in the corresponding $W$ center of mass frame, relative to the boson
direction in the lab.
\end{enumerate}

Since the original $W$ momenta
are typically only slightly off shell, the modification of the kinematics is expected, in most cases, to be minimal.
The modified momenta affect only the calculation of the weight of the event. In the LHA event file
\cite{Alwall:2006yp}
all particles are assigned the original, unprojected momenta.
This procedure can only be applied for $M_{2\ell 2\nu} >2M_W$.
In the following we will refer to the total invariant mass of the four leptons as $M_{WW}$, for simplicity.

The OSP requires a further adjustment in the computation of the projected amplitudes.
In the full calculation,
the presence of unstable, intermediate, timelike $W$ and $Z$ bosons, forces the introduction of an imaginary
part, typically $- i \Gamma M$ in tree level processes, into their
propagators. This corresponds to the partial resummation of a particular class of higher order contributions,
effectively
mixing different perturbative orders, and has the collateral effect of complicating the issue of gauge invariance
\cite{Argyres:1995ym,Beenakker:1996kn,Accomando:1999zq}.

When massive vector bosons appear only as virtual states, a simple and effective way of preserving gauge 
invariance is the
Complex Mass Scheme (CMSc) \cite{Denner:2000bj,Denner:2005fg}. 
In the CMSc, all occurrences of the vector boson mass $M$ are replaced by $\sqrt{M^2- i \Gamma M}$.
This includes the cosine of the Weinberg mixing angle and all quantities which are defined in 
terms of $\cos\theta_W$.

The CMSc, however, is not gauge invariant for amplitudes with external Weak gauge bosons,
which are implicitely regarded as stable. The simplest process which exemplifies these issues is
$ e^+ \nu_e \rightarrow W^+ \gamma$. If the polarization vector of the photon is substituted by its momentum,
the amplitude should become zero, because of the electromagnetic Ward identity.
It is easily verified that this happens only if $\Gamma_W = 0$.
In the end, gauge invariance forces all widths of intermediate vector bosons in the
OSP amplitudes to be set to zero. As a consequence all weak bosons whose
momentum can get on their mass shell must be projected on shell simultaneously.

In \append{appendixa} we discuss, in a more quantitative and detailed way, the pitfalls of the non fully
gauge invariant approximations mentioned in this section.  In the Appendix we focus on the high transverse 
momentum, high mass region where gauge violating effects are expected to be magnified, particularly when
longitudinally polarized vector bosons are involved.

All results, in the following, which are labeled as full have been obtained in the CMSc.
All OSP results have been obtained projecting on mass shell all resonant vector bosons, setting to zero all
widths in non resonant propagators and using real couplings.

\section{Setup of the simulations}
\label{sec:setup}

We have studied $p p \rightarrow j j e^-\bar{\nu}_e \mu^+ \nu_\mu$ at parton level.
All events have been generated with \Phantom \citep{Ballestrero:2007xq}, using the 
NNPDF30\_lo\_as\_0130 PDF set
\cite{Ball:2014uwa} with scale $Q =M_{WW}/\sqrt{2}$.
We consider only ElectroWeak processes at $\ordEW$. We neglect $\ordQCD$ processes and exclude the 
top--antitop background assuming  perfect $b$ quark veto.

All results shown in this paper refer to the LHC@13TeV and have been obtained with the 
following set of standard
cuts for the hadronic part:
\begin{itemize}
\item maximum jet pseudorapidity, $|\eta_j| < 5$;
\item minimum jet transverse momentum, $ p_t^j > 20$ GeV;
\item minimum jet--jet invariant mass, $M_{jj} > 600$ GeV;
\item minimum jet--jet pseudorapidity separation, $|\Delta\eta_{jj}| > 3.6$;
\item opposite sign jet pseudorapidities, $\eta_{j_1}\cdot \eta_{j_2} < 0$.
\end{itemize}

The OSP requires the mass of the four lepton system to be larger than twice the mass of the $W$, thus,
only events with $M_{WW} > 300$ GeV have been retained.
Since we are mainly interested in VBS at large invariant masses, this is not a limitation.

\begin{figure}[!tb]
\centering
\subfigure[{$\cos\theta_e$}\label{fig:distrib_thetal_nocut}]
{\includegraphics[scale=0.37]{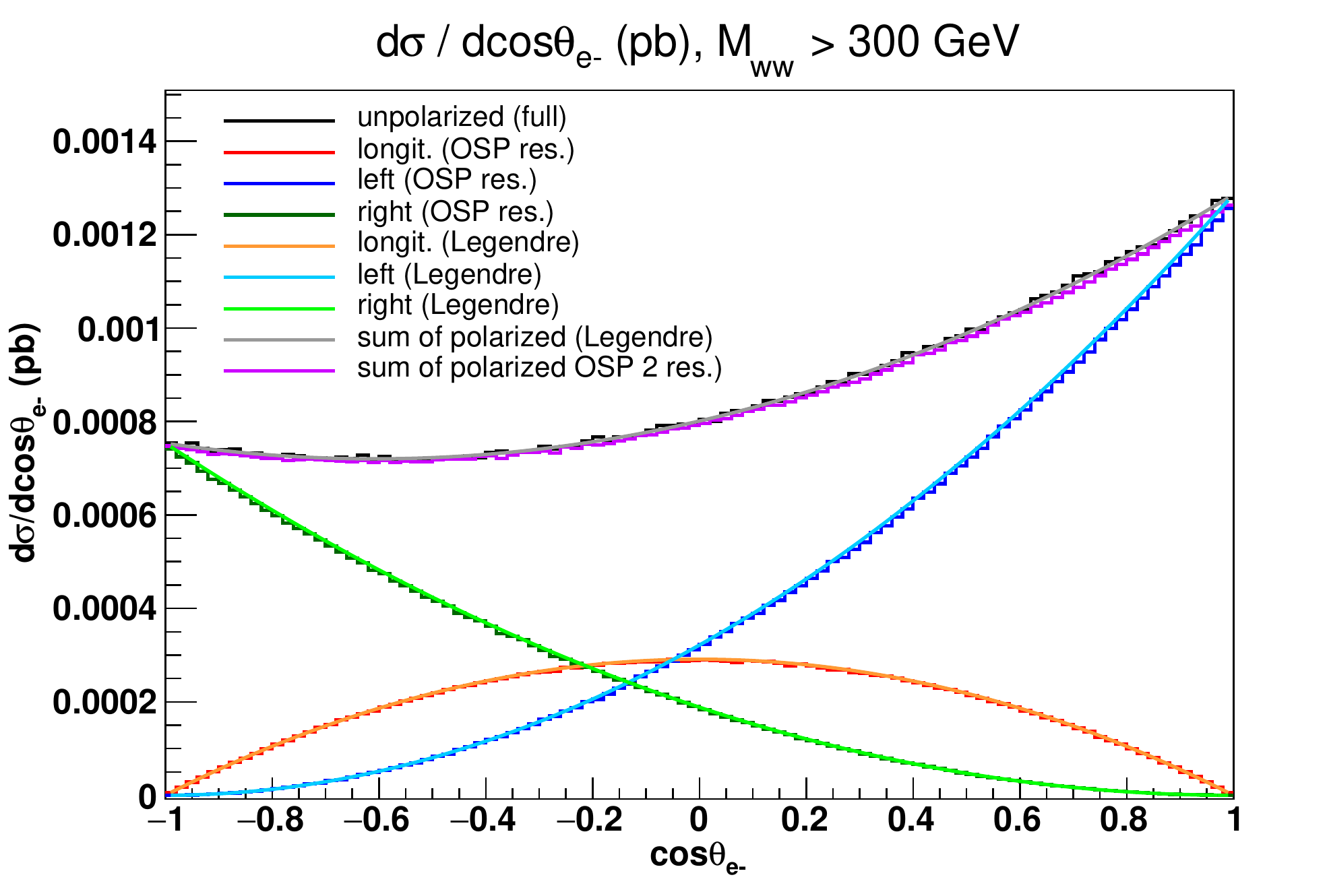}}
\subfigure[{Polarization fractions}\label{fig:polfrac_MWW_nocut}]
{\includegraphics[scale=0.37]{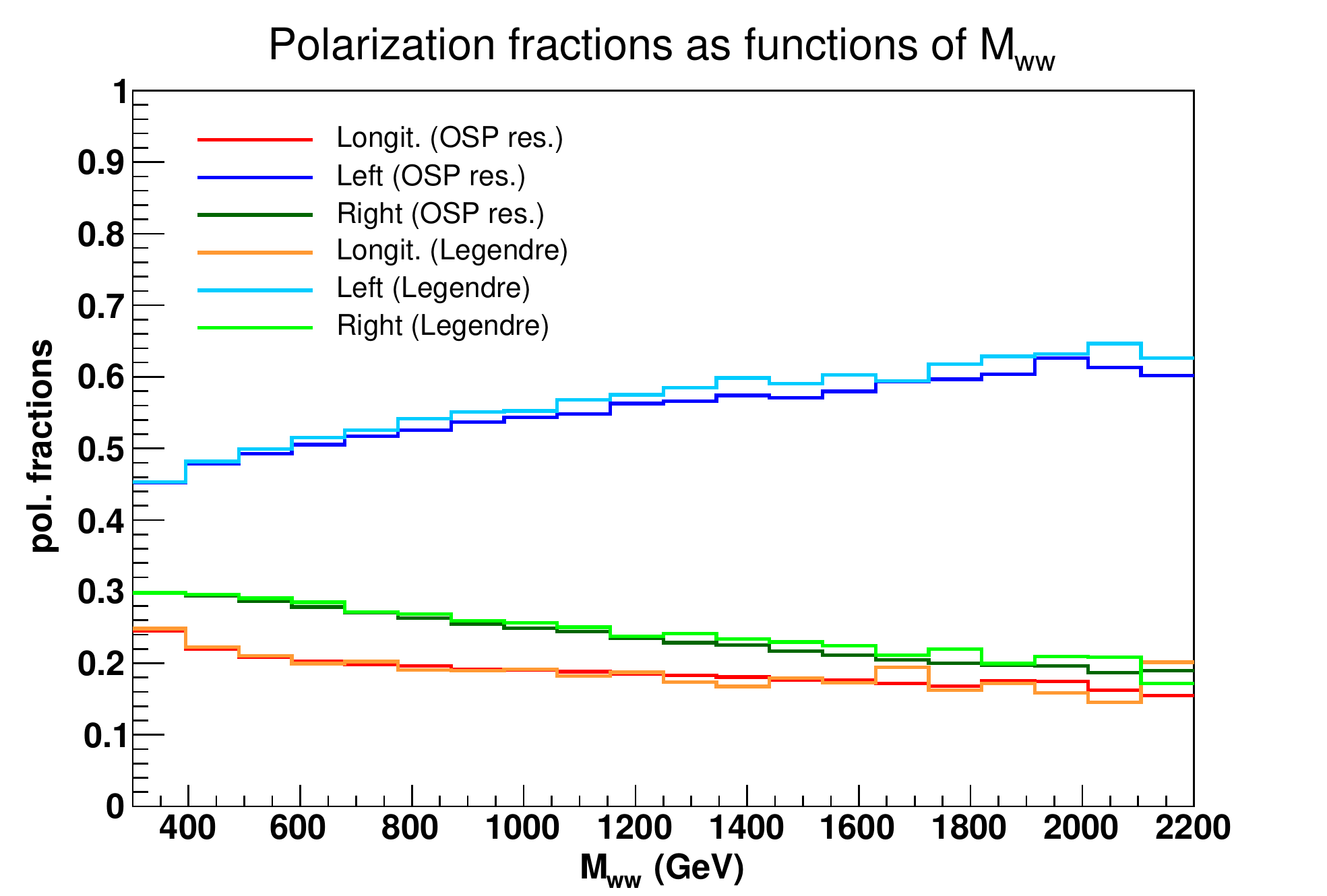}}
\caption{Distribution of electron $\cos\theta$ in the $W^-$ reference frame (left), polarization fractions as functions
of $M_{WW}$ (right).
The polarization components obtained by expanding the full angular distribution on Legendre polynomials are 
shown in lighter colors. The darker histograms are obtained integrating the polarized amplitudes squared.
The positively charged $W$ is unpolarized. }
\label{fig:distrib_thetal_polfrac_nocut}
\end{figure}

\section{Validating the approximation in the absence of cuts on the charged leptons}
\label{sec:results}

In this section we compare the differential distributions of a number of kinematic variables obtained from
the full matrix element with the incoherent sum of three OSP distributions in which the negatively charged
intermediate $W$ boson is polarized while the positively charged one remains unpolarized. Only doubly
resonant diagrams are included for polarized processes. No lepton cut is applied and, as a consequence,
the cross section from the incoherent sum of polarized cross sections coincides with the cross section obtained from
the coherent sum.

Our aim is:
\begin{itemize}
\item to demonstrate that the $W$ polarization fractions obtained with the OSP are in agreement with those 
obtained from a standard expansion in Legendre polynomials.
\item to show that the OSP reproduces well the full cross section and distributions. 
\item to explore the differences between the individual polarized distributions as a tool to separate them in the
data.
\end{itemize}

\afterpage{\clearpage}
\begin{figure}[p]
\vspace*{-2cm}
\centering
\subfigure[{$WW$ invariant mass}\label{fig:Mww_canvas_nolepcut}]
{\includegraphics[scale=0.37]{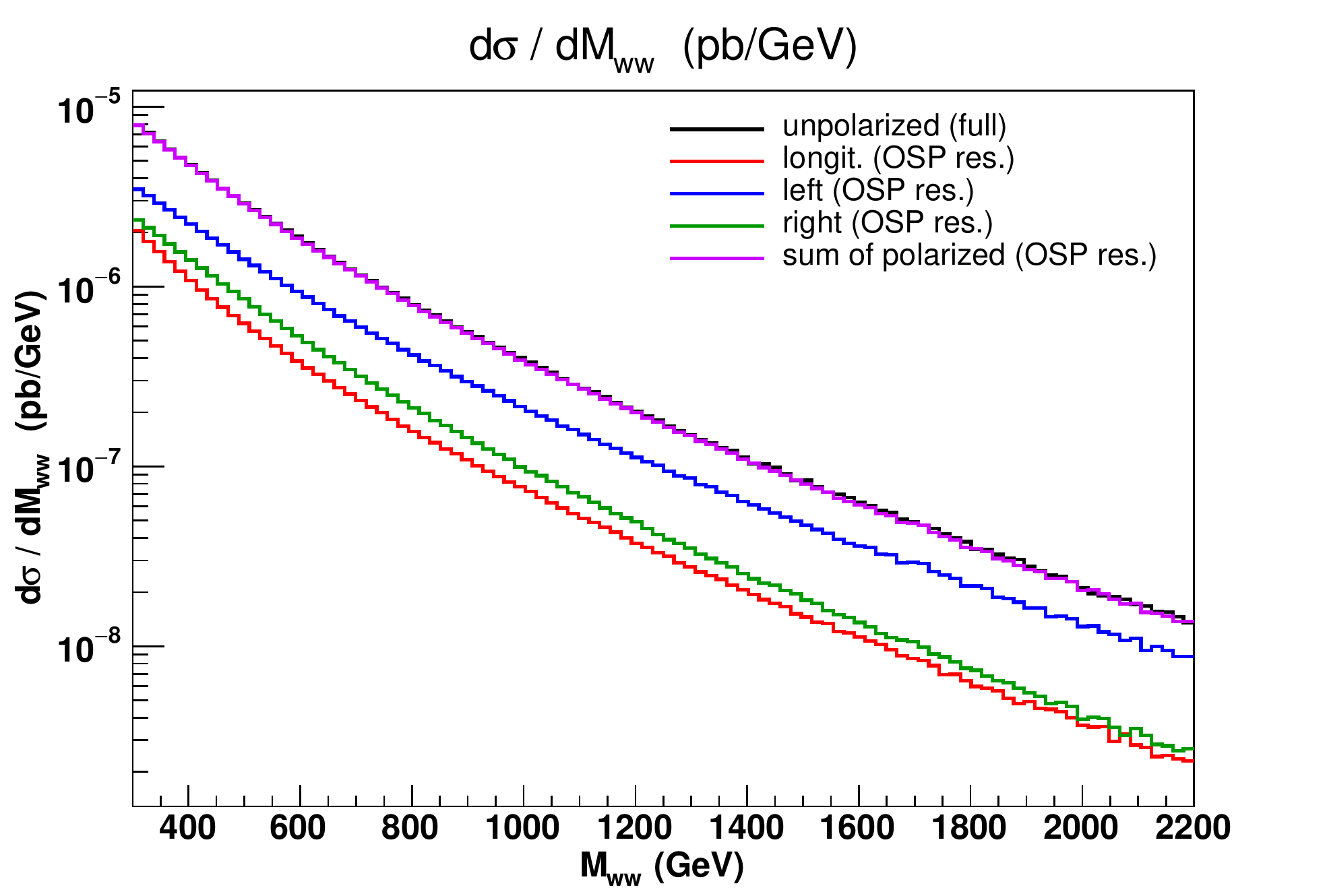}}
\subfigure[{$jj$ invariant mass}\label{fig:Mjj_canvas_nolepcut}]
{\includegraphics[scale=0.37]{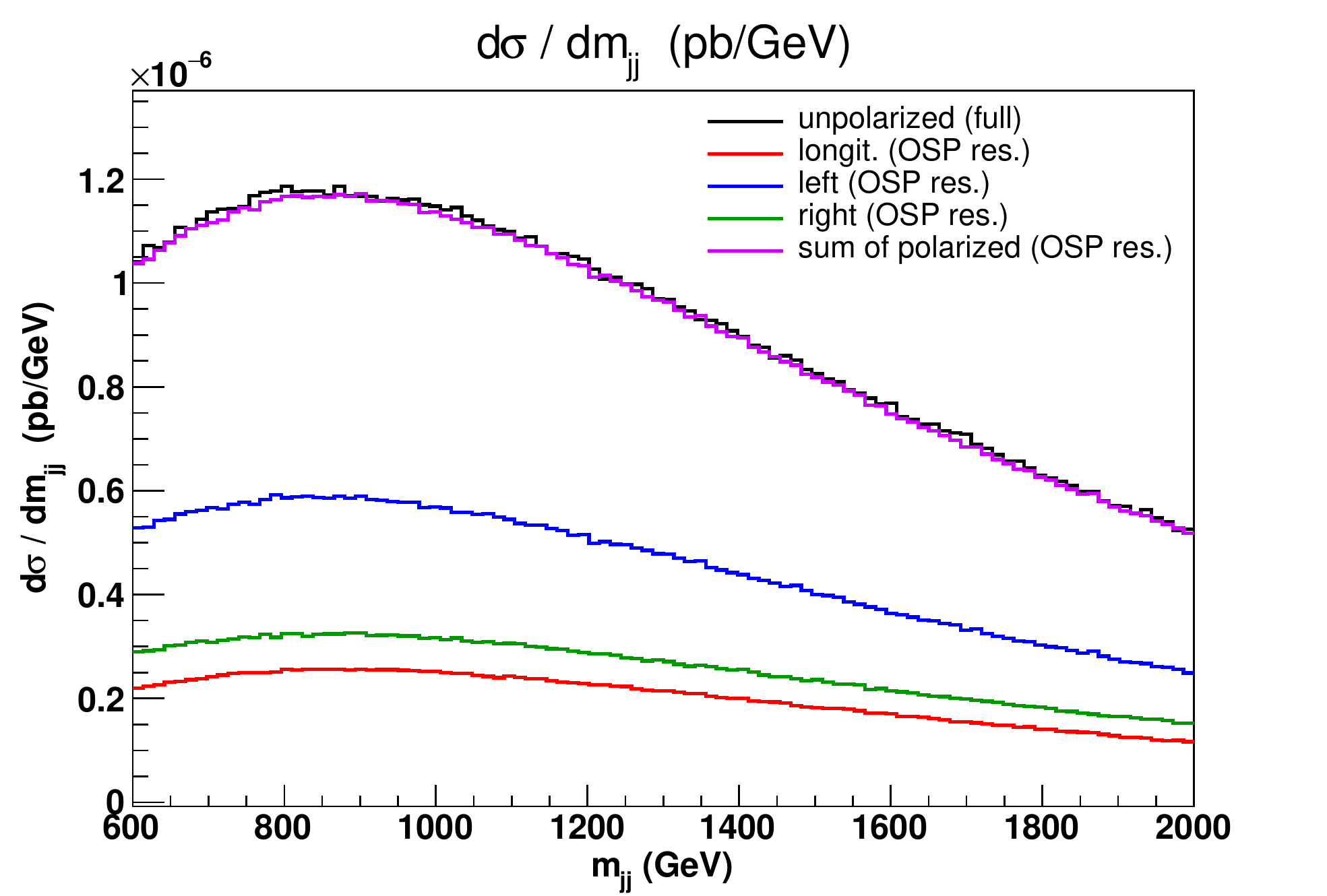}}\\
\subfigure[{$p_t^{W^-}$}\label{fig:Ptwm_canvas_nolepcut}]
{\includegraphics[scale=0.37]{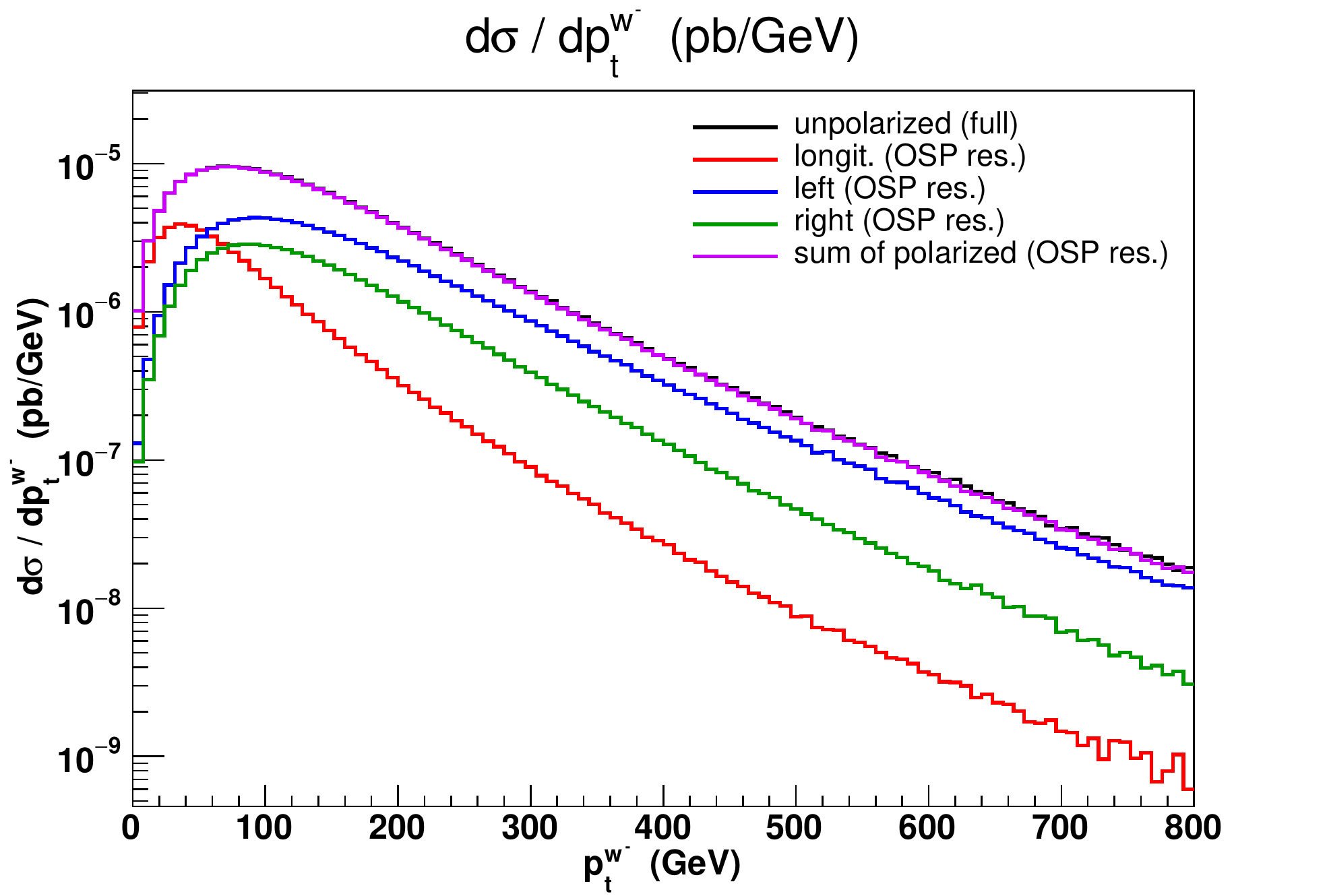}}
\subfigure[{$\eta^{W^-}$}\label{fig:Etawm_canvas_nolepcut}]
{\includegraphics[scale=0.37]{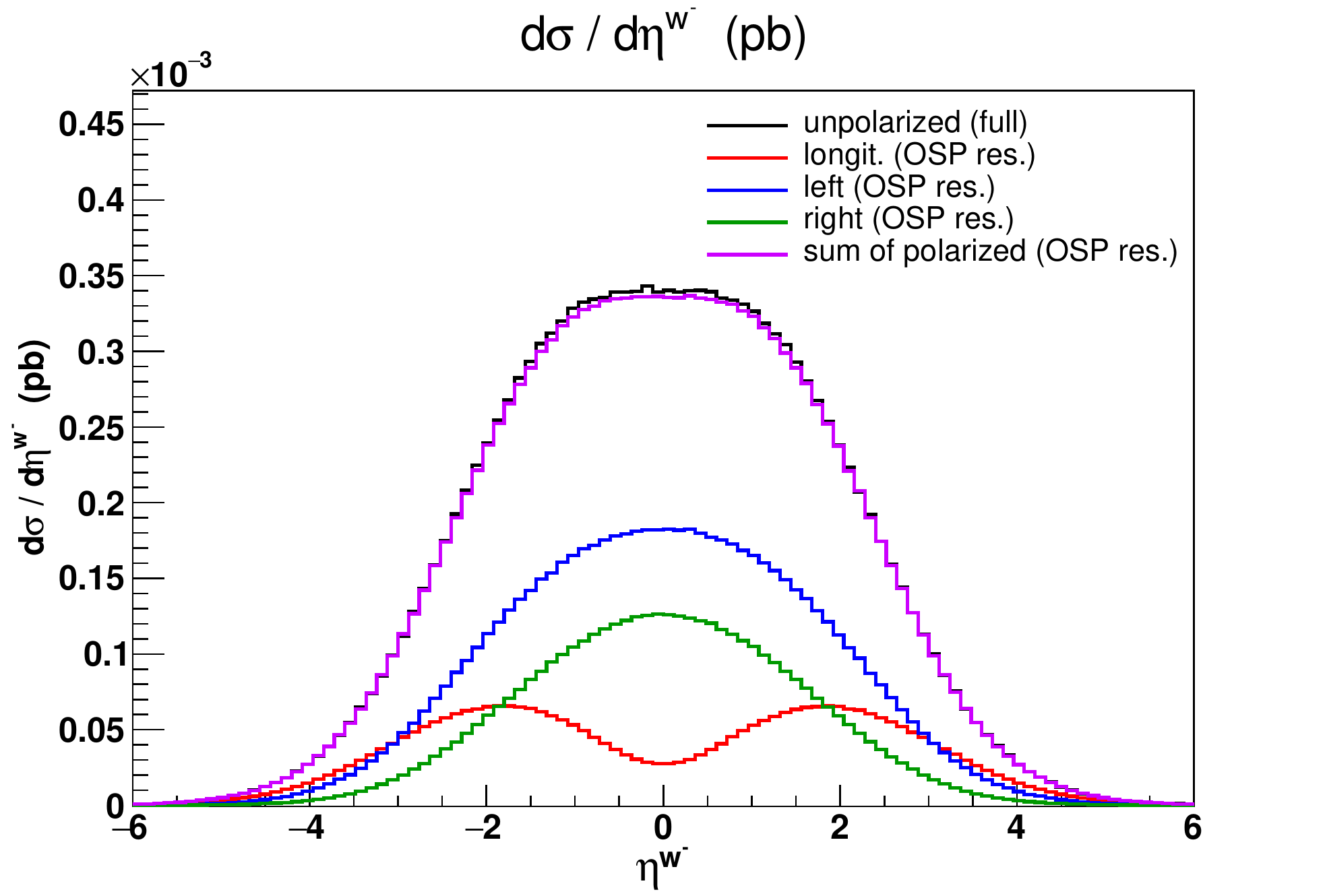}}\\
\subfigure[{$M_{ll}$}\label{fig:Mll_zoom_nolepcut}]
{\includegraphics[scale=0.37]{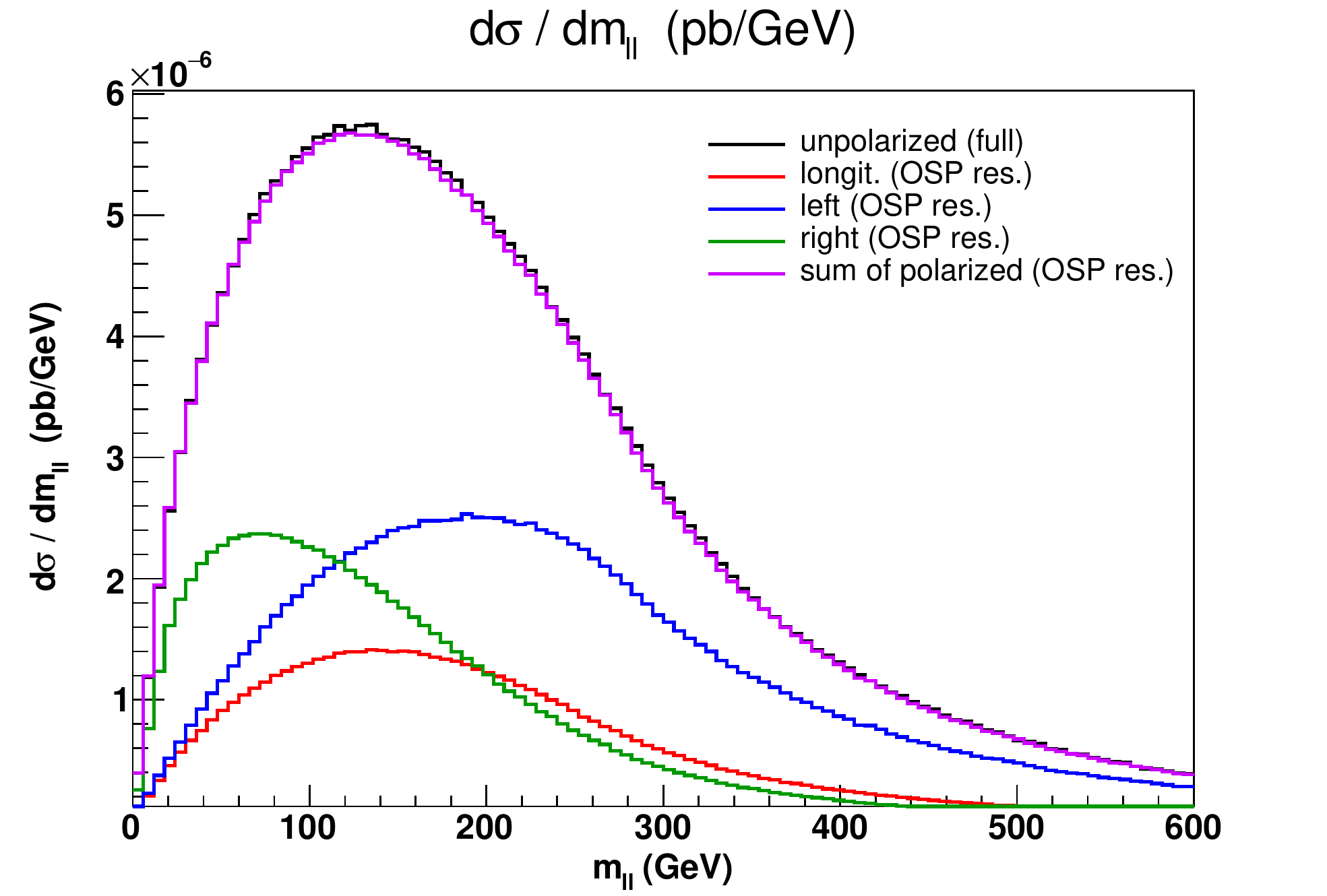}}
\subfigure[{$p_t^{e^-}$}\label{fig:pte_zoom_nolepcut}]
{\includegraphics[scale=0.37]{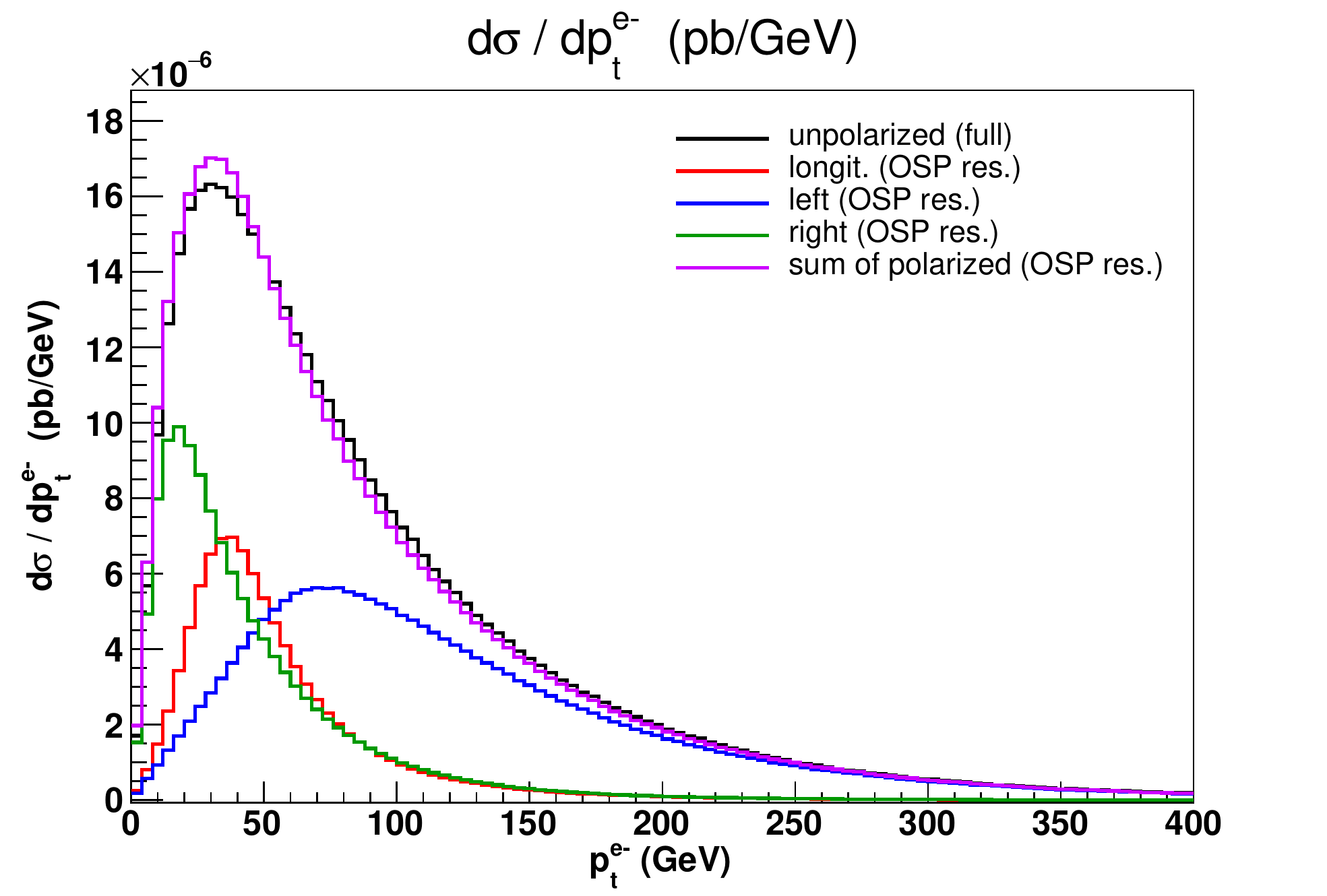}}\\
\subfigure[{$\phi_{e^-}$}\label{fig:phi_nocut}]
{\includegraphics[scale=0.38]{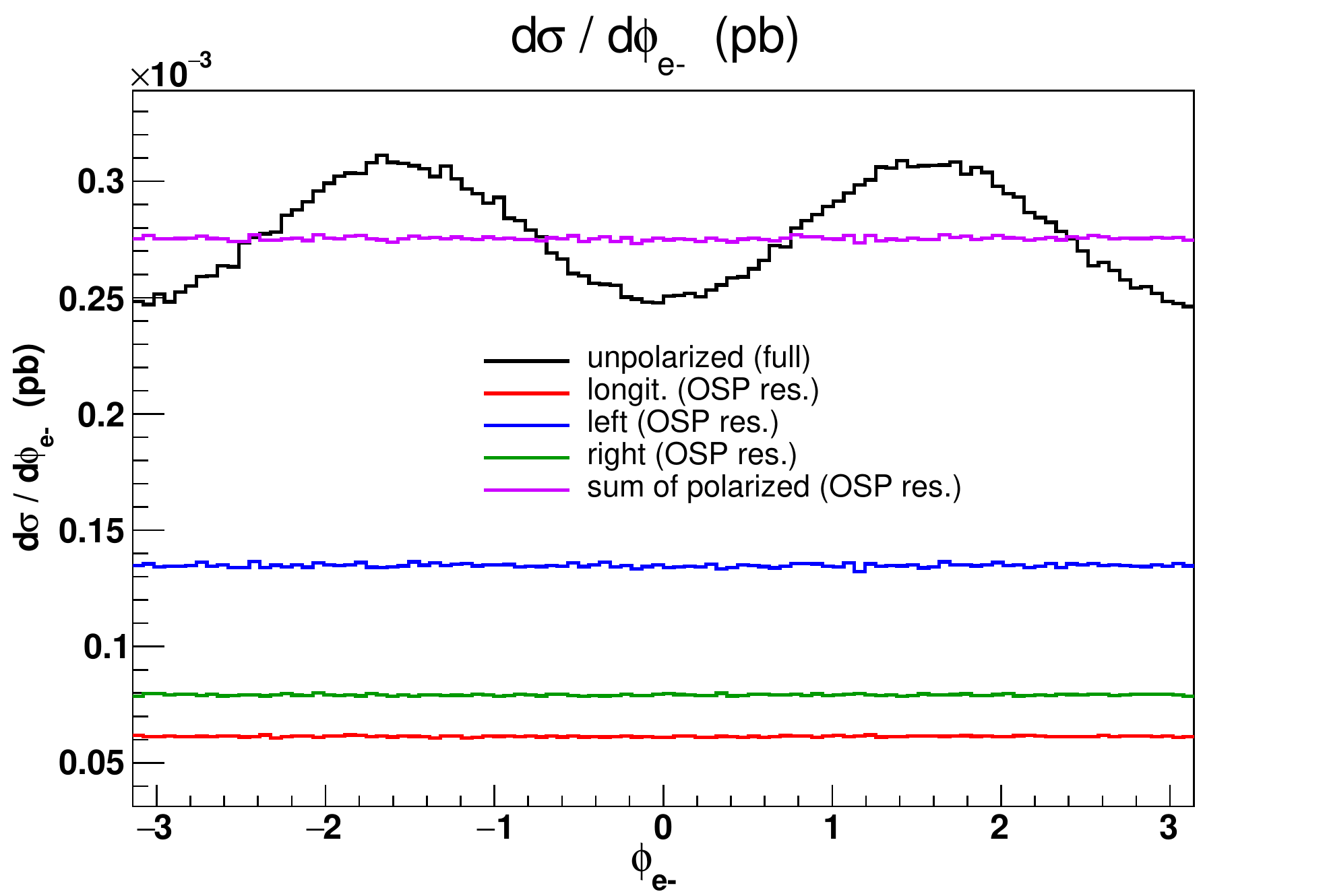}}
\caption{Differential cross sections for $p p \rightarrow j j e^-\bar{\nu}_e \mu^+ \nu_\mu$ at the LHC@13 TeV.
Comparison between unpolarized generations based on the full amplitude
and the incoherent sum of the polarized generations with OSP, which take into account only the resonant diagrams.
No cut on leptonic variables.
 }
\label{fig:var_nolepcut}
\end{figure}

The full cross section is 1.748(1) fb while
the incoherent sum of singly polarized cross sections is 1.731(1) fb, which differs from the full
result by about 1\%.

In \figsc{fig:distrib_thetal_polfrac_nocut}{fig:var_nolepcut} the black curves refers to the full differential cross sections and the violet 
ones to the sum of polarized
distributions. The individual distributions are shown in red (longitudinal), blue (left) and green (right).
We have also computed the coherent sum of the three polarized contributions, which is, in all cases, almost
indistinguishable from the full result and, therefore, is not shown.

In \fig{fig:distrib_thetal_polfrac_nocut}, on the left,
we show the distribution of the decay angle of the electron in the reference frame of the $e^-\nu_e$ pair
integrated over the full range $M_{WW}>300$ GeV.
On the right we present the polarization fractions as a function of the invariant mass of the four leptons, which
provides additional information since the distributions in \fig{fig:distrib_thetal_nocut} are dominated by events with
relatively small $M_{WW}$.

The polarization components obtained by expanding the full angular distribution on Legendre polynomials, as 
discussed in \sect{sec:Wpol_decay}, are shown as lighter shade smooth lines in
\fig{fig:distrib_thetal_nocut} and show that
the distributions from the polarized generations, in the absence of cuts on the charged leptons, have the expected
functional form, and that the polarization fractions extracted from the full results are in excellent agreement 
with those obtained from the polarized distributions.

\Fig{fig:polfrac_MWW_nocut} presents the polarization fractions of the $W^-$ as a function of $M_{WW}$.
The darker lines show the ratio of the individual polarized cross sections
to the full result in each $M_{WW}$ bin; the lighter lines are obtained from a bin by bin
expansion on Legendre polynomials of the full result.
The two methods agree over the full range. This confirms that the OSP provides reliable results and opens the way
to test it in the presence of cuts, where the Legendre expansion is known to fail.

\Fig{fig:polfrac_MWW_nocut} shows that, in the SM, $W^-$'s produced in VBS are mainly
left handed. The
fraction of left polarized $W^-$'s increases with increasing $M_{WW}$. The fraction of longitudinal and
right handed $W$'s are roughly the same at large invariant masses. The longitudinal fraction is almost constant at
about 20\%. The right handed component decreases slightly from approximatly 30\% at $M_{WW}$ = 400 GeV to
just above 20\% at $M_{WW}$ = 2000 GeV.

The colors in \fig{fig:var_nolepcut} are as in \fig{fig:distrib_thetal_polfrac_nocut}:
black refers to the full result, violet to the incoherent sum of polarized
distributions, red to the longitudinal polarization, blue and green to the left and right polarization, respectively.

The incoherent sum of three OSP distributions agrees very well with the full result for the
{$WW$ invariant mass, \fig {fig:Mww_canvas_nolepcut}, the mass of the two tag jets, \fig{fig:Mjj_canvas_nolepcut},
the transverse momentum of the $e^-\nu_e$ pair, \fig{fig:Ptwm_canvas_nolepcut}, the rapidity of the
 $e^-\nu_e$ pair, \fig{fig:Etawm_canvas_nolepcut} and the mass of the $e\mu$ pair, \fig{fig:Mll_zoom_nolepcut} .

The agreement is less satisfactory for
the transverse momentum of the electron, \fig{fig:pte_zoom_nolepcut}. Since, as already mentioned, the coherent
sum of the three contributions agrees with the full result, the discrepancy between 
the black and the violet lines in
\fig{fig:pte_zoom_nolepcut} is due to the interference among the different polarizations.
The interference is non zero because
forcing the lepton $p_t$ to a single bin restricts the angular range of the leptons, spoiling the
complete cancellation of the interferences among polarizations.

The distribution, shown in \fig{fig:phi_nocut}, of the azimuthal angle of the electron, $\phi_e$, which is defined
following \rf{Bern:2011ie}, stands apart.
Each of the three singly polarized contributions is isotropic, since the azimuthal
angle enters the decay amplitudes only as a
phase. The coherent combination of the three amplitudes produces a non trivial modulation of the differential cross
section. The incoherent sum of the polarized results reproduces well only the average value of the full distribution.

The singly polarized distributions of the transverse momentum of the $W$, of its pseudorapidity,
of the transverse momentum of the
negatively charged lepton and of the invariant mass of the two charged leptons depend significantly on the
polarization of the $W$.

\Fig{fig:Ptwm_canvas_nolepcut} shows that longitudinally polarized $W$'s have a markedly softer $p_t$ spectrum
in comparison with transversely polarized ones. In fact they dominate for $p_t^W < 50$ GeV.
\Fig{fig:Etawm_canvas_nolepcut} shows that, while transversely polarized $W$'s are predominantly produced at small
rapidities, the distribution for longitudinally polarized one presents a dip at zero and peaks at
$\eta^{W^-} = \pm\, 2$.

The $p_t^{e^-}$ and $M_{ll}$ distributions are harder for left polarized $W^-$'s than for
longitudinal polarized ones. The right handed $W^-$'s have the softest spectrum.
This behaviour is clearly related to
the distribution of decay angles for the three polarizations: the negatively charged leptons from left polarized bosons
tend to be produced along the direction of flight of the $W$, while those originating from right polarized $W$'s tend
to emerge in the opposite direction. The leptons from longitudinally polarized $W$'s fall in between the two other
cases.

\section{Joint polarization fractions for the two $W$'s}
\label{sec:double_polfrac}

The projection described in \eqns{eq:LegendreExpansion}{eq:inversion_eq} can be readily generalized to a
simultaneous expansion in products of Legendre polynomials of the two variables $\cos\theta_e$ and
$\cos\theta_\mu$. Similarly, the substitution in \eqn{eq:substitution} can be performed for each of the two final
state $W$'s.
The outcome is shown in \fig{fig:distrib_double_thetal_polfrac_nocut}. For ease of presentation,
the right and left handed contributions are summed together in the transverse component $W_T=W_R+W_L$.

The polarization components obtained by expanding, in each bin, the full angular distribution on Legendre 
polynomials are shown in lighter colors. The darker histograms are obtained integrating the
amplitudes squared with definite polarization for each $W$.
The two independent determinations of the joint polarization fractions agree extremely well
over the full range in $M_{WW}$.
This implies that the method we propose can be relied on for analyzing double polarized cross sections.

\begin{figure}[!tb]
\centering
\includegraphics[scale=0.39]{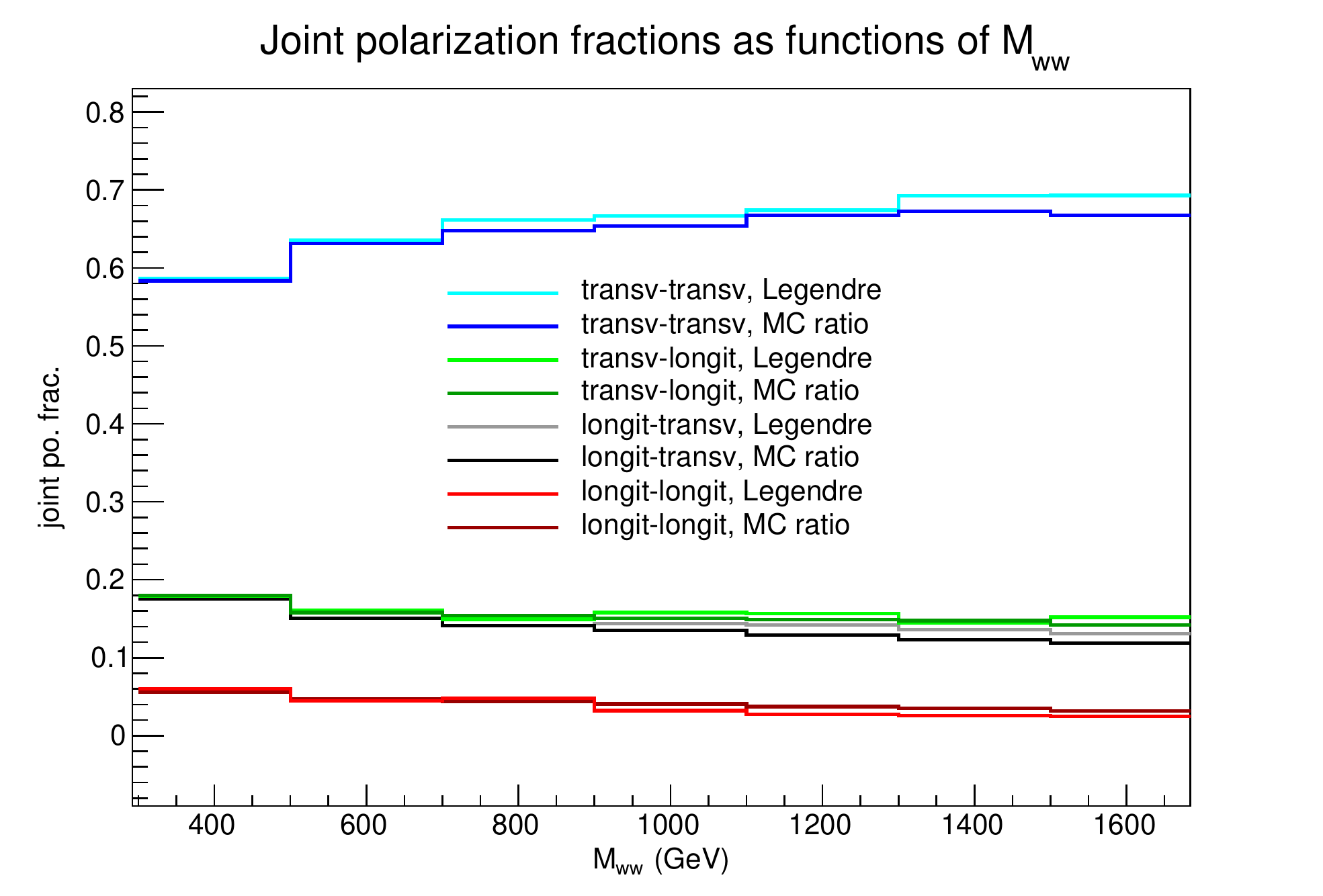}
\caption{Double polarization fractions as functions of $M_{WW}$.
The right and left handed contributions are summed together in the transverse component $W_T=W_R+W_L$.
The polarization components obtained by expanding, in each bin, the full angular distribution on Legendre 
polynomials are 
shown in lighter colors. The darker histograms are obtained integrating the polarized amplitudes squared.
}
\label{fig:distrib_double_thetal_polfrac_nocut}
\end{figure}

\Fig{fig:distrib_double_thetal_polfrac_nocut} shows that
the $W^+_T W^-_T$ fraction is always the largest one and dominates at large invariant masses, comprising
about 70\% of the total cross section.
The $W^+_T W^-_0$ and $W^+_0 W^-_T$ components are essentially equal and almost constant at about 18\%.
The longitudinal--longitudinal fraction is the smallest one, of the order of a few percent.
This implies that measuring the scattering with two longitudinally polarized $W$'s in the final state,
will require determining the
polarization of both vector bosons, since a longitudinal $W$ is expected in most cases to be produced in association
with a transversely polarized companion.

\section{Leptonic cuts and their effects}
\label{sec:cuts}

In this section we document how the distributions presented
in \sect{sec:results} are modified by the introduction of realistic acceptance cuts on the electron which is
the decay product of the $W^-$, whose polarization we wish to determine. 
Our results confirm that the polarization fractions of the $W$ cannot be determined anymore by a projection on
the first three Legendre polynomials.
We show that, in the presence of standard leptonic cuts, the interference among the polarized amplitudes is small.
Therefore, the incoherent sum of the three OSP results approximates fairly well,
in most cases, the full distribution.  

 \begin{figure}[!tb]
\centering
\subfigure[{$\cos\theta_e$}\label{fig:distrib_thetal_lepcut}]
{\includegraphics[scale=0.37]{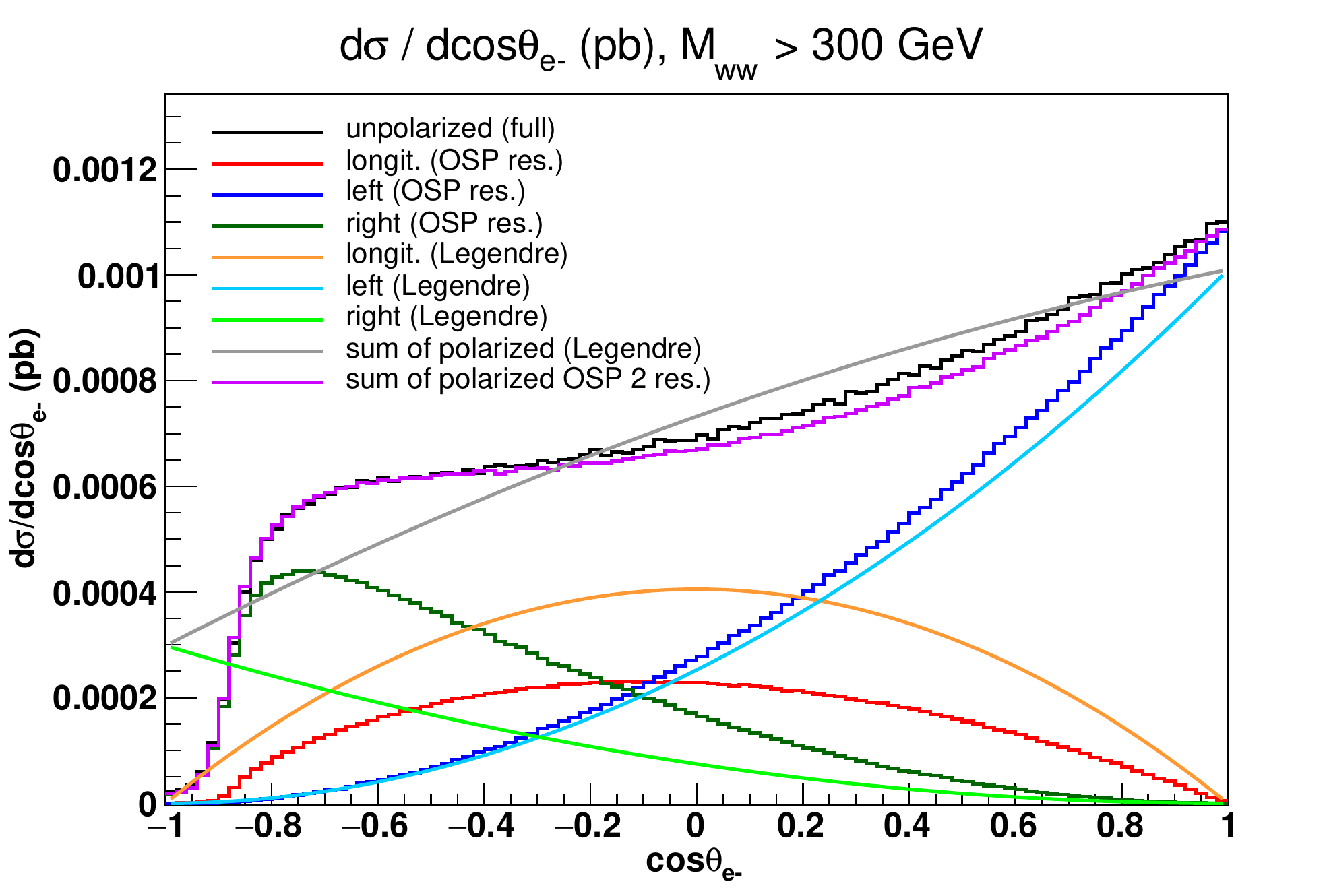}}
\subfigure[{Polarization fractions}\label{fig:polfrac_MWW_lepcut}]
{\includegraphics[scale=0.37]{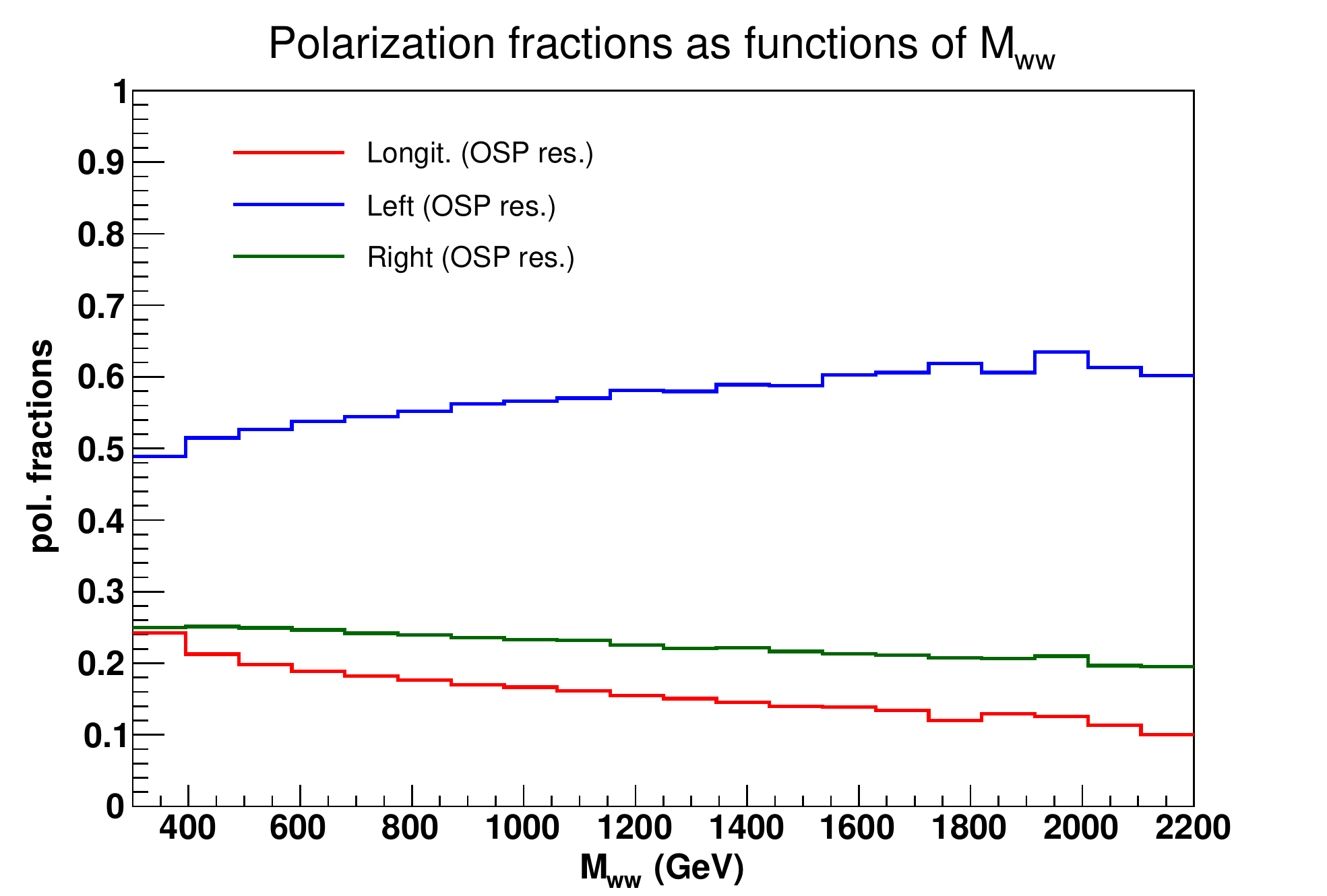}}
\caption{On the left, distributions of $\cos\theta_e$ in the $W^-$ center of mass frame; on the right, the
 polarization fractions as functions of $M_{WW}$.
$p_t^{e}>20$ GeV, $|\eta^{e}| < 2.5$.}
\label{fig:distrib_thetal_polfrac_lepcut}
\end{figure}

We require:
\begin{equation}
 p_t^{e}>20 \,\,\mathrm{GeV}, \qquad |\eta^{e}| < 2.5.
\label{eq:lepton_cuts}
\end{equation}

The full cross section is 1.411(1) fb, the coherent sum of OSP polarized amplitudes gives 1.401(1) fb, while
the incoherent sum of singly polarized cross sections is 1.382(1) fb, which differs from the full result by about 2\%.

In \fig{fig:distrib_thetal_polfrac_lepcut} and \fig{fig:var_lepcut} we show the same set of results presented
\sect{sec:results}.
A number of features are worth noticing. The cross section from the coherent sum of polarized amplitudes is very
close to the incoherent sum of cross sections, which indicates that the interference among polarizations is
generally small. The rigorous cross section is well reproduced by both approximations.

In \fig{fig:distrib_thetal_polfrac_lepcut}, on the left,
we show the distribution of the decay angle of the electron in the reference frame of the $e^-\nu_e$ pair.
On the right we present the polarization fractions as a function of the invariant mass of the four leptons.

The full angular distribution is approximated within a few percent, over the full range, by the sum of the
unpolarized results.
The full result, shown by the black histogram, however, is not of the form of \eqn{eq:dcdist} and cannot be
described in terms of the three
first Legendre polynomials. This becomes clear expanding the full result as in
\eqns{eq:LegendreExpansion} {eq:inversion_eq},
which yields the blue, green and orange smooth
curves in \fig{fig:distrib_thetal_polfrac_lepcut}. Their sum is the smooth gray curve which fails to describe the
correct distribution.
As already noticed in \rf{Stirling:2012zt}, when acceptance cuts are imposed, the polarization fractions cannot
be reliably extracted from the angular distribution of charged leptons by comparing it to the functional form
expected when no cuts are set.

The polarization fractions in \fig{fig:polfrac_MWW_lepcut} are computed as the ratio of the individual polarized
cross sections to the full result in each $M_{WW}$ bin.
An expansion on Legendre polynomials would be meaningless.

\afterpage{\clearpage}
\begin{figure}[p]
\vspace*{-1.2cm}
\centering
\subfigure[{$WW$ invariant mass}\label{fig:Mww_canvas_lepcut}]
{\includegraphics[scale=0.37]{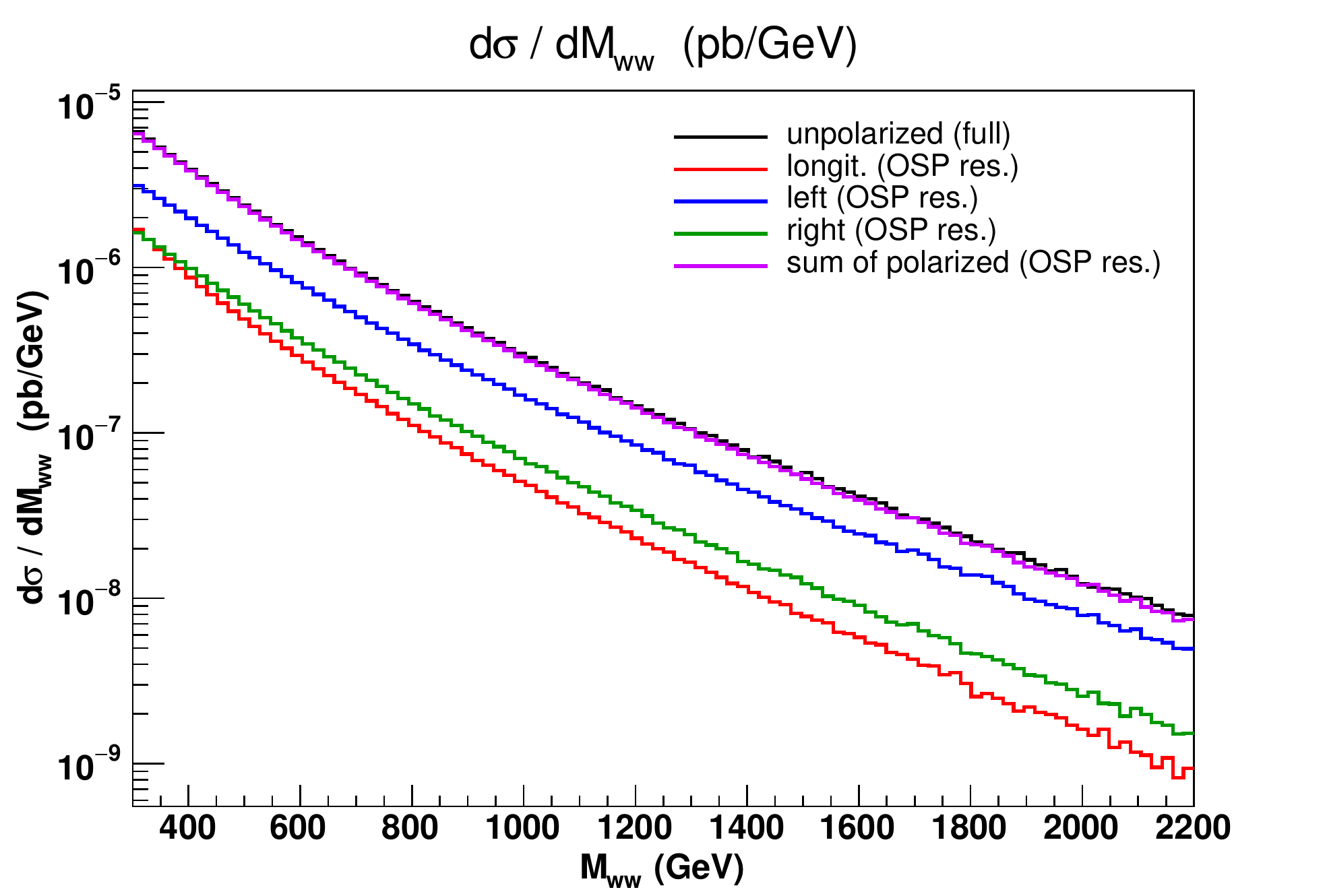}}
\subfigure[{$jj$ invariant mass}\label{fig:Mjj_canvas_lepcut}]
{\includegraphics[scale=0.37]{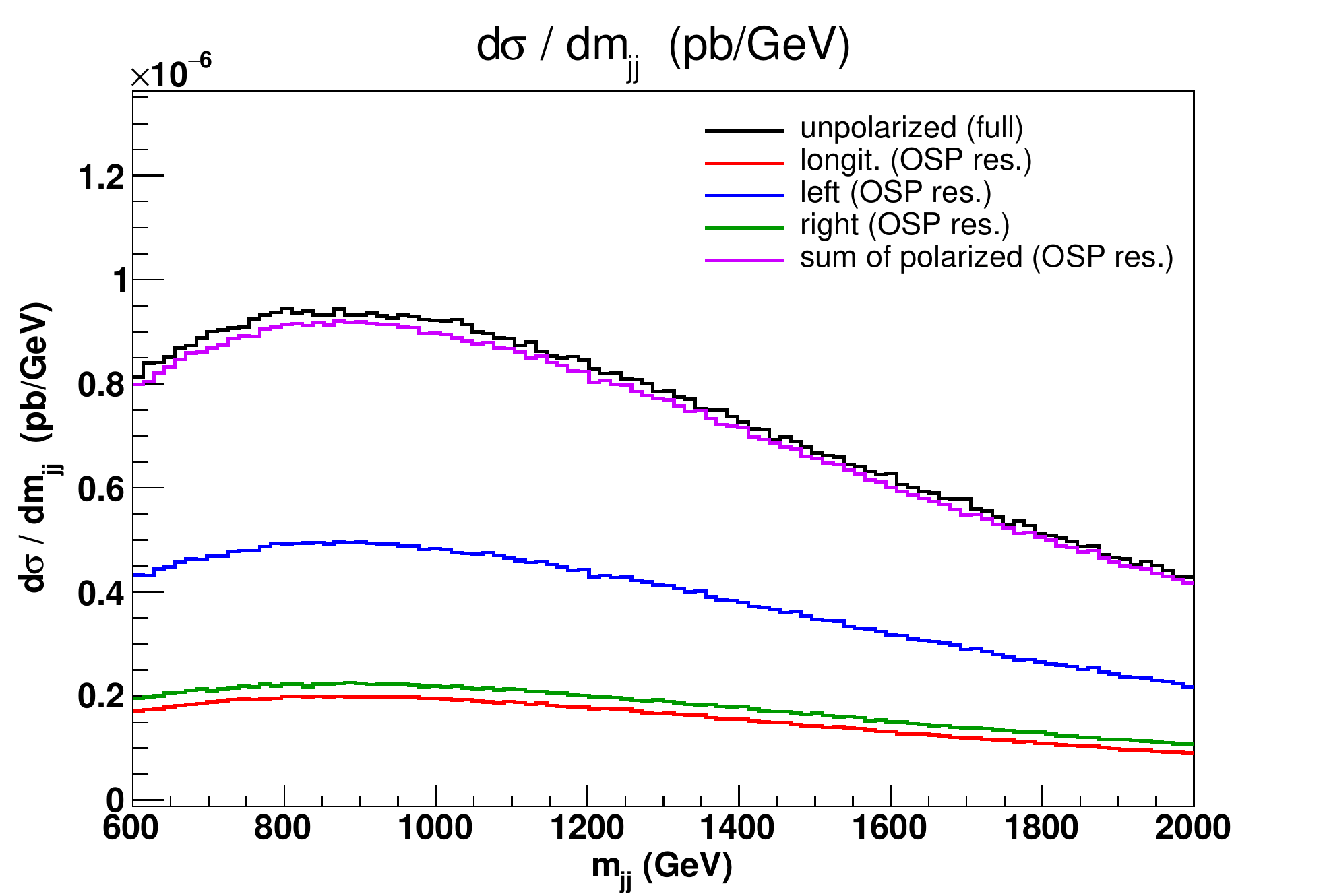}}\\
\subfigure[{$p_t^{W^-}$}\label{fig:Ptwm_canvas_lepcut}]
{\includegraphics[scale=0.37]{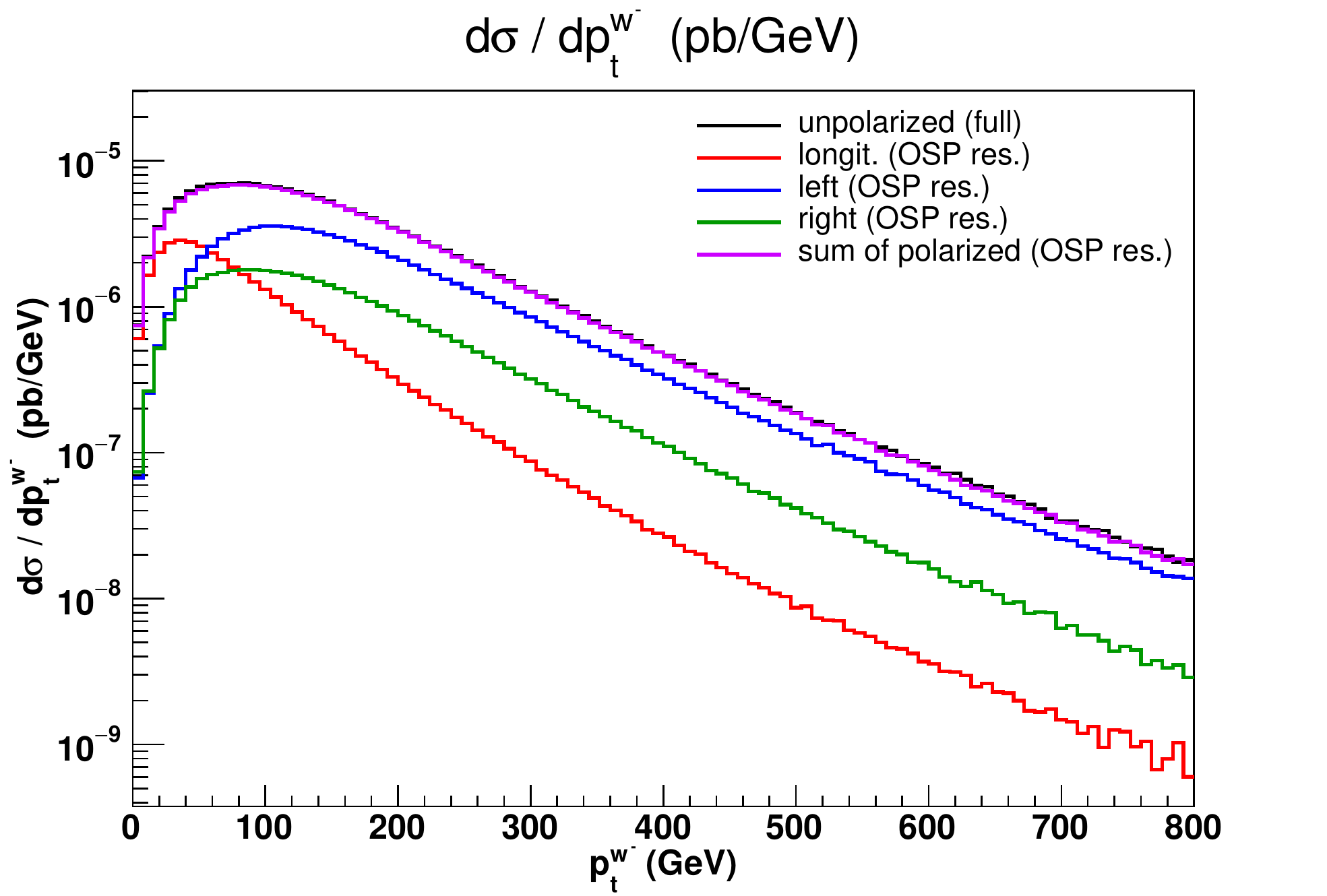}}
\subfigure[{$\eta^{W^-}$}\label{fig:Etawm_canvas_lepcut}]
{\includegraphics[scale=0.37]{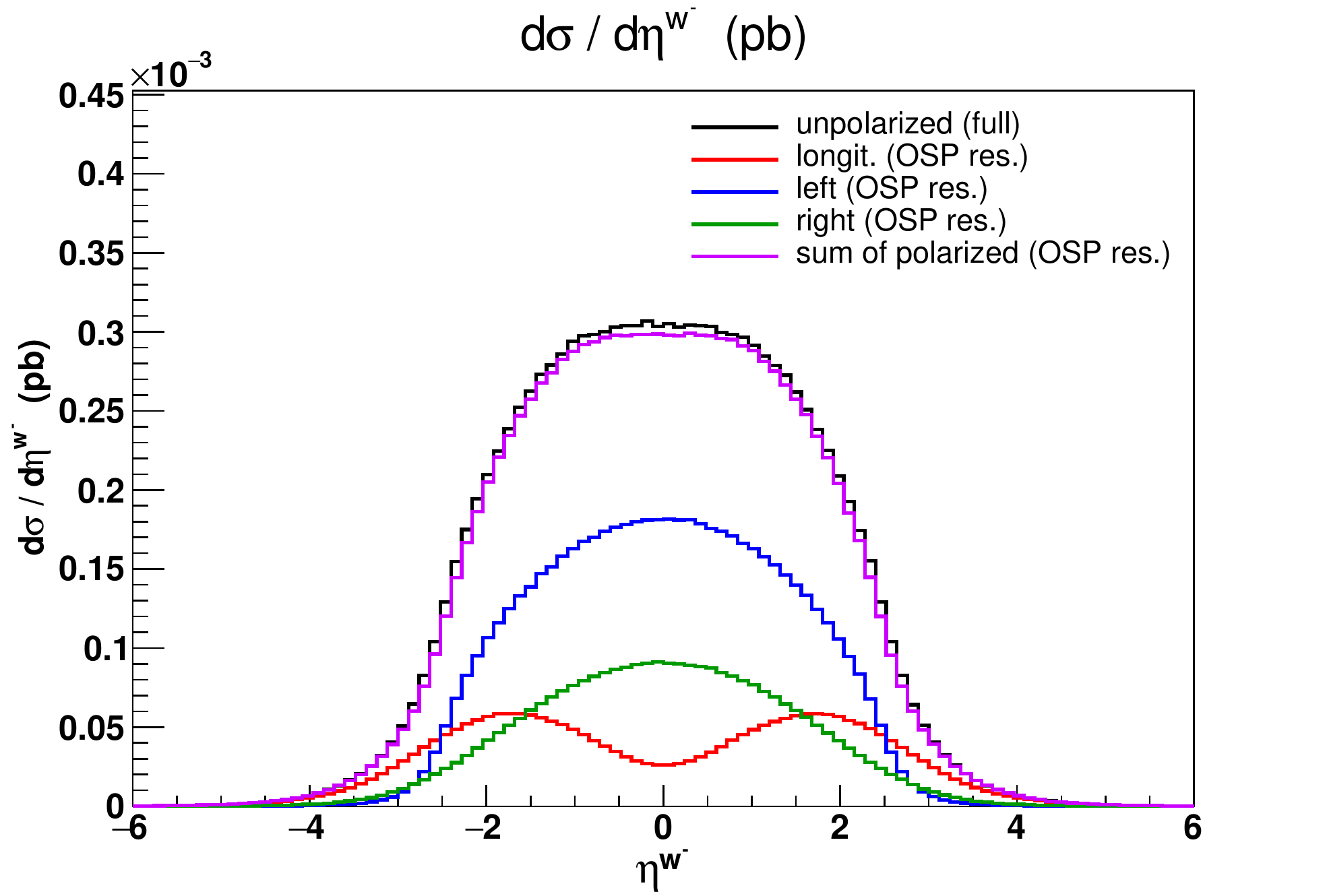}}\\
\subfigure[{$M_{ll}$}\label{fig:Mll_zoom_lepcut}]
{\includegraphics[scale=0.37]{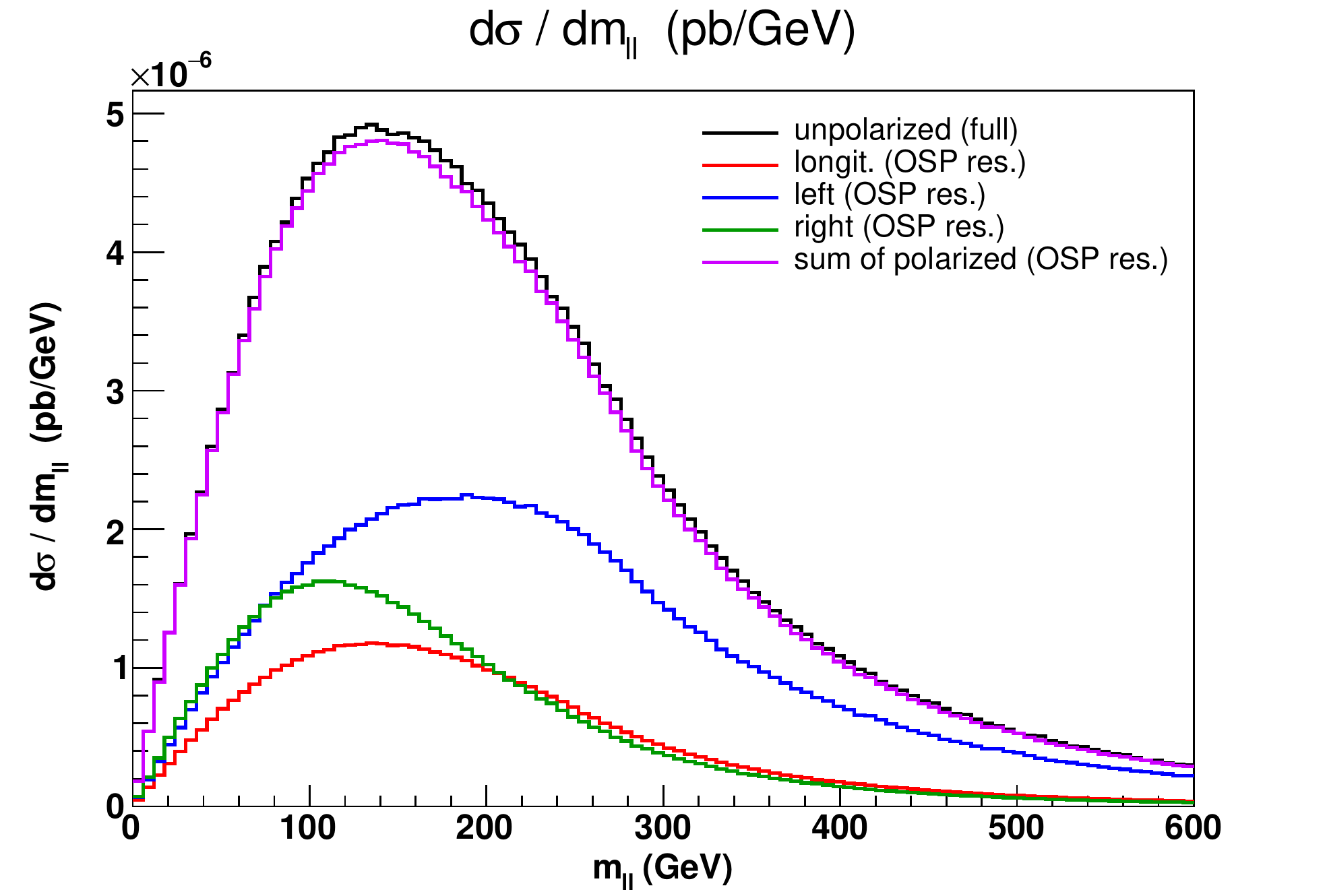}}
\subfigure[{$p_t^{e^-}$}\label{fig:pte_zoom_lepcut}]
{\includegraphics[scale=0.37]{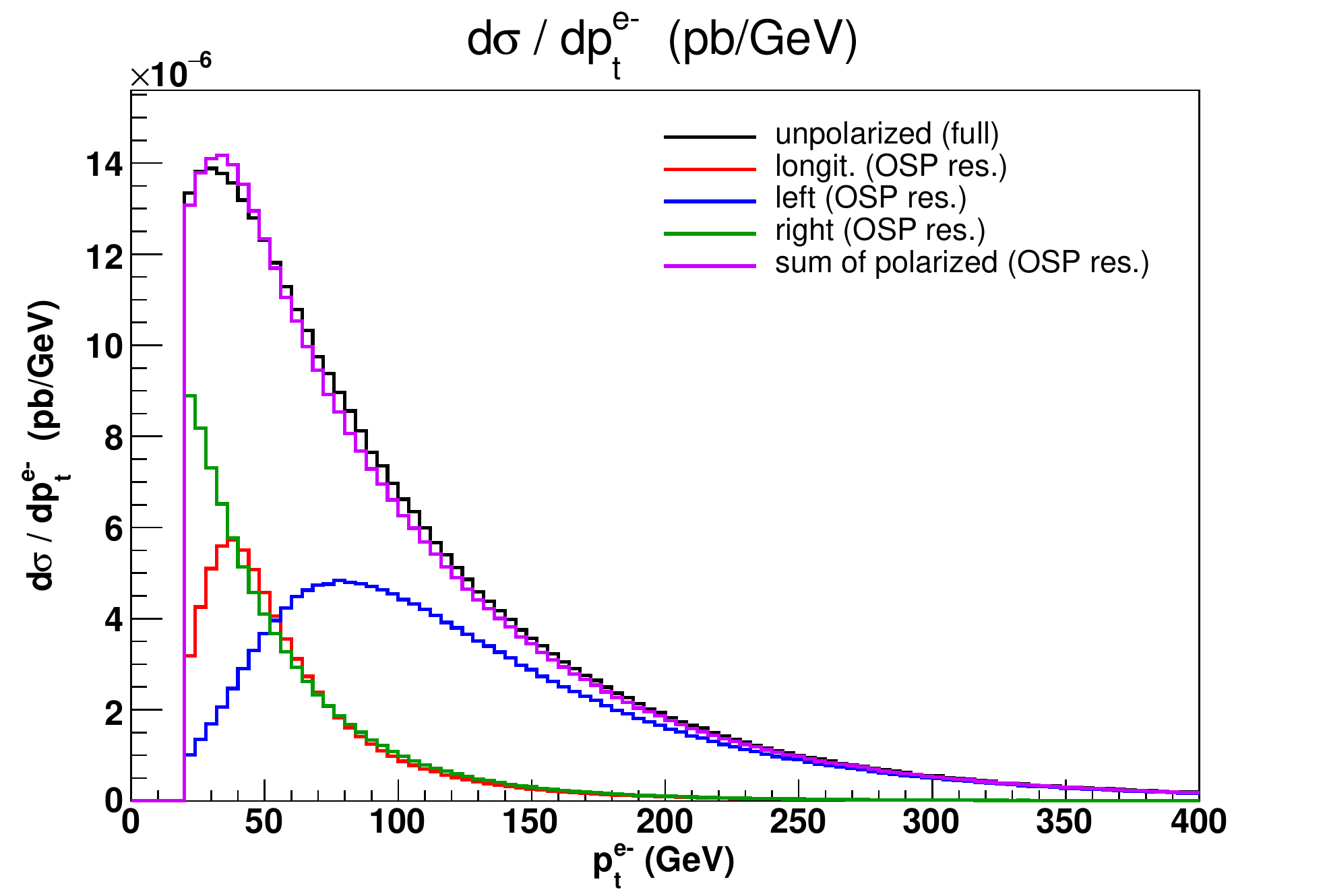}}\\
\subfigure[{$\phi_{e^-}$}\label{fig:phi_lepcut}]
{\includegraphics[scale=0.38]{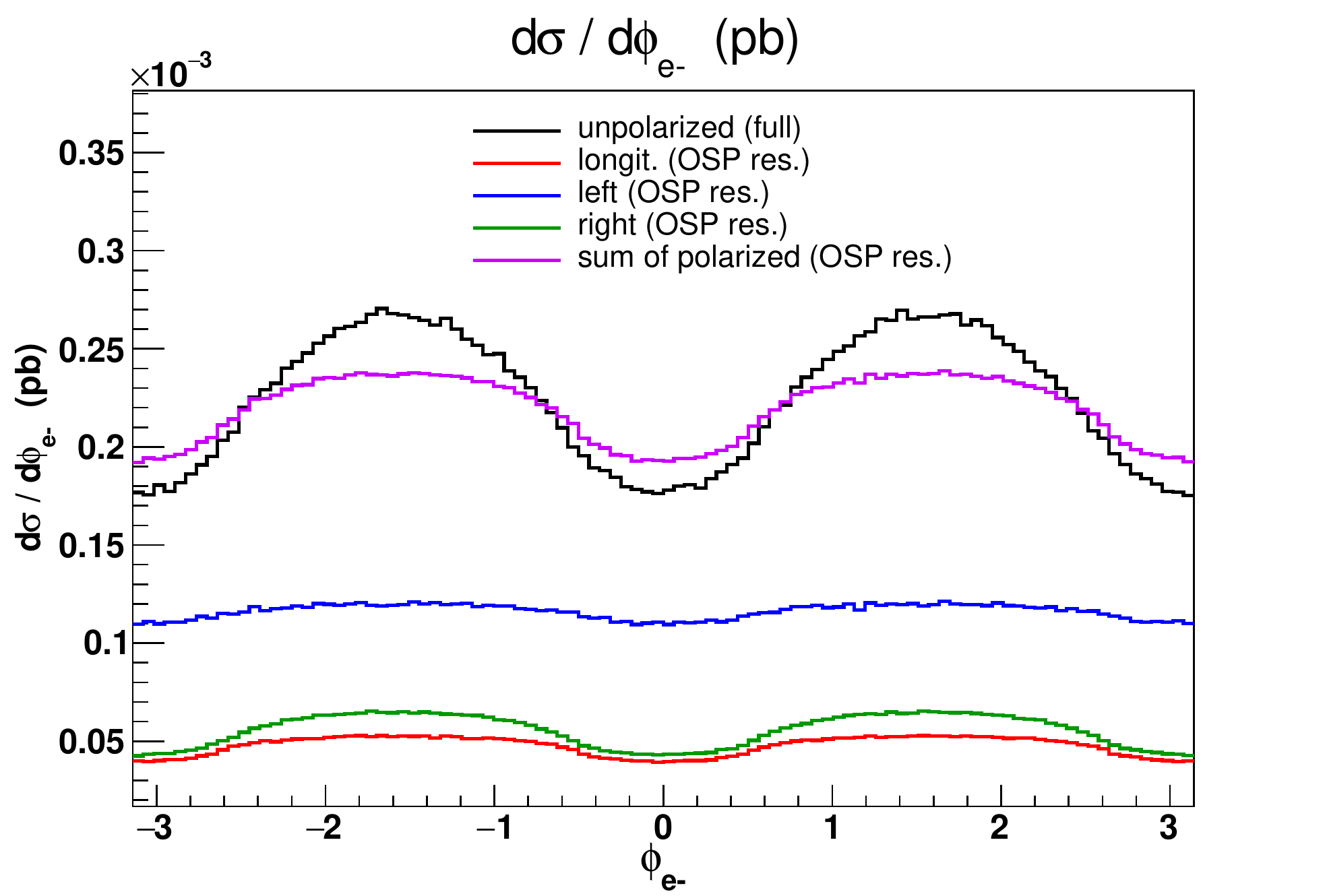}}
\caption{Differential cross sections: comparison between unpolarized and sum of the polarized
distributions. $p_t^{e}>20$ GeV, $|\eta^{e}| < 2.5$.}
\label{fig:var_lepcut}
\end{figure}

The most prominent feature in comparison with the curves on the left hand side of
\fig{fig:distrib_thetal_polfrac_nocut} is the depletion at $\cos\theta_e$ = -1 in the angular distribution. The
right handed component is the one most affected, with the resulting shape substantially different from the
$(1-\cos\theta)^2$ behaviour displayed in the absence of cuts. The bulk of the effect is again related to the
preferred
direction of emission of the charged leptons from right handed $W^-$'s, which tend to produce leptons with
smaller transverse momentum.

There are, however, subtler effects into play, as can be seen from the polarization fractions as a function of
$M_{WW}$ in
\fig {fig:polfrac_MWW_lepcut}. One notices that the longitudinal component decreases faster than the right handed
one for increasing
$M_{WW}$. The two curves, which are very close at $M_{WW} <$ 400 GeV, separate at large invariant masses, in
contrast with
the trend they display in \fig{fig:polfrac_MWW_nocut}. This is due to the different distribution in $W$ rapidity of the
two polarized cross sections. With increasing $M_{WW}$ the average absolute value of the $W$ rapidity with
longitudinal polarization increases and a larger number of the corresponding leptons fail the rapidity acceptance cut.

Even in the presence of leptonic cuts, the fraction of longitudinally polarized $W^-$'s is well above 10\%.

The incoherent sum of three OSP distributions agrees well with the full result for the 
$WW$ invariant mass, \fig {fig:Mww_canvas_lepcut}, the mass of the two tag jets, \fig{fig:Mjj_canvas_lepcut},
the transverse momentum of the $e^-\nu_e$ pair, \fig{fig:Ptwm_canvas_lepcut}, the rapidity of the
 $e^-\nu_e$ pair, \fig{fig:Etawm_canvas_lepcut} and the mass of the $e\mu$ pair, \fig{fig:Mll_zoom_lepcut} .

As before, the transverse momentum of the electron,
\fig{fig:pte_zoom_lepcut}, is affected by interferences, but the effect is not particularly
enhanced by the presence of cuts.

The distribution of the azimuthal angle of the electron in \fig{fig:phi_lepcut} displays again peculiar features.
Each of the three singly polarized contributions develops a dip at $\phi_e = 0$ and $\phi_e = \pm \pi$.
The incoherent combination of the three amplitudes follows more
closely the full result than when no cuts are applied but does not reproduce the correct curve.

The individual polarizations are not affected equally by the cuts. Typically, the cross section for right handed $W$'s
is reduced the most, followed by the cross section for longitudinally polarized $W$'s. Left handed $W$ bosons
seem to be the least sensitive to acceptance cuts.

\section{Determining the polarization fractions}
\label{sec:measuring}

\begin{figure}[!b]
\centering
\subfigure[{$\cos\theta_e$}\label{fig:distrib_thetal_lepcut_noH}]
{\includegraphics[scale=0.37]{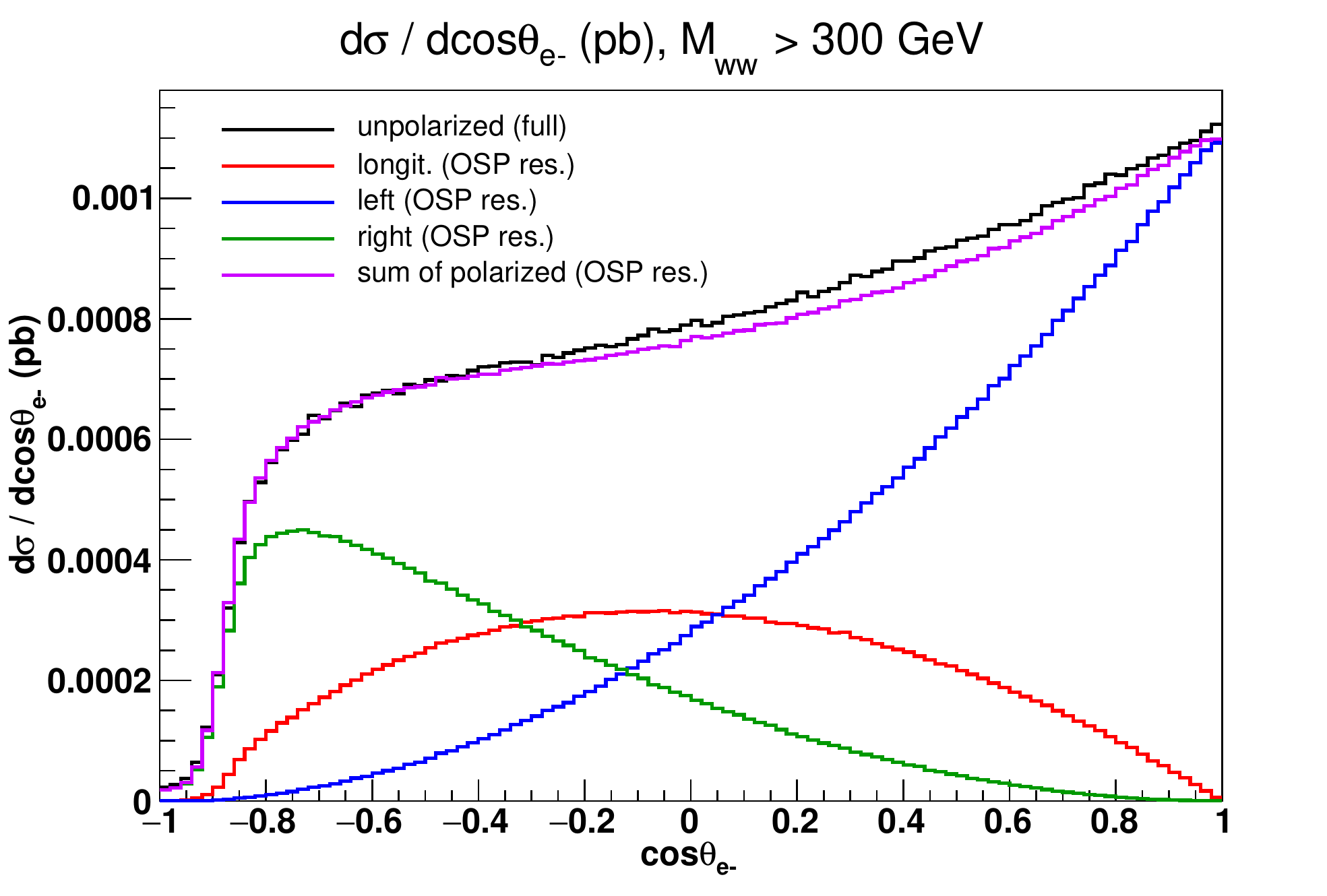}}
\subfigure[{Polarization fractions}\label{fig:polfrac_MWW_lepcut_noH}]
{\includegraphics[scale=0.37]{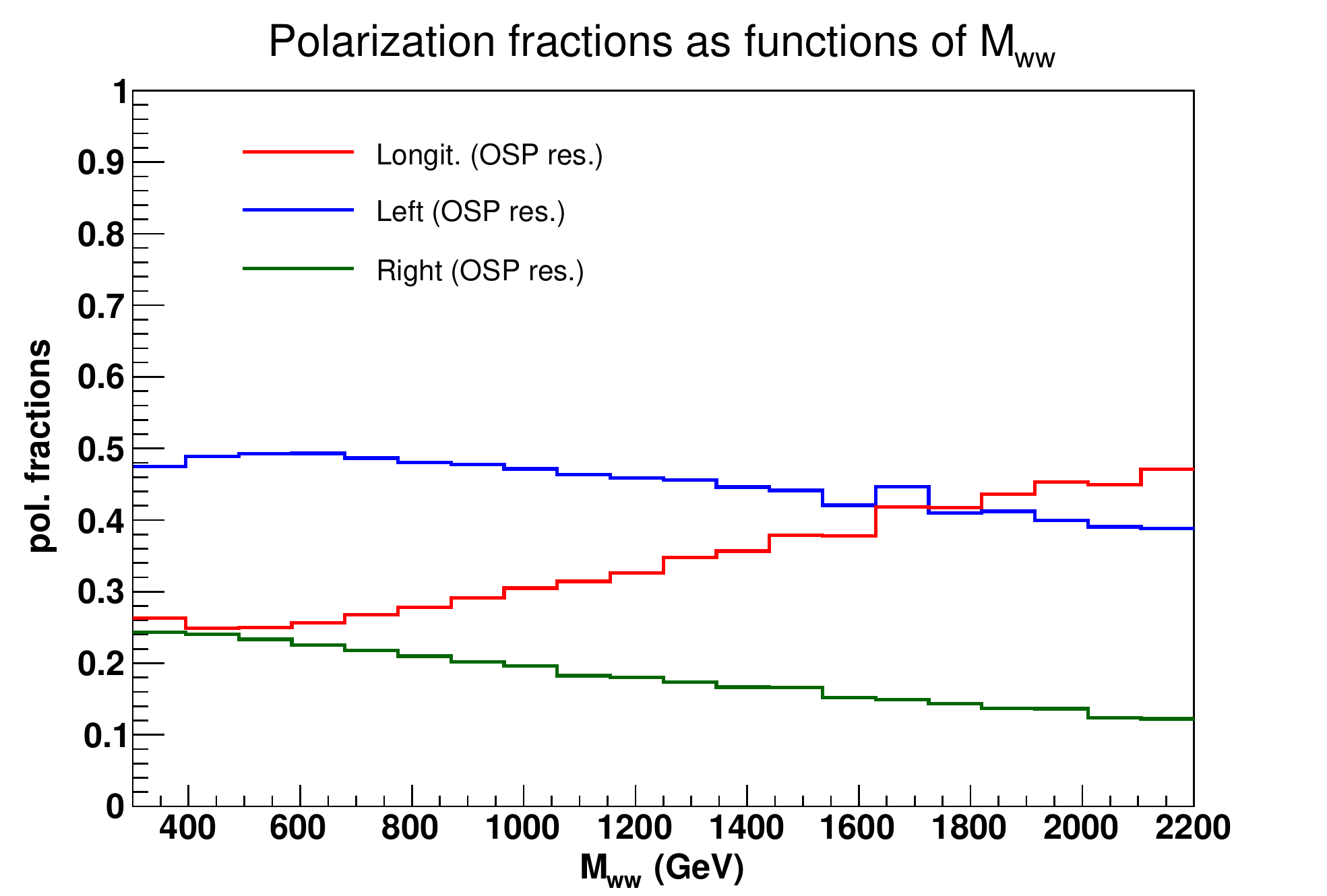}}
\caption{$\cos\theta_e$ distribution (left) and polarization fractions as functions of $M_{WW}$ (right) in the
Higgsless model. $p_t^{e}>20$ GeV, $|\eta^{e}| < 2.5$.}
\label{fig:distrib_thetal_polfrac_lepcut_noH}
\end{figure}

\begin{figure}[!b]
\centering
{\includegraphics[scale=0.39]{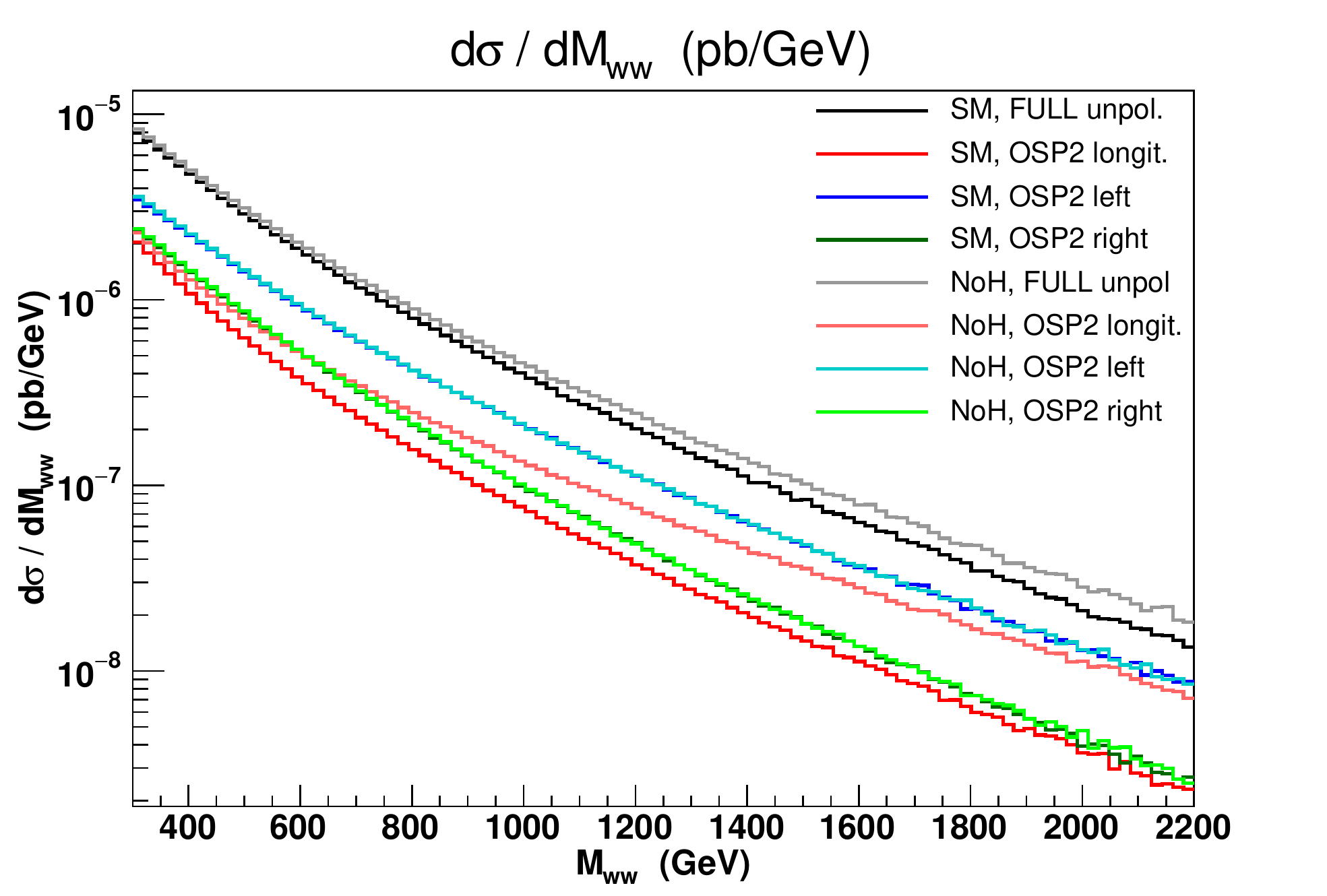}}
\caption{$M_{WW}$ distribution in the Higgsless model compared with the SM results. 
$p_t^{e}>20$ GeV, $|\eta^{e}| < 2.5$.}
\label{fig:Mww300_noHvsSM}
\end{figure}

The fact that the full distribution is well described by the incoherent sum of the polarized differential 
distributions
allows the determination of the polarization fractions within a single model,
even in the presence of cuts on the charged leptons. The measured angular distribution can 
be fitted to a linear
combination of the normalized shapes obtained from Monte Carlo simulations of VBS events with
final state $W$'s of definite polarization. 
We have verified that this procedure works well in the SM for one polarized vector boson.
The results shown in \sect{sec:double_polfrac} suggest that it should be straightforward to extend the 
method to double polarized events.

However, it would be inconvenient to have to generate templates for all extensions of the SM, 
being unknown
which specific model is realized in nature, should the SM need extending.
A natural question is whether this method can provide a model
independent, practical way of extracting the polarization fractions from the data, comparable with 
the Legendre
expansion, which is applicable when the $W$ decay is unrestricted by cuts.
The essential condition for this approach, is that the shapes of the polarized distributions are 
sufficiently universal.
On one hand, once a polarized $W$ transverse momentum and rapidity are fixed, its decay is completely 
specified.
Therefore, in any sufficiently small bin in phase space, the angular distribution of the charged lepton
will be the same, irrespective of the underlying dynamics.
On the other hand, the influence of cuts depends on the transverse momentum and rapidity of the $W$.
Different models yield different distributions of these quantities.
Hence, for finite size bins, the details of the averaging over phase space will vary, and some
degree of model dependence is to be expected.

In this section we pursue this possibility comparing the distributions produced in the SM with those
obtained in different models.

To this purpose,
we have studied the Higgsless model obtained from the SM by sending the Higgs mass to infinity. Such model,
while not experimentally viable any more, can be viewed as the most extreme modification of the SM in the VBS
domain, in
which the unitarity violating growth of the scattering of longitudinally polarized vector bosons is completely
unmitigated by Higgs exchange.

In \figsc{fig:distrib_thetal_polfrac_lepcut_noH}{fig:Mww300_noHvsSM}
 we present some results for the Higgsless model.
In \fig{fig:distrib_thetal_polfrac_lepcut_noH}, on the left,
we show the distribution of the decay angle of the electron in the reference frame of the $e^-\nu_e$ pair.
On the right we present the polarization fractions as a function of the invariant mass of the four leptons.

Comparing \fig{fig:distrib_thetal_lepcut_noH} with \fig{fig:distrib_thetal_lepcut}, it is immediately evident that
the two angular distributions of the charged lepton, in the center of mass of the $W$, for $M_{WW} > 300$ GeV,
are markedly different. The region around $\cos\theta = 0$ is clearly more populated in the
Higgsless model as a consequence of a larger fraction of longitudinally polarized bosons.
This is confirmed by the comparison betwen polarization fractions on the right hand side of the two figures.
In \fig{fig:polfrac_MWW_lepcut_noH} the growth of the longitudinal component is quite prominent.
Notice, however, that, even in this extreme model, the left handed fraction is the largest one for all $WW$ 
invariant masses up to about 1800 GeV.

In \fig{fig:Mww300_noHvsSM} we compare the differential cross sections as functions of $M_{WW}$ for the 
Higgsless model (lighter curves) with the corresponding SM results (darker curves).
The full, unpolarized distribution is shown in black/gray. The curves for polarized $W^-$ are given in 
blue, green and red for the left, right and longitudinal polarization, respectively.
The unpolarized cross section for the 
Higgsless model shows the expected enhancement at large invariant mass with respect to the SM.
The comparison of the polarized distributions provides additional information. The differential cross
sections for a left or right polarized $W^-$ are identical in the SM and in the  Higgsless model.
The difference between the full SM result and the Higgsless one 
is fully accounted for by the difference in the
longitudinal component.

The results discussed in the previous sections show that the differential cross section with a definite $W^-$
polarization can be interpreted, up to corrections of a few percent, as the sum of three differential
cross sections in which the positively charged $W$ assumes the three possible polarizations.
Since the results for a left and right polarized $W^-$ indicate that the absence of the Higgs does not 
increase the cross section for a transversely polarized $W^-$ and a longitudinally polarized $W^+$,
the large difference in the cross section for a longitudinally polarized $W^-$ must be attributed to the 
component in which both $W$'s are longitudinal.

\begin{figure}[!tb]
\centering
\subfigure[{Shapes $M_{WW} > 300$ GeV.}\label{fig:shape_SMvsNoH_300}]
{\includegraphics[scale=0.37]{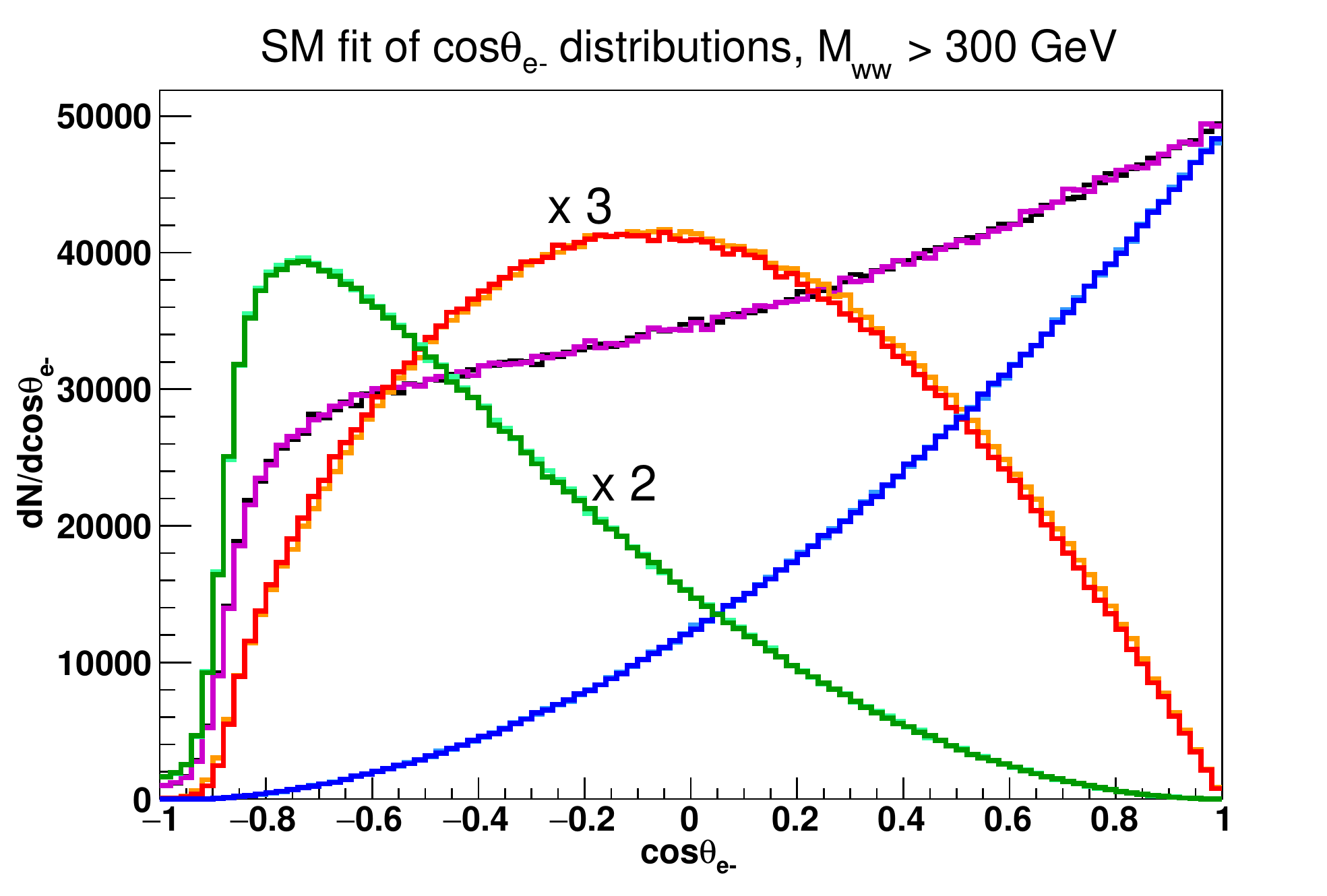}}
\subfigure[Shapes $M_{WW} > 1000$ GeV.\label{fig:shape_SMvsNoH_1000}]
{\includegraphics[scale=0.37]{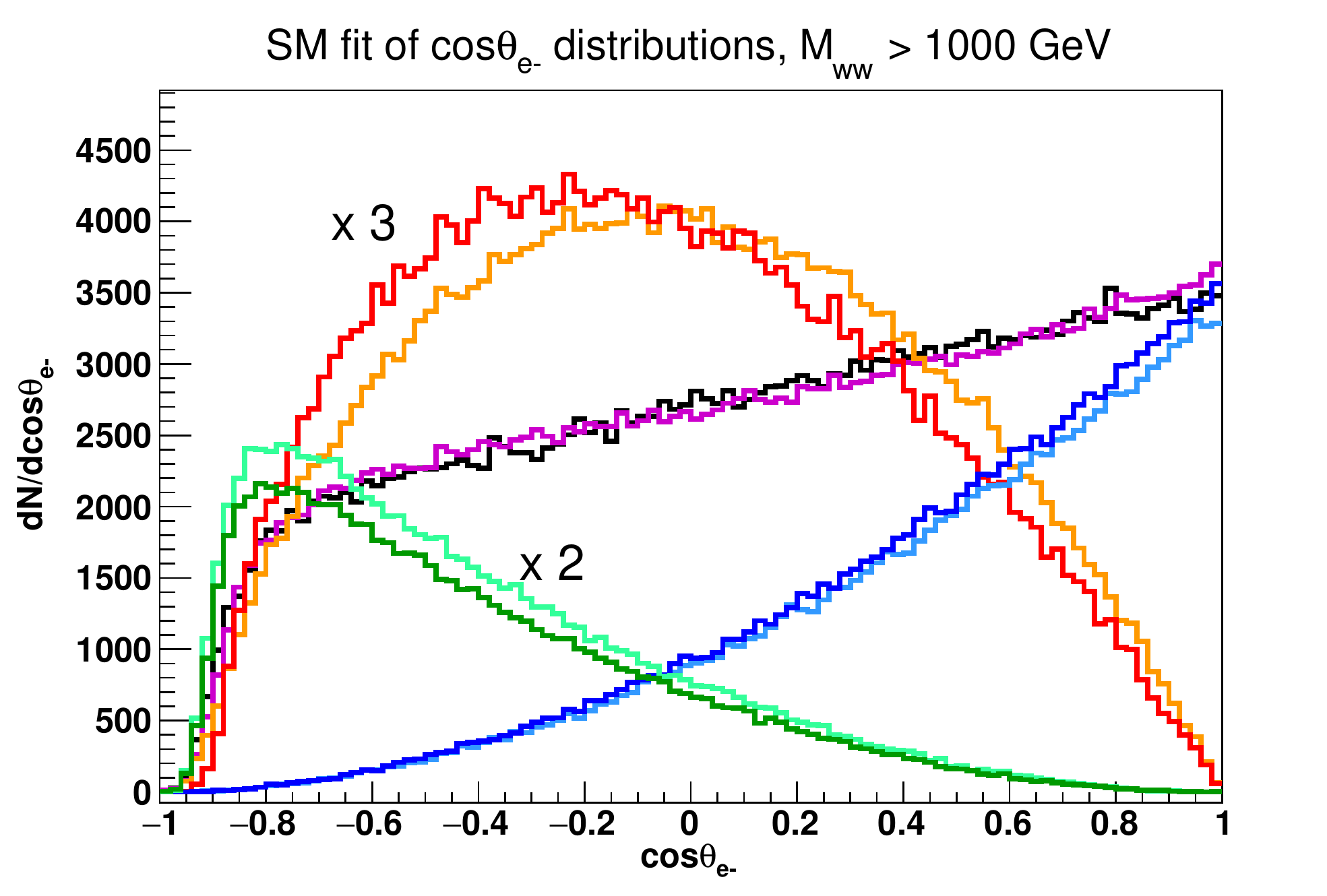}}\\
\subfigure[Shapes 1000 GeV $< M_{WW} < 1100$ GeV.\label{fig:shape_SMvsNoH_1000_1100}]
{\includegraphics[scale=0.37]{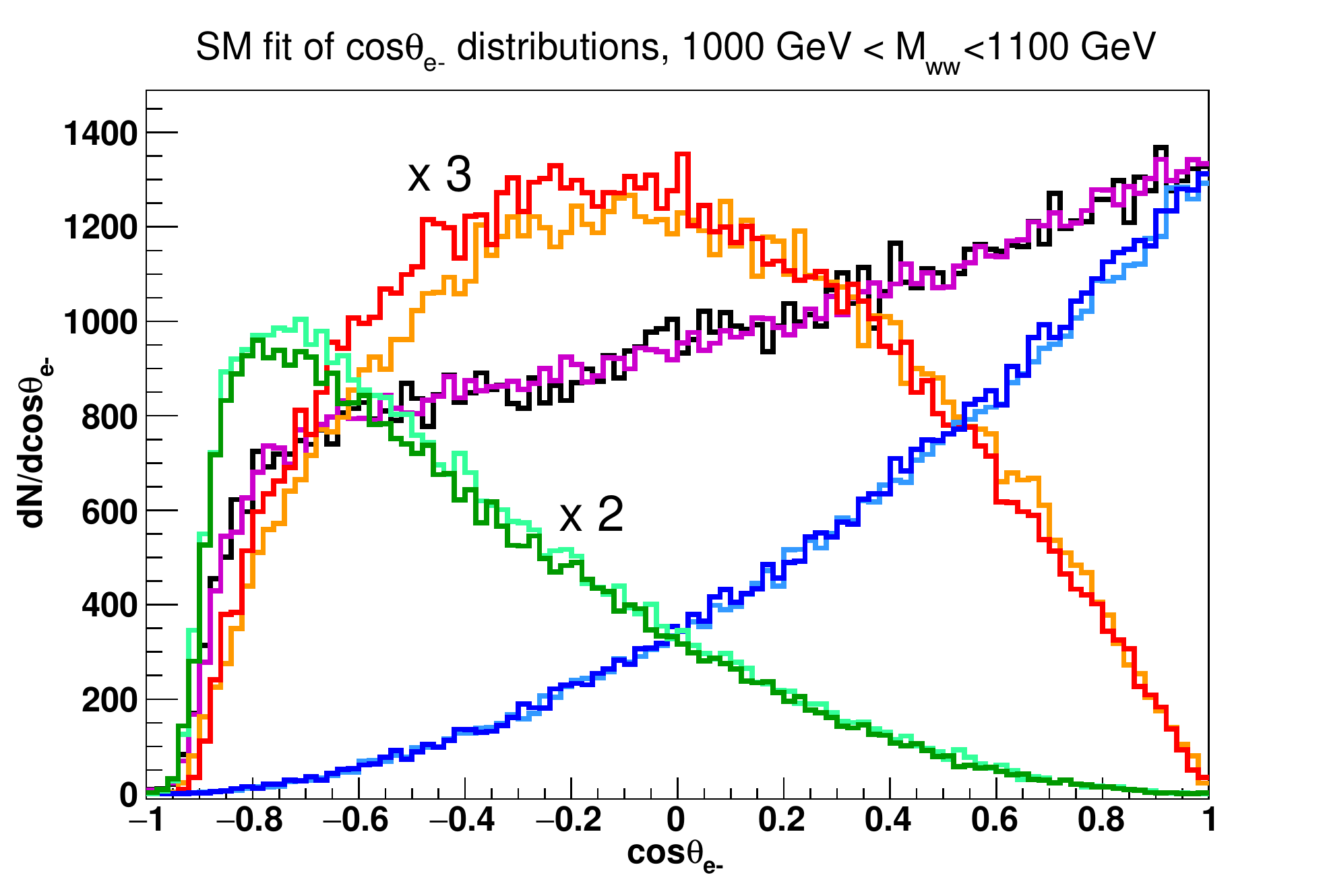}}
\subfigure{\includegraphics[scale=0.37]{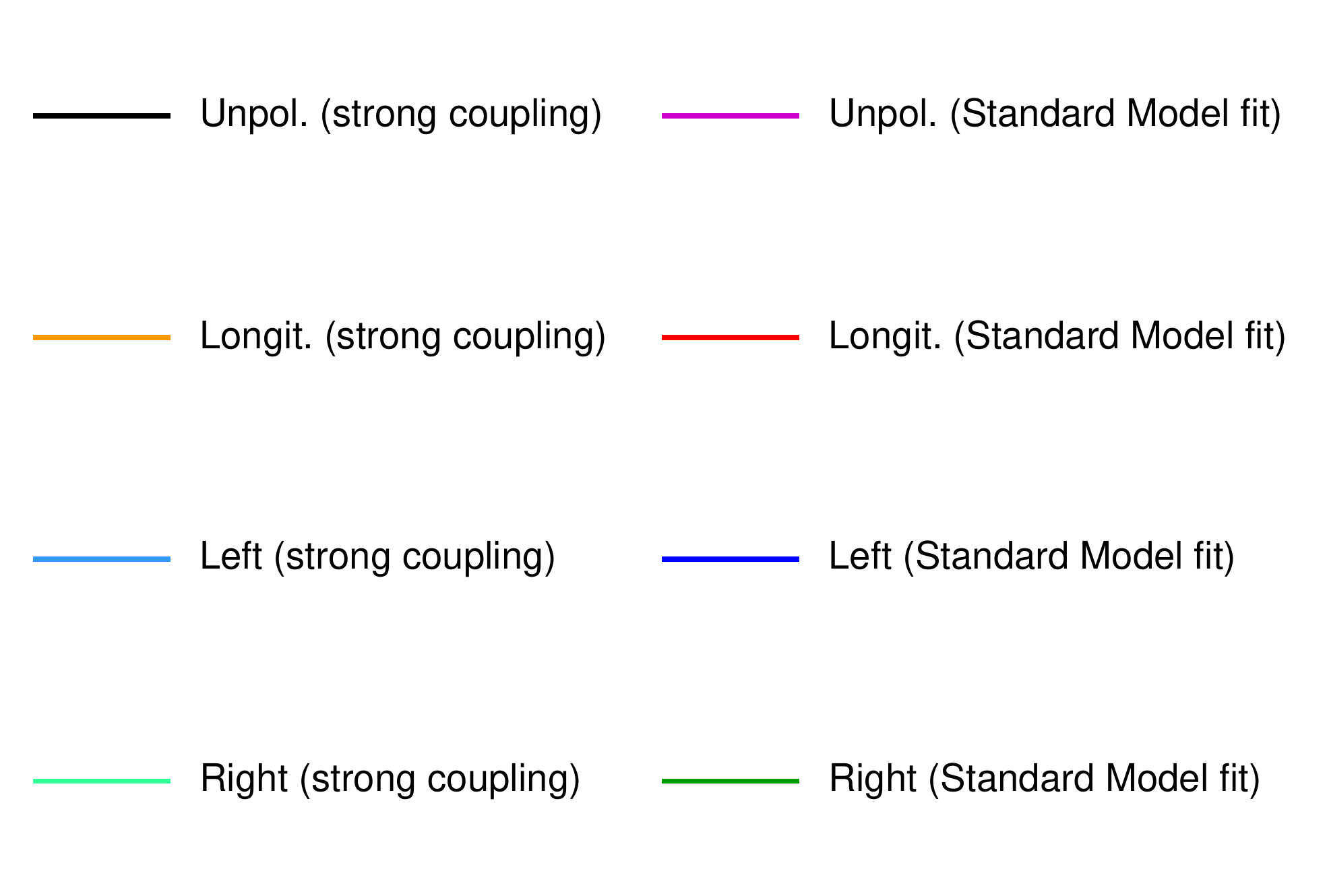}}
\caption{Shape comparison between SM and Higgsless
generation. $p_t^{e}>20$ GeV, $|\eta^{e}| < 2.5$.
The longitudinal and right handed components have been multiplied by three and two, respectively.
}
\label{fig:shape_SMvsNoH}
\end{figure}

We have also examined a $\mathcal{Z}_2$--symmetric Singlet extension of the SM with an additional heavy scalar
 \cite{Silveira:1985rk,Schabinger:2005ei,O'Connell:2006wi,BahatTreidel:2006kx,Barger:2007im,
 Bhattacharyya:2007pb,Gonderinger:2009jp,Dawson:2009yx,Bock:2010nz,Fox:2011qc,Englert:2011yb,
 Englert:2011us,Batell:2011pz,Englert:2011aa,Gupta:2011gd}.
In this model the couplings of the two Higgses to SM particles are proportional to the SM couplings of the Higgs,
multiplied by universal factors:

\begin{alignat}{5}
 g_{xxs} = g_{xxh}^{\text{SM}}(1 + \Delta_{xs}) \qquad \text{with} \qquad 1+\Delta_{xs} =
 \begin{cases}\cos\alpha& s\,=\,\hzero\\ \sin\al&s=\Hzero \end{cases} \label{eq:coupling},
\end{alignat}

\begin{alignat}{5}
 g_{xxs_1s_2} = g_{xxhh}^{\text{SM}}(1 + \Delta_{xs_1}) (1 + \Delta_{xs_2}) \label{eq:coupling2},
\end{alignat}

where $xx$ represents a pair of SM fermions or vectors, and $\alpha$ is the mixing angle.
As a consequence, since only small values of $\sin\alpha$ are allowed
\cite{Pruna:2013bma,Lopez-Val:2014jva,Robens:2015gla}, the width of the heavy Higgs is much
smalller than the width of a SM Higgs of the same mass.
We have taken $\sin\alpha = 0.2$, $m_H = 600$ GeV and $\tan\beta = 0.3$, where $\tan\beta$ is the 
ratio of the
vacuum expectation values of the two neutral scalar fields, which yields
$\Gamma_H = 6.45$ GeV.

With the exception of the mass window in the vicinity of the heavy Higgs resonance all results for the Singlet 
model follow closely those of the SM.

\begin{figure}[!tb]
\centering
\subfigure[{Shapes $M_{WW} > 300$ GeV.}\label{fig:shape_SMvsSingl_300}]
{\includegraphics[scale=0.37]{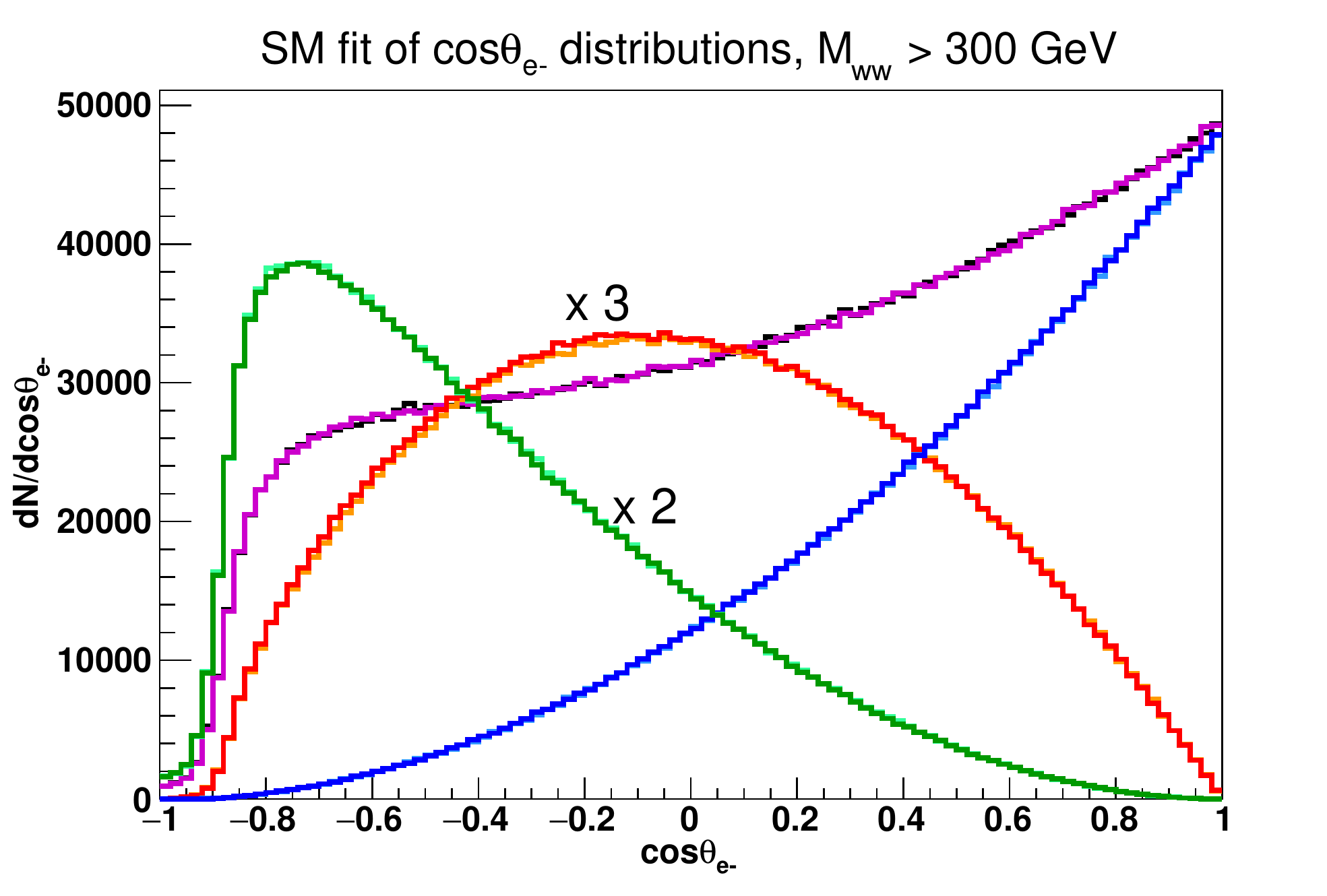}}
\subfigure[Shapes 580 GeV $< M_{WW} < 620$ GeV.\label{fig:shape_SMvsSingl_580_620}]
{\includegraphics[scale=0.37]{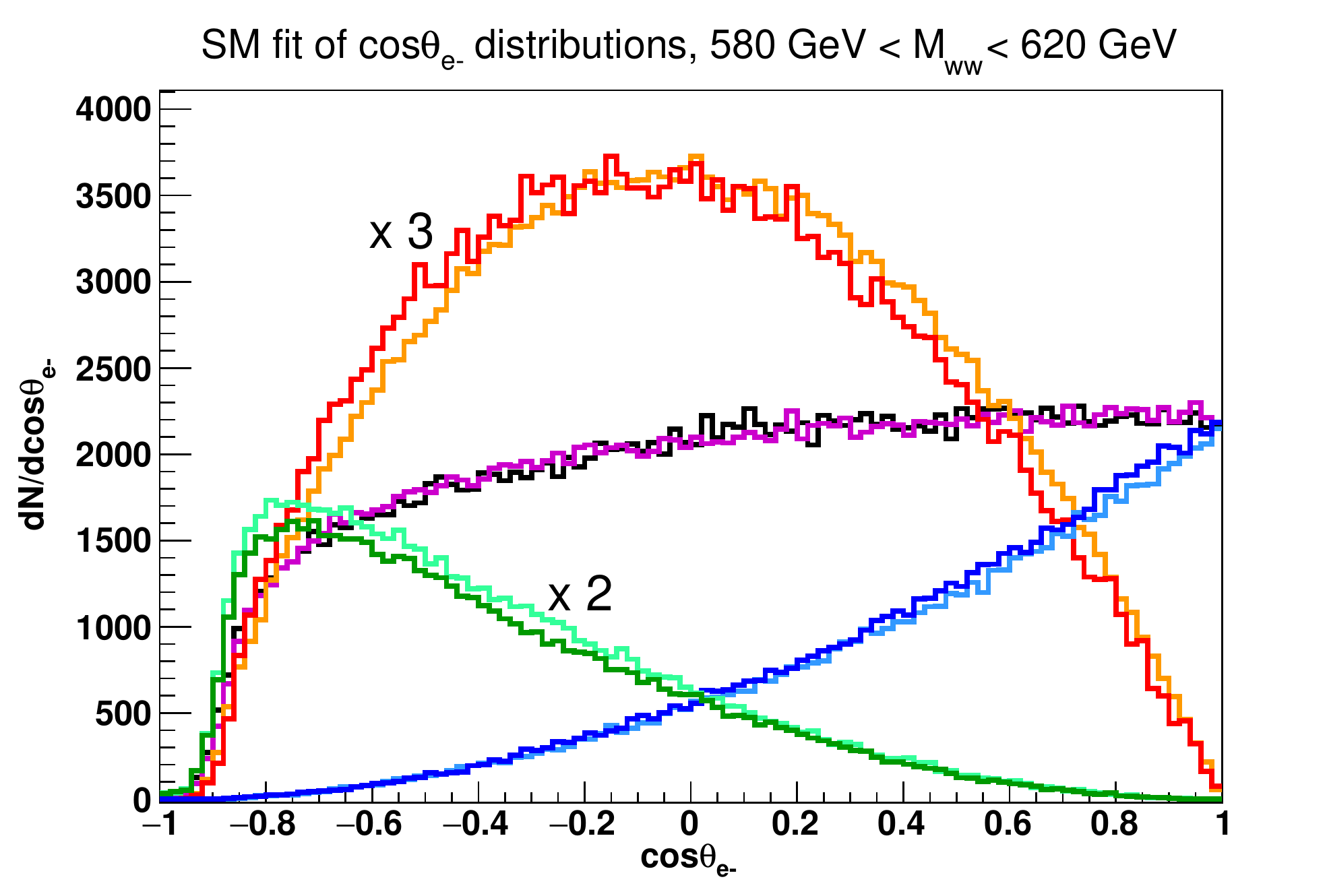}}
\caption{Shape comparison between SM and Singlet
generation. Lines as in \fig{fig:shape_SMvsNoH}. $p_t^{e}>20$ GeV, $|\eta^{e}| < 2.5$.
The longitudinal and right handed components have been multiplied by three and two, respectively.
}
\label{fig:shape_SMvsSingl}
\end{figure}

We have checked that, for each $W$ polarization,
the shape of charged lepton angular distributions, $1/\sigma \cdot  d\sigma/d\cos\theta$,
in both the singlet and the noHiggs models, are in reasonable agreement with
the corresponding SM shape in all 100 GeV
intervals in the $WW$ invariant mass, even though the normalizations can be quite different.
We have also verified the agreement between the sum of singly polarized
distributions and the full result in both cases.

The results are shown in \fig{fig:shape_SMvsNoH} for the Higgsless model,
and in \fig{fig:shape_SMvsSingl} for the Singlet one.
In both figures the black histogram shows the result of the full calculation. The lighter red, green and blue curves
are the singly polarized distributions in the Higgsless and Singlet models.
The darker red, green and blue curves show the results for the longitudinal, right and left handed polarizations,
respectively, of the fit of the full results using the SM templates.
The longitudinal and right handed components have been multiplied by three and two, respectively, to improve 
the overall readability of the plots.

\Fig{fig:shape_SMvsNoH_300} and \fig{fig:shape_SMvsSingl_300} show that, when a large range of invariant masses
is taken into account, the shapes of the polarized components are indeed very similar in the three models.
This is not surprising since the distributions are dominated by low invariant mass events, for which the differences
between the various models are expected to be small.

\Fig{fig:shape_SMvsNoH_1000}, \fig{fig:shape_SMvsNoH_1000_1100} and \fig{fig:shape_SMvsSingl_580_620}
show that, when restricting the comparison
to large invariant masses, the shapes of the left and right handed components
remain quite similar in the various models,
while the shape of the longitudinal component in the Higgsless and Singlet model appears slightly
shifted to large values of $\cos \theta$ in comparison with the SM distribution.
The effect seems to increase for increasing $M_{WW}$, in fact, it is more pronounced in
\fig{fig:shape_SMvsNoH_1000}, which contains events of higher average mass, than in
\fig{fig:shape_SMvsNoH_1000_1100}.

Despite these differences, the fit of both models using SM templates is quite good. 
We have fitted the full result for each model, by means of a $\chi^2$ minimization,
with a linear combination of four normalized functions extracted from the SM distributions:
the shapes for the longitudinal, right and left handed components together with the
shape of the SM interference. The latter is defined as the normalized difference between the unpolarized, full
distribution and the sum of the three singly polarized ones.

In \tbn{table:polfrac}, we show a selection of results.
The rows labeled SM, no Higgs and Singlet show the percent ratio of the singly polarized cross 
sections to the full, unpolarized cross section for each model. The rows labeled Fit give the 
coefficients, in percent, of the four SM shapes obtained from the fit of the angular distribution
of the electron in the corresponding model.

When the fit is extended to the full range of invariant masses,  $M_{WW} > 300$ GeV,
the polarization fractions returned by the fit reproduce the results obtained from the ratio of
the OSP polarized cross sections to the full one.  

\begin{table}[bt]
\begin{center}
\begin{tabular}{@{}ccccc@{}}
\toprule
& \hspace*{2mm} Long. \hspace*{2mm}
& \hspace*{5.8mm} L \hspace*{5.8mm}
& \hspace*{5.8mm} R \hspace*{5.8mm}
& \hspace*{3.8mm} Int. \hspace*{3.8mm}
\\
\midrule
\multicolumn{5}{c}{$M_{WW} > 300$ GeV} \\
\midrule
SM & 21 & 52 & 25 & 2 \\
\hspace*{1mm} no Higgs \hspace*{1mm} & 27 & 48 & 23 & 2 \\
Fit no Higgs & 26 & 48 & 23 & 2\\
\hspace*{1mm} Singlet \hspace*{1mm} & 23 & 51 & 24 & 2 \\
Fit Singlet & 23 & 51 & 24 & 2\\
\midrule
 \multicolumn{5}{c}{$M_{WW} > 1000$ GeV} \\
\midrule
SM & 15 & 58 & 22 & 4 \\
\hspace*{1mm} no Higgs \hspace*{1mm} & 35 & 45 & 17 & 3 \\
Fit no Higgs & 35 & 47 & 15 & 2\\
\midrule
 \multicolumn{5}{c}{580 GeV $ <M_{WW} < 620$ GeV} \\
\midrule
SM & 19 & 53 & 25 & 3 \\
\hspace*{1mm} Singlet \hspace*{1mm} & 42 & 38 & 18 & 2 \\
Fit Singlet & 42 & 40 & 17 & 2\\
\bottomrule
\end{tabular}
\end{center}
\caption{The rows labeled SM, no Higgs and Singlet show the percent ratio of the polarized cross sections to the full
result. The rows labeled Fit give the 
coefficients, in percent, extracted from a $\chi^2$ fit of a linear combination of the SM shapes for the longitudinal,  
left, right handed components and the interference to the full unpolarized angular distribution in the corresponding
model.
}
\label{table:polfrac}
\end{table}

For the Higgsless model, we have investigated the agreement between these two determinations of
the polarization fractions in each 100 GeV interval in the $WW$ invariant mass.
The accord is quite good in each bin,
even though the normalizations become progressively different as the invariant mass grows.

The $M_{WW} > 1000$ GeV range, where the longitudinal polarization fractions in the Higgsless model is more 
than twice the SM one, gives the poorest agreement. However,
the discrepancy between the polarization fractions from the fit and those from the ratio of
cross sections is of the order of 10\%  at most.

The fitted polarization fractions for the Singlet model in the heavy Higgs peak region, 
580 GeV $ <M_{WW} < 620$ GeV, are in very good agreement with those estimated from the 
ratio of the singly polarized cross sections to the full result in that bin.

The results presented in this section suggest that it is possible to extract the polarization fractions of the $W$ in an 
almost model--independent way, exploiting the 
similarity of the shapes of polarized $\cos\theta_e$ distributions for different underlying theories, in most 
kinematic regions.

The angular distributions we have discussed, are difficult to measure, particularly when both $W$'s decay
leptonically. However, the same approach can be applied to other kinematic variables whose distributions discriminate
among the different
polarizations, as for instance the invariant mass of the charged lepton pair and the transverse momentum of the charged
lepton, as shown in \fig{fig:var_lepcut}.
In order to enhance the accuracy of the obtained results, 
more refined multivariate fit methods could be investigated.

\section{Conclusions}
In this paper we have investigated the possibility of defining cross sections for processes with polarized $W$ bosons, 
including off shell and non resonant effects.

We have proposed a method which
is based on the observation that the set of doubly resonant diagrams, suitably projected on shell to preserve
gauge invariance, approximates well the full result and allows an expansion in terms of amplitudes
in which each final state $W$ has a definite polarization. This procedure agrees with the standard approach based
on Legendre polynomials in the absence of cuts on the decay leptons. When acceptance cuts are imposed on the
 leptons, and the Legendre polynomials procedure fails, it is possible to extract the polarization fractions using
singly polarized SM Monte Carlo templates.
These same templates can be used to measure the polarization of the $W$ with reasonable accuracy even if new
physics is present.

\section*{Acknowledgements}
Discussions with Pietro Govoni have been invaluable and are gratefully ac\-knowledged. 
The authors would like to acknowledge the contribution of the COST Action CA16108.

\begin{appendices}

\section{Numerical effects of 
different approximations in the large mass, large transverse momentum region. }
\label{appendixa}

In this appendix we exemplify how different, gauge invariance violating  approximations
in the calculation of amplitudes may produce unreliable results in the large energy regime.

We study two instances. The first one is the naive way to isolate the resonant contribution to $W^+W^-$
production in VBF, by simply dropping all other diagrams and requiring the invariant mass of each $\ell \nu$ pair to 
be close to the $W$ mass.
In the second one, we examine quantitatively the relevance of non null vector boson widths in the 
OSP projected amplitudes.
All the results in this Appendix refer to the Standard Model.

\subsection{Resonant contributions and gauge cancellations}
\label{appendixa1}

\begin{figure}[!tbh]
\centering
\subfigure[{$M_{WW}$}
\label{fig: MWW_RESnoLepCuts}]{\includegraphics[scale=0.37]{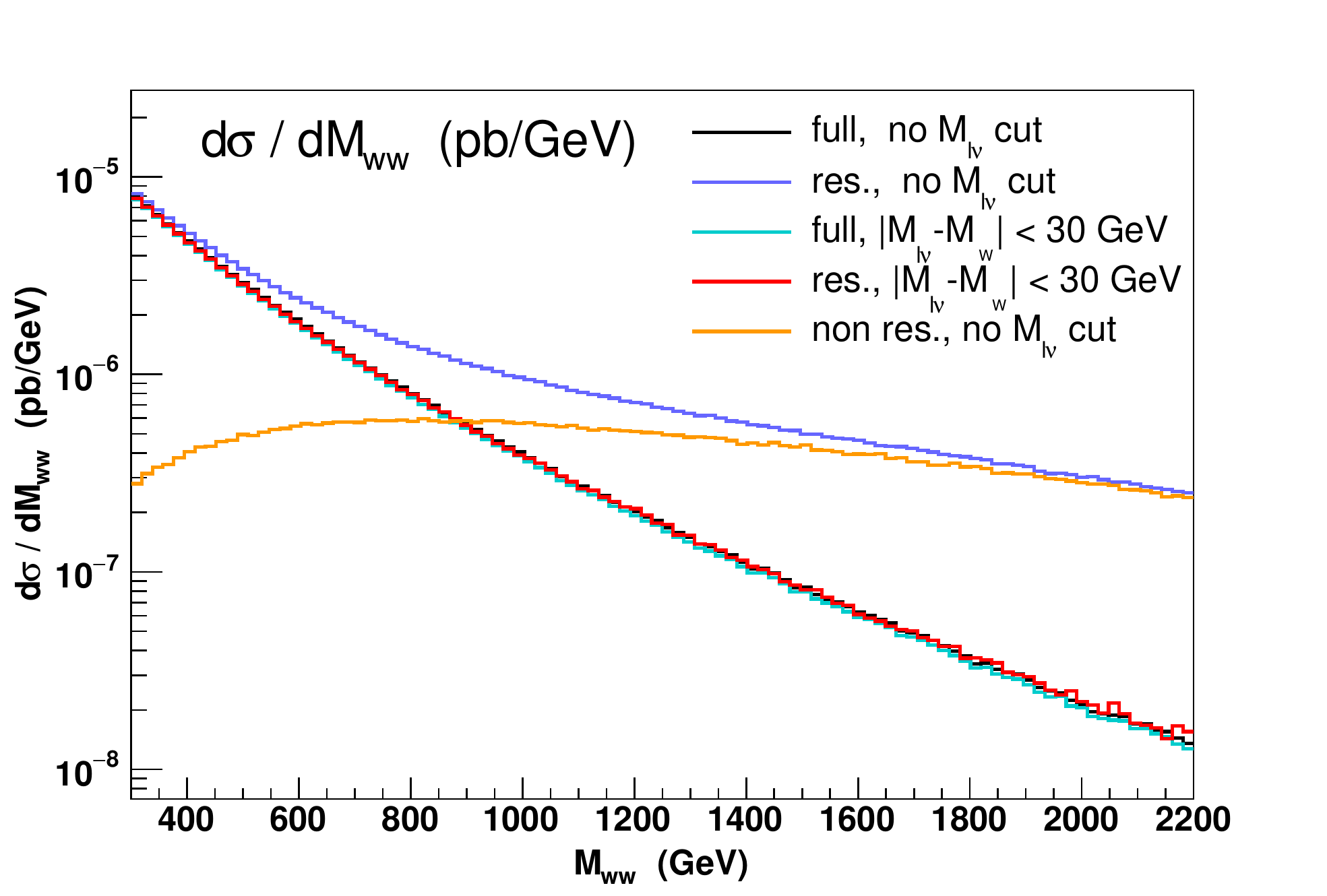}}
\subfigure[{$p_t^{W^-}$}
\label{fig: ptwm_RESnoLepCuts}]{\includegraphics[scale=0.37]{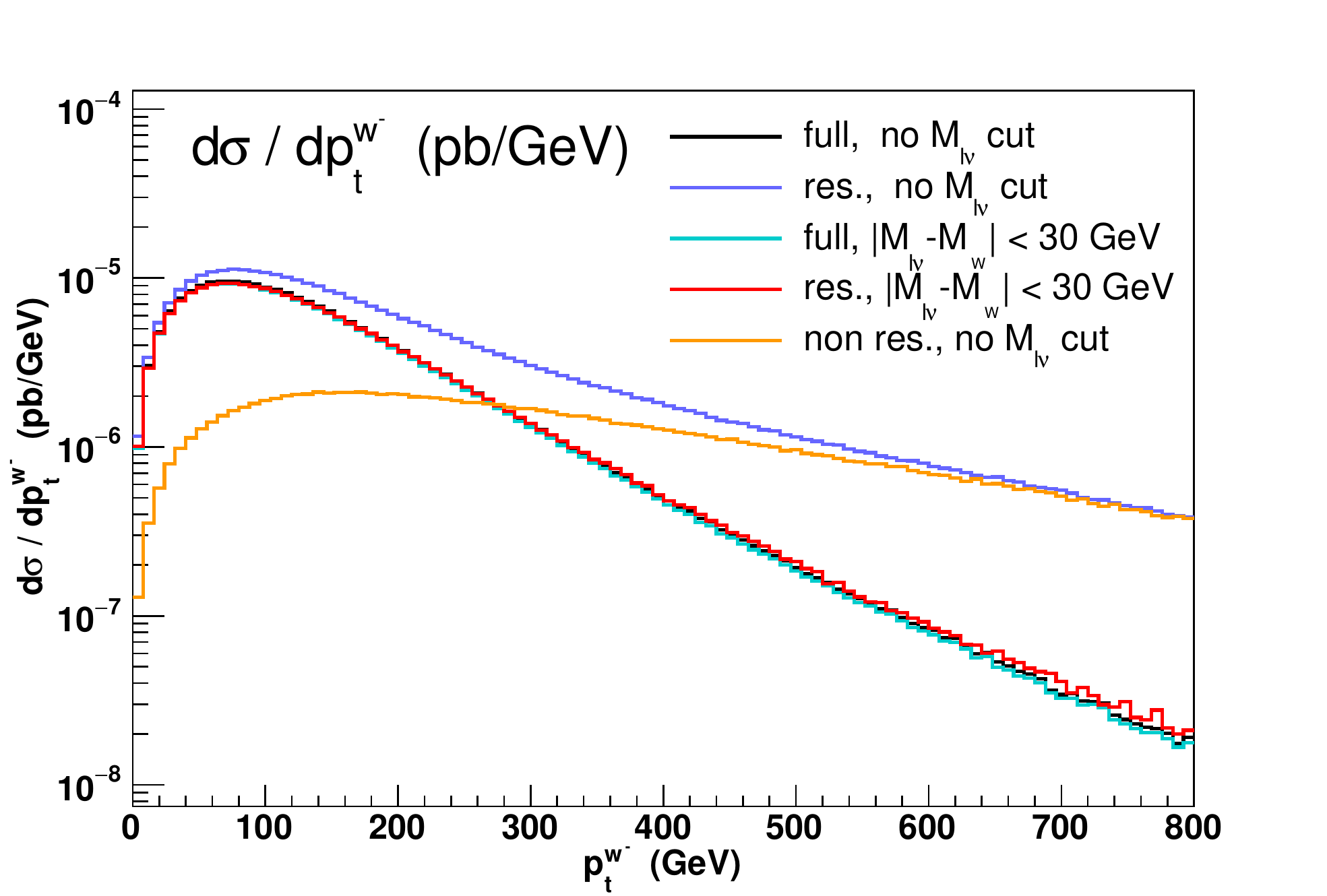}}\\
\subfigure[{$M_{WW}$}
\label{fig: MWW_OSPnoLepCuts}]{\includegraphics[scale=0.37]{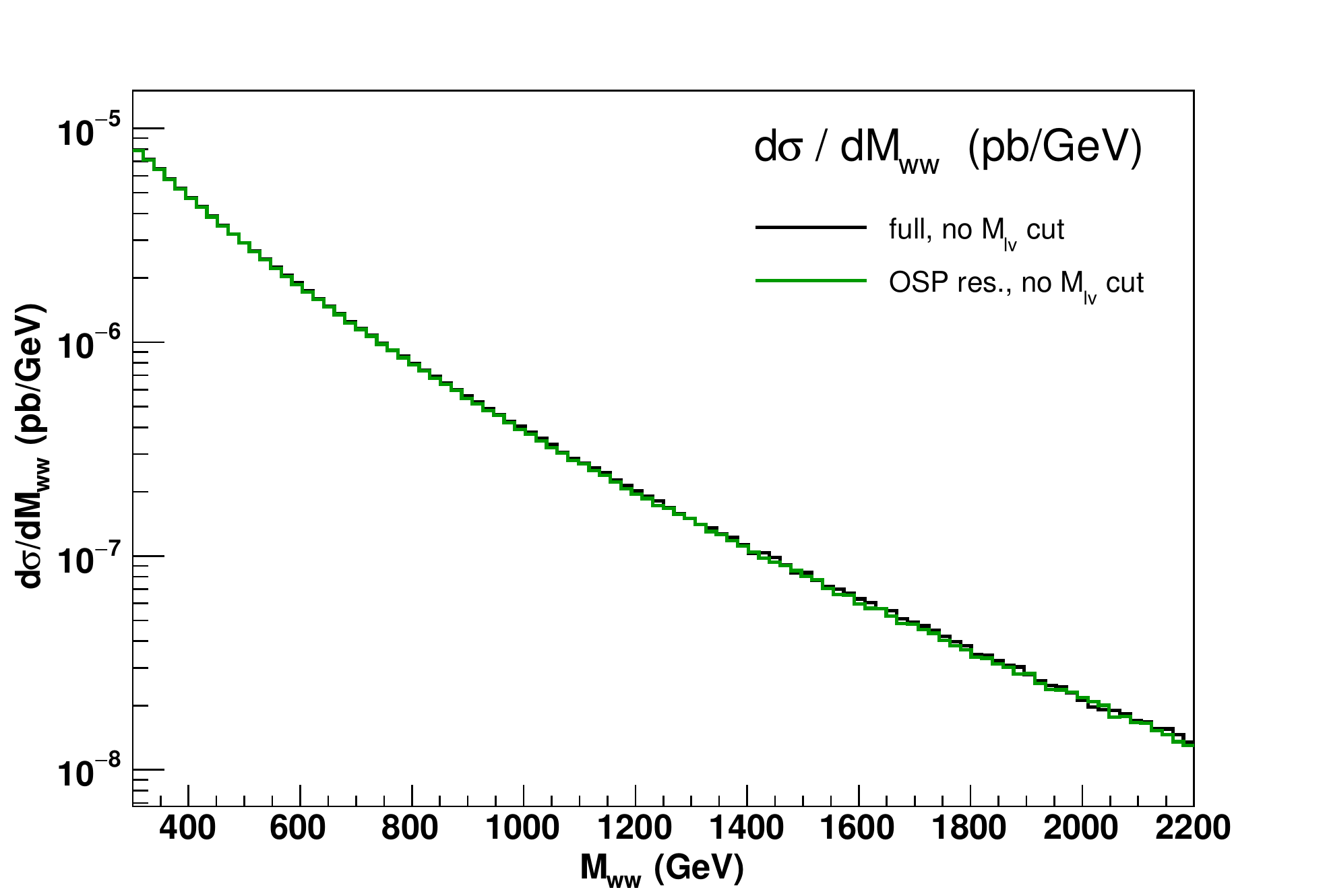}}
\subfigure[{$p_t^{W^-}$}
\label{fig: ptwm_OSPnoLepCuts}]{\includegraphics[scale=0.37]{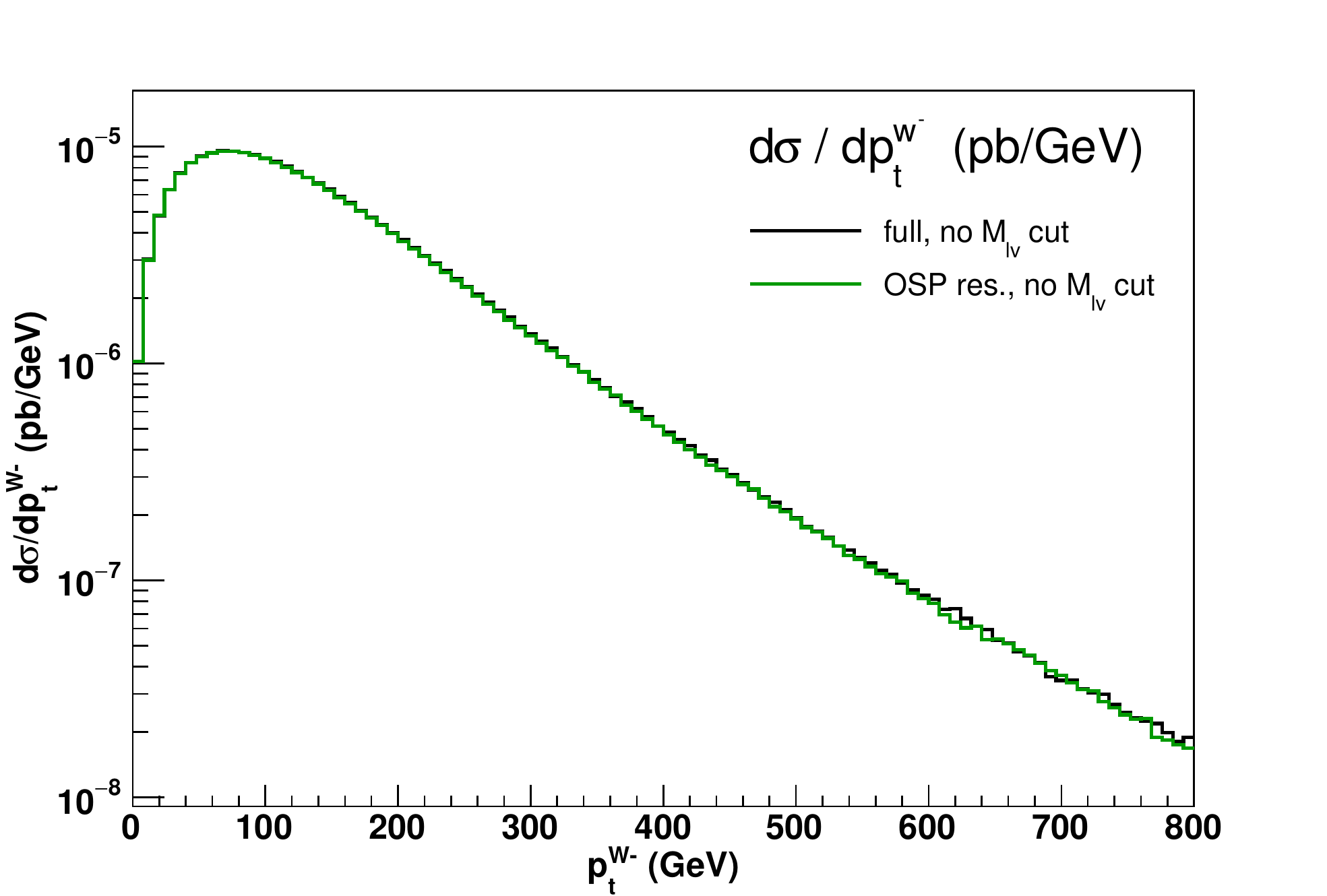}}
\caption{Differential cross sections as a function of $M_{WW}$ and $p_t^{W^-}$ under different assumptions.
In black the full results without cuts on the lepton system; in light blue the full result with
the constraint $\vert M_{\ell\nu}-M_W\vert <$ 30 GeV for each ${\ell\nu}$ pair. The blue--violet, red and green
histograms are computed using only the resonant diagrams. The blue--violet curves are obtained
without lepton cuts; the red ones requiring $\vert M_{\ell\nu}-M_W\vert <$ 30 GeV;
the green plots are the result of the OSP without any restriction on the lepton--neutrino system.
}\label{fig:DistNoLepCuts}
\end{figure}

\begin{figure}[!b]
\centering
{\includegraphics[scale=0.39]{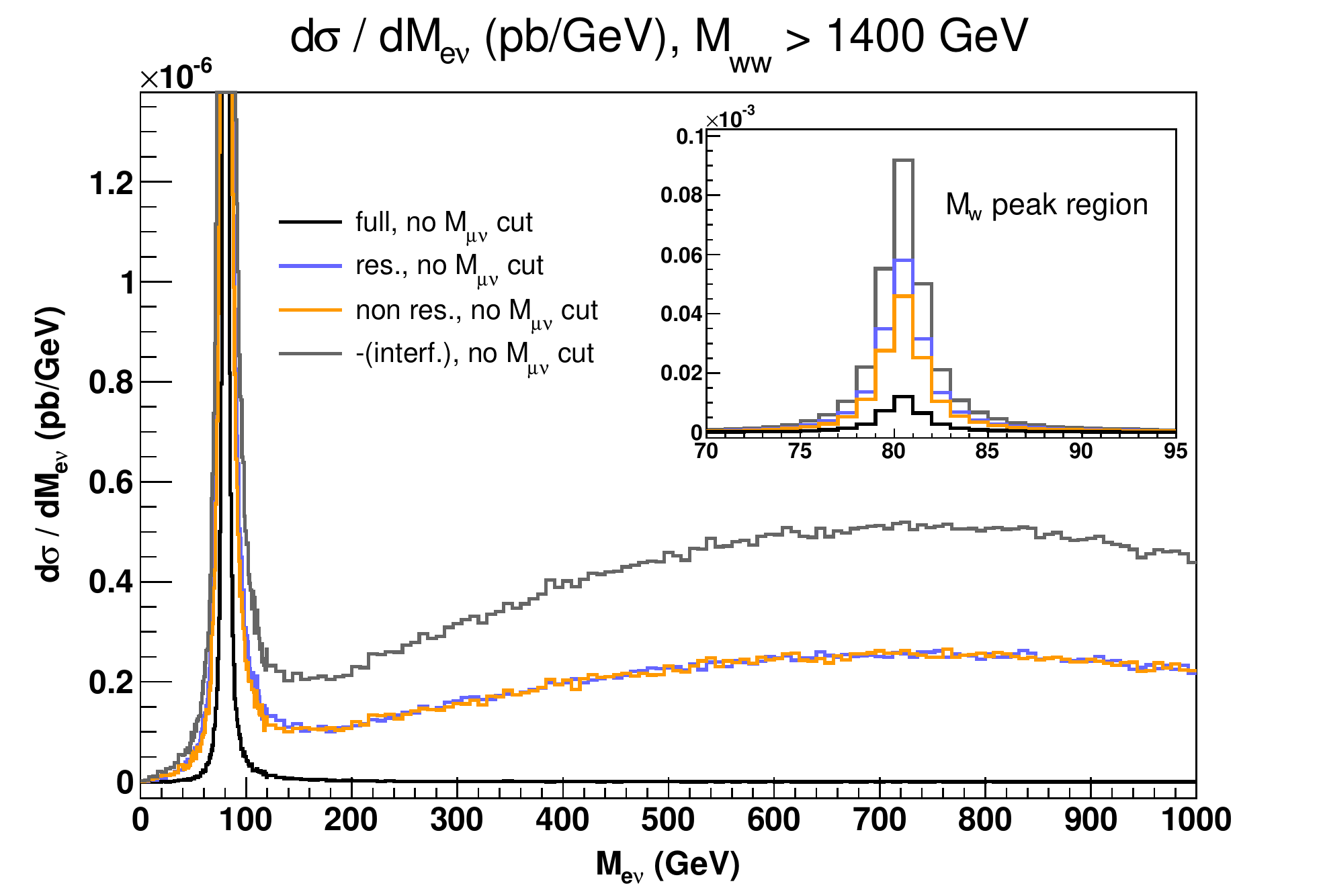}}
\caption{Invariant mass distribution of the $e^-\bar{\nu}_e$ pair for $M_{WW} > 1400$ GeV in the SM results. No cuts on leptons.}
\label{fig:Mlv_v3}
\end{figure}

The first approximation is based on the intuitive notion that the closer the invariant
mass of the lepton--neutrino system to the mass of the $W$ boson, the more dominant the resonant diagrams are
\footnote{This contribution can be computed in MadGraph5 generating the process:\\
\hspace*{0.6cm}
\texttt{p p $>$ j j w+ w-, w+ $>$ mu+ vm, w- $>$ e- ve{\raise.17ex\hbox{$\scriptstyle\mathtt{\sim}$}}}. }.
Therefore, it is reasonable to expect that by restricting the invariant mass of each $\ell \nu$ pair around $M_W$,
the non resonant diagrams can be neglected.

In order to have a reference point for the size of  possible effects, we have computed separately,
without any constraint on the invariant masses of the lepton--neutrino pairs,
the contribution of the doubly resonant diagrams and of the complementary set of diagrams, that is, those
which are singly resonant or non resonant, which we will refer to as non resonant for simplicity.

Then, we have restricted the mass of  electron neutrino pair to a neighborhood of the $W$ mass.
Specifically, we have required
$\vert M_{\ell\nu}-M_W\vert <$ 30 GeV as a compromise between minimizing the non resonant contribution and
constraining too much a variable which cannot be measured precisely at hadron--hadron colliders.
We have also tried narrower intervals, finding similar results.

In the upper row of \fig{fig:DistNoLepCuts}
we present the differential distribution of the four lepton mass (left) and of the
transverse momentum of the $\ell^-\nu$ pair (right). The result obtained from the full
calculation, without any restriction on the lepton--neutrino system is shown in black.
We compare it with the full results obtained
restricting the mass of each $\ell\nu$ pair to $\vert M_{\ell\nu}-M_W\vert <$ 30 GeV (light blue). Clearly, the two
distributions agree quite well
over the full range in $M_{WW}$ and $p_t^{W^-}$, showing that, as expected, in most of the
events, the mass of the $\ell\nu$ system is close to $M_{W}$.
In the same plots we also show the results, obtained over the full leptonic phase space,
taking into account only the doubly resonant diagrams (blue--violet)
and those obtained taking into account only the non resonant ones (yellow).
Finally, the distribution
obtained from the doubly resonant diagrams with the additional constraint 
$\vert M_{\ell\nu}-M_W\vert <$ 30 GeV is given in red.

For completeness, in the lower part of  \fig{fig:DistNoLepCuts} we show that the distributions from the full 
calculation agree very well, for both variables, 
with those produced by the coherent sum of OSP projections.

The comparison between the black histogram, the blue-violet and the yellow curves demonstrates
that the resonant and non resonant diagrams interfere strongly and that, in the absence of cuts on $M_{\ell\nu}$, 
the set of resonant diagrams does not reproduce
well the correct result, the discrepancy increasing with increasing invariant mass or transverse
momentum of the lepton--neutrino pair.

\begin{table}
\centering
\begin{tabular}{@{}p{48mm}cccc@{}}
\toprule
 \centering Region &  \centering Full & Resonant &  Non-resonant &  Interference \\
\midrule
$50\, \textrm{GeV} < M_{e^-\nu} < 110\, \textrm{GeV}$,\newline $50\, \textrm{GeV} < M_{\mu^+\nu} < 110\, \textrm{GeV}$ & 3.975$\cdot 10^{-5}$ & 4.233$\cdot 10^{-5}$ & 2.670$\cdot 10^{-6}$ & -5.244$\cdot 10^{-6}$ \\
\cmidrule{1-1}
$50\, \textrm{GeV} < M_{e^-\nu} < 110\, \textrm{GeV}$,\newline $M_{\mu^+\nu} > 110\, \textrm{GeV}$  &1.050$\cdot 10^{-6}$ &1.574$\cdot 10^{-4}$ &  1.554$\cdot 10^{-4}$& -3.118$\cdot 10^{-4}$\\
\cmidrule{1-1}
$M_{e^-\nu} > 110\, \textrm{GeV}$,\newline $50\, \textrm{GeV} < M_{\mu^+\nu} < 110\, \textrm{GeV}$  &1.065$\cdot 10^{-6}$ & 1.587$\cdot 10^{-4}$& 1.584$\cdot 10^{-4}$& -3.161$\cdot 10^{-4}$\\
\cmidrule{1-1}
$M_{e^-\nu} > 110\, \textrm{GeV}$,\newline $M_{\mu^+\nu} > 110\, \textrm{GeV}$  & 3.751$\cdot 10^{-8}$& 1.693$\cdot 10^{-4}$& 1.693$\cdot 10^{-4}$&-3.386$\cdot 10^{-4 }$\\
\bottomrule
\end{tabular}
\caption{Cross--sections (pb) in different $M_{e^-\nu},\,M_{\mu^+\nu}$ regions, for $M_{ww} > 1400$ GeV.}
\label{tab:M_lv_numbers}
\end{table}

However, as shown by the red histogram, if the mass of the individual $\ell\nu$ pair is forced to be 
sufficiently close to the $W$ mass, the set of resonant diagrams describes rather well the full results, 
at least on a logarithmic scale.

Nevertheless, if  the comparison between the full calculation and the naive approximation is pushed to the 
high invariant mass, high $p_t$, region, the latter fails to reproduce the full result. 

In \fig{fig:Mlv_v3} we show the distribution of the invariant mass of the $M_{e^-\nu}$ pair for 
$M_{WW} > 1400$ GeV. No additional cut is imposed on any lepton. The region close to $M_W$ is 
highlighted in the insert.
The color code is as in \fig{fig:DistNoLepCuts}:
the black curve refers to the full result; the blue--violet 
histogram is computed using only the resonant diagrams; the yellow curve
is obtained from the non resonant set of diagrams. The gray curve shows the interference contribution, with 
opposite sign with respect to its actual value, and refers to the difference between the full result and the
sum of the resonant and non resonant one.

Two features leap to the eye: the large interference between the resonant and non resonant diagrams, even 
in the vicinity of the $M_W$ peak and the enhancement at large invariant masses which, while about two 
orders of magnitude smaller than the peak, extends over a very large region.

Additional information is provided in \tbn{tab:M_lv_numbers} which shows the total cross section in 
different zones in the $M_{e^-\nu}$, $M_{\mu^+\nu}$ plane, separating the region close to the mass of the 
$W$ from the large mass one.

\Tbn{tab:M_lv_numbers} shows a discrepancy of about 5\% between the full result and the doubly 
resonant cross section already when both lepton neutrino pairs are close to the $W$ mass shell.
This shows, as anticipated, that, in this regime, restricting the mass of both pairs to within 30 GeV from the
$W$ mass is not enough to reproduce the full results using only the resonant contribution.

Furthermore, for both the resonant diagrams and the non resonant ones, the cross section in the  
regions in which one of the pairs is in the vicinity of the $W$ peak while the 
other is outside this region and in the region in which both pairs are off shell are large, about four times
larger than the full cross section when both lepton--neutrino pairs are nearly on shell.
However, in each of these three regions, the contribution of the resonant diagrams and the contribution
from non resonant ones cancel each other to better than 1 percent when only one of the pairs has a mass 
close to $M_W$, and to about 2 per mill when both are off shell, leading to the expected
physical distribution dominated by the Breit--Wigner peak.

The enhancement at large invariant masses in the $M_{e^-\nu}$ distribution in \fig{fig:Mlv_v3} can be
qualitatively
understood as follows. Let's consider the subset of the doubly resonant diagrams in which vector bosons 
scatter among themselves as in 
\fig{fig:VBSsgn_diag}. These are in one to one correspondence with the set of diagrams which describe the 
scattering among on shell vector bosons. Furthermore let's keep fixed the initial and final quark momenta,
so that the space--like invariant mass of each of the two virtual $W$'s entering the scattering and 
the total mass of the two bosons which decay to the final state leptons are also fixed. 

In the scattering between longitudinally polarized vector bosons, the leading term of each diagram without 
Higgs exchange grows as $s^2$.  The sum of all these diagrams however grows only like $s$,
because the sum of the leading terms turns out to be proportional to $s + t + u$, which corresponds
to  the sum of the masses squared of the bosons partecipating in the scattering, decreasing by one unit the 
degree of divergence. 
 
The dominant behaviour for the corresponding diagrams with off shell vector bosons can be extracted
substituting 
\begin{equation}
-g^{\mu\nu} + \frac{k^{\mu}k^{\nu}}{M^2} \rightarrow  \varepsilon^{\mu}_0
\varepsilon^{\nu^*}_0
\end{equation}
in each propagator, that is, taking into account the contribution with all intermediate vector bosons 
longitudinally polarized.

At high energy, $\varepsilon^{\mu}_0(k) = k^\mu/M + \mathcal{O}(1)$. When the polarization vectors 
act on the external fermion lines, the term proportional to the momentum gives zero and no enhancement 
is produced.
The remaing part of each diagram in the set under consideration, has the same analytic expression of the 
corresponding on shell diagram,
with the difference that the boson momenta are not on their mass shell.

The sum of the leading contributions again reduces to $s + t + u$, but in this case
the external ``masses''  do not coincide
with their on shell values, and can become large when the time--like vector bosons are highly off shell,
compensating the effect of the large denominators in the vector propagators.
Eventually this enhancement will be cut off by phase space constraints for very large values of 
$M_{e^-\nu}$.

\begin{figure}[!tb]
\centering
\includegraphics[scale=0.39]{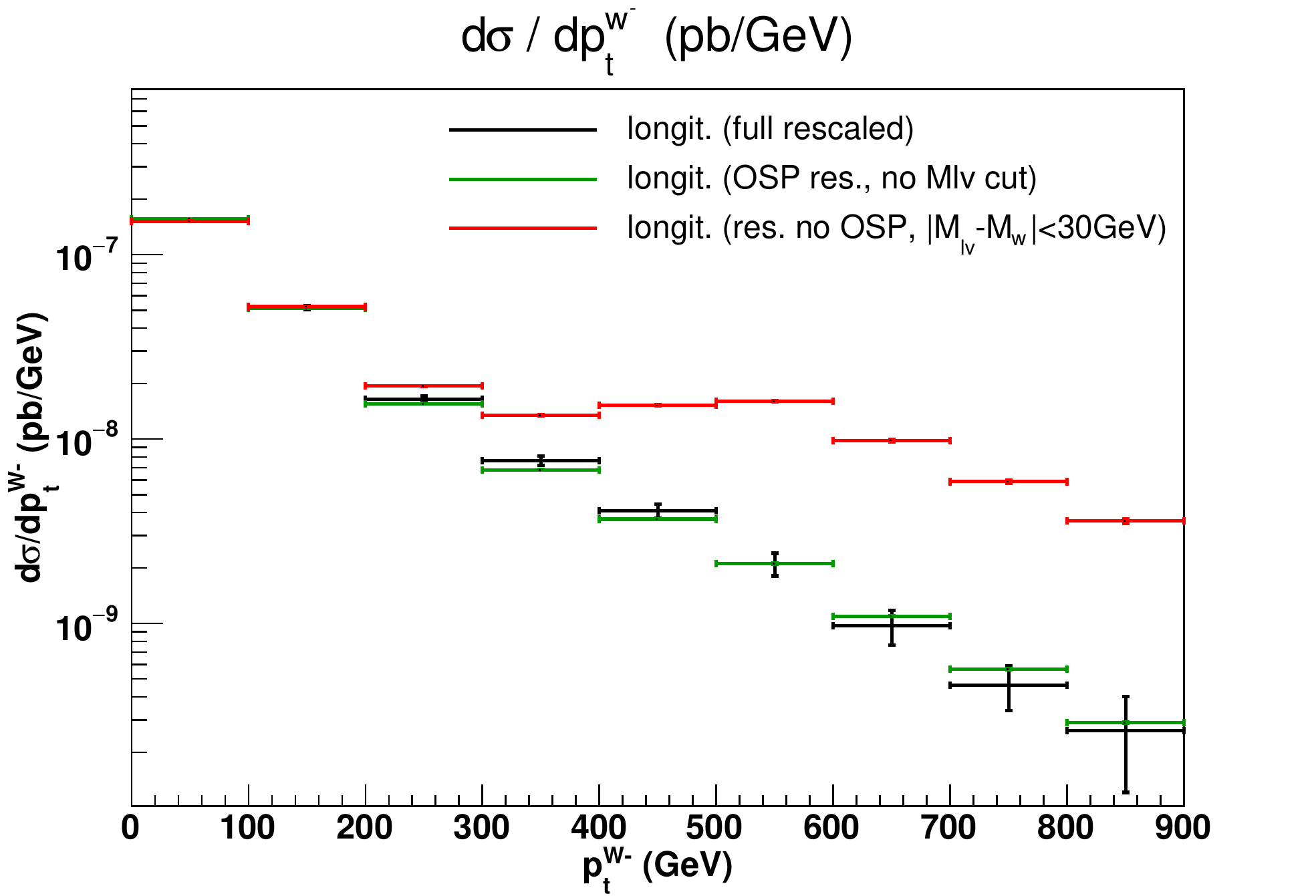}
\caption{Transverse momentum distribution of longitudinally polarized $W^-$ for $M_{WW}>$ 1000 GeV. 
The full result is shown in black; the OSP one, in green.
The red points are obtained neglecting non resonant diagrams and requiring
 $\vert M_{\ell\nu}-M_W\vert <$ 30
GeV.
The red and green results have been obtained with a fixed polarization for the negatively charged $W$.
The black points have been computed from a generation of the full process, extracting the 
longitudinal fraction $f_0$ in each transverse momentum bin through a Legendre expansion,
cfr \sect{sec:Wpol_decay}.
}
\label{fig:distrib_ptwm_longit_mww1000}
\end{figure}

As a further proof that all issues related to anomalous behaviours at high energy are connected with the 
longitudinal polarization of the vector bosons, we present, in \fig{fig:distrib_ptwm_longit_mww1000}, the 
transverse momentum distribution of  the longitudinally polarized $W^-$
for events with $M_{WW}>$ 1000 GeV. 
The full result is shown in black; the OSP one, in green.
The red points are obtained neglecting non resonant diagrams and requiring
 $\vert M_{\ell\nu}-M_W\vert <$ 30
GeV. In all cases the $W^+$ is unpolarized and no cut on the leptonic variables is applied.
While the OSP prediction agrees with the full result over the whole range,
the curve obtained
requiring $\vert M_{\ell\nu}-M_W\vert <$ 30 GeV overshoots the true result by a large factor for
transverse momenta above 300 GeV, 
confirming that this naive approximation cannot be relied on in the 
high energy regime.

\subsection{Vector boson widths in the OSP method}
\label{appendixa2}

We now move to the second example, namely computing the OSP amplitudes in the CMSc scheme, instead of setting
the weak bosons widths to zero and using real couplings.
Typically, for an amplitude without internal vector bosons which can
go on mass shell, the difference between the
CMSc and setting all widths to zero is small, of order $\mathcal{O}(\Gamma/M)= \mathcal{O}(\alpha)$.
However, when the energy of the vector bosons grows large, these differences can become significant, 
because any deviation from a fully gauge invariant treatment of the boson widths spoils 
the cancellation of terms which are proportional to the energy.

In order to examine quantitatively the relevance of these gauge violating effects in VBS,
we study the single process $uu \rightarrow uu \,e^-\bar{\nu_e} \,\mu^+\nu_\mu$,
for large invariant masses of the four lepton system, $M_{WW} > 1000$ GeV, large transverse momentum
of the $e^-\bar{\nu_e}$ pair, $ p_t^{W^-} > 800$ GeV. 
The mass of each pair is required to be very close to the $W$
mass, $\vert M_{\ell v}-M_W \vert < 0.2$ GeV.
No further cut on the charged leptons are applied.

A full calculation in the Complex Mass Scheme gives a cross section of $0.641(3)\cdot 10^{-8}$ pb, 
which we regard as the true result.
We then recompute the cross section restricting the calculation to the doubly resonant diagrams and 
projecting them on mass shell,
as discussed in \sect{sec:resonant}, employing two different prescriptions for masses and couplings.
Using complex masses and couplings, as in the CMSc, yields
$0.754(2)\cdot 10^{-8}$ pb, which is not compatible with the correct value.
However, when all masses and couplings are taken to be real, as advocated in \sect{sec:resonant},
the result is
$0.650(2)\cdot 10^{-8}$ pb, in excellent agreement with the cross section obtained from
the full calculation.
This proves again, in our opinion, that failing to maintain gauge invariance could indeed lead to significantly
flawed results. 
Obviously, it does not prove that any gauge invariant calculation would reproduce the correct answer. 
In particular, the correctness and usefulness of the projection procedure needs to be verified in each case.

Gauge effects are particularly strong for longitudinally polarized vector bosons.
Therefore, we have also computed
the on shell projected cross sections for a longitudinally polarized $W^-$:
using complex masses and couplings we obtain $1.282(5)\cdot 10^{-9}$ pb, while,
taking them to be real gives
$0.202(1)\cdot 10^{-9}$ pb, which differ by a factor of about six.
Notice that the difference between these two cross sections is equal, within errors,
to the difference between the
corresponding unpolarized results, suggesting that the discrepancy between the results for longitudinally
polarized $W^-$ accounts for the totality of the difference in the unpolarized case.

Since no cuts on the charged lepton is imposed, the cross section for a longitudinally polarized
$W$ boson can also be extracted from the full distribution, projecting on Legendre polynomials. This yields
$0.211(5)\cdot 10^{-9}$ pb, confirming the OSP result with all internal widths set to zero.

\end{appendices}

\newpage


\bibliographystyle{JHEP}

\bibliography{Wpol_VBF_jhep}

\end{document}